\RequirePackage{fix-cm}
\documentclass[smallextended]{svjour3}       
\UseRawInputEncoding
\smartqed  
\usepackage{graphicx}
\usepackage{lipsum}
\usepackage{pdfpages}
\usepackage{float}
\usepackage{caption}  
\usepackage{subcaption} 
\usepackage{nicefrac}
\usepackage{url}
\usepackage{booktabs}
\usepackage{comment}
\usepackage{color,soul} 
\usepackage{array}
\usepackage{multirow}
\usepackage{bigstrut}
\usepackage{subcaption}
\usepackage{units}
\usepackage{flushend}
\usepackage{url}
\usepackage{natbib}
\usepackage{hyperref}
\usepackage{colortbl}
\usepackage{booktabs}
\usepackage{natbib}
\usepackage{makecell}
\usepackage{mathptm}
\usepackage{tcolorbox}

\usepackage{fancyvrb} 
\usepackage{fvextra} 
\usepackage{appendix}
\usepackage{microtype}

\usepackage{tabularx}

\setboolean{@twoside}{false}

\newcommand{\RQone}{RQ1: Are merged pull requests released more quickly using a CI service?}
\newcommand{\RQtwo}{RQ2: Does the increased number of PR submissions after adopting a CI service increase the delivery time of pull requests?}
\newcommand{\RQthree}{RQ3: What factors impact the delivery time after adopting a CI service?}
\newcommand{\RQfive}{RQ4: What is the perceived influence of CI on the time to deliver merged PRs?}
\newcommand{\RQfour}{RQ5: What are the perceived causes of delay in the delivery time of merged PRs?}
\newcommand{\RQsix}{RQ6: What is the perceived influence of CI on the software release process?}
\newcommand{\RQseven}{RQ7: What is the perceived influence of CI on the code review process?}
\newcommand{\RQeight}{RQ8: What is the perceived influence of CI on attracting more contributors to open-source projects?}

\journalname{Empirical Software Engineering}

\begin{document}
	
	\title{
	The Impact of a Continuous Integration Service on the Delivery Time of Merged Pull Requests
	}
	
	
	\author{Jo{\~a}o Helis Bernardo         \and
		Daniel Alencar da Costa \and 
		Uir{\'a} Kulesza \and 
		Christoph Treude 
	}
	
	
	\institute{
		Jo\~{a}o Helis Bernardo \at
		Federal Institute of Rio Grande do Norte (IFRN) \\            
		Federal University of Rio Grande do Norte (UFRN) \\
		Natal, Brazil \\
		\email{joao.helis@ifrn.edu.br}           
		\and
		Daniel Alencar da Costa \at
		University of Otago \\
		Dunedin,  New Zealand\\
		\email{danielcalencar@otago.ac.nz}
		\and
		Uir{\'a} Kulesza \at
		Federal University of Rio Grande do Norte (UFRN) \\
		Natal, Brazil \\
		\email{uira@dimap.ufrn.br}
		\and
		Christoph Treude \at
		University of Melbourne \\
		Melbourne, Australia \\
		\email{christoph.treude@unimelb.edu.au}		
	}
	
	\date{Received: date / Accepted: date}
	
	\maketitle

\begin{abstract}

Continuous Integration (CI) is a software development practice that builds and tests software frequently (e.g., at every push). One main motivator to adopt CI is the potential to deliver software functionalities more quickly than not using CI. However, there is little empirical evidence to support that CI helps projects deliver software functionalities more quickly. Through the analysis of 162,653 pull requests (PRs) of 87 GitHub projects, we empirically study whether adopting a CI service (\textsc{TravisCI}) can quicken the time to deliver merged PRs. 
We complement our quantitative study by analyzing 450 survey responses from participants of 73 software projects.
Our results reveal that adopting a CI service may not necessarily quicken the delivery of merge PRs. Instead, the pivotal benefit of a CI service is to improve the decision making on PR submissions, without compromising the quality or overloading the project's reviewers and maintainers. The automation provided by CI and the boost in developers' confidence are key advantages of adopting a CI service. Furthermore, open-source projects planning to attract and retain developers should consider the use of a CI service in their project, since CI is perceived to lower the contribution barrier while making contributors feel more confident and engaged in the project.
		
 
\end{abstract}

\section{Introduction}

Development teams are required to deliver software functionalities more quickly than ever to
improve the time-to-market and success of their software projects \citep{Debbiche2014}. The quick delivery of functionalities may keep customers engaged with the project while providing valuable feedback. 
To improve the processes of software integration and packaging, Continuous Integration (CI) has been proposed as part of the Extreme Programming (XP) methodology \citep{Beck2000-ja}, which claims that CI can provide more confidence for developers and quicken the delivery of software functionalities \citep{Laukkanen2015-ab}. 

Continuous Integration is a set of practices that enables development teams to integrate software more frequently \citep{Fowler2006-zc}. The increased number of integrations is possible by using automated tools such as automated tests. Ideally, CI should automatically compile, test, and package the software whenever code modifications occur. Nowadays, several vendors (e.g., \textsc{Apache} or \textsc{GitHub}) have developed tools to provide {\em CI as a service} for developers to implement their CI pipelines. Examples of these tools are \textsc{CloudBees}, \textsc{GitHub Actions}, and \textsc{TravisCI}. The main philosophy behind CI is that the software must always be in a working state, which is constantly put to test at each integration \citep{Duvall2007-tb}. CI has been widely adopted by the software development community in both open-source and corporate software projects. 

Existing research has analyzed the usage of CI in open-source projects hosted on \textsc{GitHub} \citep{Vasilescu2015-tj,Vasilescu2015-tn,Hilton2016-xy,bernardo2018studying,nery2019empirical,soares2022effects,santos2022investigating}. For instance, \citet{Vasilescu2015-tn} investigated the productivity and quality outcomes of projects that use CI services on \textsc{GitHub}. They found that projects that use CI merge pull requests (PRs) more quickly if they are submitted by core developers. Also, core developers discover significantly more bugs when they use CI. 
Although existing research has demonstrated that CI may provide benefits for development teams, \citet{soares2022effects} revealed that several studies investigate the benefits of using a {\em CI service} instead of studying the benefits of CI as a whole practice. For example, instead of checking whether studied projects adopt the full set of practices required by CI~\citep{felidre2019continuous}, most studies have assumed that projects use CI solely because a CI service was employed (such as \textsc{TravisCI}). We as a community should be clear about such scenarios when reporting our results. According to \cite{Fowler2006-zc}, adopting CI is not using a CI service alone but also adopting and maintaining specific development practices. 
To adopt CI appropriately, projects must maintain short build durations, fix broken builds as immediately as possible, check-in code frequently, and maintain high code coverage \citep{Duvall2007-tb,felidre2019continuous,santos2022investigating}.

In this regard, a common claim about adopting CI is that projects are able to release more frequently \citep{Stahl:2014:MCI:2562355.2562828,Hilton2016-xy}, implying that software updates would be delivered more quickly to their end-users. However, there is no sufficient empirical evidence to show that CI can indeed be associated with a quicker delivery of software functionalities to end users. Studying whether CI can quicken the delivery of software functionalities is important because release delays are frustrating to end users~\citep{costa2014empirical,Da_Costa2016-cb}.

In our prior work \citep{bernardo2018studying}, we quantitatively analyzed whether the use of a CI service (\textsc{TravisCI}) is correlated with the
time to deliver merged {\em Pull Requests} (PRs) of GitHub projects.
Our study investigated 162,653 PRs from 87 GitHub projects, which were implemented in 5 programming languages.\footnote{\url{https://prdeliverydelay.github.io/\#studied-projects}} We found that the time-to-deliver PRs is shorter \textit{after} adopting \textsc{TravisCI} in only 51.3\% of the projects. As we have observed that the use of a CI service is not necessarily associated with a quicker delivery of pull requests, we designed a qualitative study to obtain deeper explanations for our results while deepening our understanding of the potential influence of CI {\em as a whole practice} on the time-to-market of merged PRs. For example, do developers believe that CI, as a whole, influence the delivery of PRs in their projects? We designed a qualitative study because we could not find answers to such questions in our previous quantitative analyses. To sum up, our qualitative study complements our previous study by providing more explanations and context for the results we observed in the previous study.
Therefore, we survey 450 participants from 73 GitHub projects (out of the initial 87 projects of our quantitative study). 
Our qualitative analysis is composed of:

\begin{itemize}
    \item Data collection from survey responses of 450 participants of 73 popular open-source projects from GitHub. 
    \item An open-coding analysis of the answers to the open-ended questions of our survey using a thematic analysis technique.
    \item An analysis of the extent to which the survey participants are in accordance with the quantitative results of our prior work.
\end{itemize}

\subsection{\textbf{Quantitative Study}}

Our quantitative study addresses the following
research questions:

\begin{itemize}
    \item \textit{\textbf{\RQone}} 
    The wide adoption of CI is often motivated by the perceived benefits of this practice. For instance, higher confidence in the software product \citep{Duvall2007-tb}, higher release frequency \citep{Stahl:2014:MCI:2562355.2562828}, and the prospect of delivering software updates more quickly \citep{Laukkanen2015-ab}. However, there is a lack of studies that empirically investigate the association between using a CI service and the time-to-deliver of merged PRs. In $RQ1$, we study the delivery time of merged PRs \textit{before} and \textit{after} the use of \textsc{TravisCI}.
	
    \item \textit{\textbf{\RQtwo}} 
    In $RQ1$, we find that only 51.3\% of the projects deliver merged PRs more quickly \textit{after} adopting \textsc{TravisCI}. This result contradicts the assumption that merged PRs would be delivered more quickly {\em after} the adoption of a CI service in most of the projects.
    We then ask the following question: is there another key factor influencing the delivery time of merged PRs {\em after} \textsc{TravisCI} is adopted, such as a significant increase in workload?
		
    \item \textit{\textbf{\RQthree}} 	
    In $RQ1$ and $RQ2$, we study the impact of adopting a CI service on the delivery time of merged PRs. Nevertheless, it is also important to understand what are the characteristics of the delivery time of merged PRs \textit{before} and \textit{after} the use of a CI service. Such information may help decision makers to track and avoid a high delivery time.
\end{itemize}

\subsection{\textbf{Qualitative Study}}

In our qualitative study, we address the following research questions:
    	
\begin{itemize}
    \item \textit{\textbf{\RQfive}}
    In this RQ, we aim to deepen our understanding of how CI may impact the delivery time of PRs. We consult contributors of projects that use CI to obtain qualitative data, which can provide relevant insights to the research community and practitioners. 
    
    \item \textit{\textbf{\RQfour}} 		
    In $RQ4$ we observe that 42.9\% of participants are skeptical regarding the impact of CI on the delivery time of merged PRs. Therefore, in this RQ, we further discuss {\em indirect factors} (i.e., factors that are not necessarily related to CI) that participants believe may also impact the delivery time of PRs.

    \item \textit{\textbf{\RQsix}}
    Given that in $RQ2$ we observe a substantial increase in the number of delivered PRs per release (\textit{after} the adoption of \textsc{TravisCI}), 
    we aim to obtain further insights as to why the increase in the number of delivered PRs occurs. For this purpose, we consult our participants regarding their perceived influence of CI on the release process of their project. For example, is it the case that, because CI encourages the constant packaging of the software, preparing a release is no longer a challenge?	
    
    \item \textit{\textbf{\RQseven}} 
    Intriguingly, in $RQ1$, we find that PRs are merged faster \textit{before} the adoption of \textsc{TravisCI} in 73\% (\nicefrac{46}{63}) of the projects. This result motivates us to further investigate factors that influence the merge time when CI is adopted. 
    
    \item \textit{\textbf{\RQeight}}
    In $RQ1$ and $RQ2$, we observe that there exist a higher number of contributors and PR submissions \textit{after} the adoption of \textsc{TravisCI}. In $RQ8$, we consult our participants to better understand whether CI has any influence on attracting more contributors to open-source projects.	
    
\end{itemize}
	
\textbf{Paper organization.}
The remainder of this paper is
organized as follows. 
In Section~\ref{sec_related_work}, we discuss the related work.
In Section~\ref{sec_empirical_study}, we explain the design of our quantitative and qualitative studies.
In Sections~\ref{sec_quantitative_study_results} and~\ref{sec_qualitative_study_results}, we present the results of our quantitative and qualitative studies, respectively. 
We discuss the practical implications of our observations for the research and practice in software engineering in Section~\ref{sec_discussion}. 
In Section~\ref{sec_threats_to_the_validity}, we discuss the threats to the validity and limitations of our study.
Finally, we draw conclusions in Section~\ref{sec_conclusions}. 

\section{Related work}
\label{sec_related_work}

Through a systematic literature review of {\em Agile Release Engineering} practices, \cite{Karvonen201787} highlighted that empirical research in software engineering is crucial to better understand the impact of adopting CI on software development. 

\subsubsection*{\textbf{CI and team productivity}}

The study by \cite{Hilton2016-xy} revealed that 70\%
of the most popular \textsc{GitHub} projects use CI. The authors identified that CI helps projects to release more often, whereas the CI build status may foster a faster integration of PRs. \cite{Vasilescu2015-tn} studied the potential impact of CI on the quality and productivity of software projects. They found that projects that use CI merge PRs more quickly if these PRs are submitted by core developers. The authors found that core developers identify significantly more bugs when using CI. Regarding the acceptance and latency of PRs in CI (where latency is the time taken to merge a PR), \cite{Yu2016-cy} found that the likelihood of rejecting a PR increases by 89.6\% when the PR breaks the build. The results also show that the more succinct the PR is, the greater the probability that the PR is reviewed and merged earlier. 
Furthermore, \cite{zhao2017impact} investigated the transition to \textsc{TravisCI}\footnote{\url{https://travis-ci.org/}} in open-source projects. 
According to their study, the following changes may occur when \textsc{TravisCI} is adopted: (i) a small increase in the number of merged commits; (ii) a statistically significant decrease in the number of merge commit churn; (iii) a moderate increase in the number of closed issues; and (iv) a stationary behavior in the number of closed PRs. 

In our study, we use an approach similar to \cite{Vasilescu2015-tn} to identify projects that use \textsc{TravisCI}.  
Our goal with our quantitative study is to understand the association between \textsc{TravisCI} and the time taken for PRs to be delivered to end users. Furthermore, our qualitative study investigates how contributors of open-source projects perceive the impact of CI on the review and release processes of their projects.
Our work is complementary to prior studies, contributing to a larger understanding of how CI can impact several development activities in software projects (i.e., code review and project release).

\subsubsection*{\textbf{CI and code review}}

Recent studies have investigated the impact of CI on code review. The study by \cite{zampetti2019study} 
found that PRs that generate successful builds have 1.5 more chances of being merged. Our qualitative investigation corroborates the results of \cite{zampetti2019study}, showing that the CI build status can influence the decisions of code reviewers. Furthermore, \cite{cassee2020silent} found that the discussion held before the acceptance of a PR reduced considerably \textit{after} CI was adopted. Conversely, the number of changes developers performed during code review remained roughly the same. The work of \cite{zhang2022a_pull_latency} investigated the influence of various factors on PR latency. They found that, when using CI, the build status and duration are moderately relevant factors for accepting PRs. In a follow-up study, \cite{zhang2022b_pull_decision} observed that CI assists the acceptance of PRs by automating the code review process and replacing part of the code inspection work, accelerating the review process.
Indeed, our qualitative study reveals that, among the list of CI factors that impact the code review process, the most cited factors were related to an improvement in \textit{automation} and \textit{confidence}.
The participants of our study highlighted that CI facilitates the understanding of code decisions, accelerating the code review process. 

\subsubsection*{\textbf{Adherence to CI best practices}}

\cite{Vasilescu2015-tj} studied the use of \textsc{TravisCI} in a sample of 223 \textsc{GitHub} projects.
They found that the majority of projects (92.3\%) are configured to use \textsc{TravisCI} but less than half actually use the CI service. \cite{felidre2019continuous} analyzed 1,270 open-source projects using \textsc{TravisCI} to understand the adherence of projects to the recommended CI practices. 
The authors observed that 748 (∼60\%) projects perform infrequent check-ins. The study by \cite{nery2019empirical} studied the relationship between the use of CI and the evolution of software tests. The authors found that the overall test ratio and coverage of projects improved after CI was adopted.
In our work, our participants mention that CI impacts the delivery time of merged PRs by improving \textit{project quality}, \textit{automation}, and the \textit{release process}. According to our participants, CI improves the code quality and stability, making developers more confident to ship releases. The confidence in developers can be fostered by comprehensive automated testing, especially when the code coverage is high.

\cite{gallaba2018use} studied 9,312 open-source projects using \textsc{TravisCI} to understand how projects are using or misusing the features of \textsc{TravisCI}. The authors found that the majority (48.16\%) of \textsc{TravisCI} configurations is specifying job processing nodes. Furthermore, explicit deployment code is rare (2\%), which indicates that developers rarely use \textsc{TravisCI} to implement Continuous Delivery. In our qualitative study,  our participants often emphasize the relevance of automated tasks in CI to improve the project release process. \textit{Automated tests} and \textit{release automation} (i.e., Continuous Deployment) are frequently mentioned when our participants explain the influence of CI on project releases. However, in addition to many developers understanding CI as a Continuous Deployment enabler, such a feature is misused by many projects that use CI \citep{gallaba2018use}.

\section{Empirical Study Design}
\label{sec_empirical_study}

We perform two complementary studies: one quantitative and one qualitative. For each study, we describe their studied projects, their data collection process, and their methodology.

\subsection{\textbf{Quantitative Study---Study I}}
\label{sec_quantitative_study}

In Study I, we divide our projects in two time periods, \textit{before} and \textit{after} the adoption of \textsc{TravisCI}. Segmenting the data into these two time periods is necessary to study the association between the adoption of \textsc{TravisCI} and the delivery time of PRs.

\subsubsection{Studied Projects}
\label{subsec_quant_studided_projects}

Our goal is to identify projects with substantial historical data that adopted \textsc{TravisCI} eventually. We use such projects to better understand the potential influence of adopting a CI service on the delivery time of merged PRs. We use an approach similar to \cite{Vasilescu2015-tn} and \cite{Hilton2016-xy} to select projects that use \textsc{TravisCI}. We use the date of the first build on \textsc{TravisCI} to determine when \textsc{TravisCI} was introduced in a project. 

We selected a set of 87 popular \textsc{GitHub} projects (33 JavaScript, 23 Python, 11 Java, 10 Ruby, and 10 PHP). We collect metrics related to the PRs and releases of each project. 
The detailed information about all computed metrics for each PR is described in Tables \ref{tab_explanatory_variables_1} and \ref{tab_explanatory_variables_2}. We believe these metrics can be correlated with the \textit{delivery time} of merged PRs. A total of 162,653 delivered PRs were collected (123,543 PRs were delivered \textit{after} the adoption of \textsc{TravisCI}, whereas 39,110 were delivered \textit{before} the adoption of \textsc{TravisCI}). The unbalanced number of PRs across time periods is a reflection of the duration of the adoption of \textsc{TravisCI} in different projects. The median age of our projects is 5.1 years, where the use of \textsc{TravisCI} accounts for 60.8\% (3.1 years) of the age of our projects. Table \ref{tab:pulls_per_language} shows the number of PRs per programming language \textit{before} and \textit{after} the adoption of \textsc{TravisCI}. Our project selection and data collection processes are explained in more detail in an earlier publication of this work \citep{bernardo2018studying}.

\bgroup
\def\arraystretch{1.1}
\begin{table}[t]
	\centering
	\footnotesize
	\caption{Summary of the number of projects and released pull requests grouped by programming language.}
	\begin{tabular}{|c|c|c|c|c|}
		\hline
		\textbf{Language} & \textbf{Projects} & \textbf{PRs total}  & \textbf{PRs \textit{before} CI} & \textbf{PRs \textit{after} CI} \\
		\hline \hline
		JavaScript   & 33 & 57,104  & 17,556 & 39,548  \\ 
		Python       & 23 & 55,003 & 9,107 & 45,896  \\ 
		Java           & 11 & 7,700   & 3,433 & 4,267    \\ 
		Ruby          & 10 & 22,864 & 3,197 & 19,667  \\ 
		PHP            & 10 & 19,982 & 5,817 & 14,165  \\ 
		\hline
		\textbf{Total} & \textbf{87} & \textbf{162,653}  & \textbf{39,110} & \textbf{123,543} \\ 
		\hline
	\end{tabular}
	\label{tab:pulls_per_language}
\end{table}
\egroup

\bgroup
\def\arraystretch{1}

\begin{table*}[htbp]
	\caption{Metrics that are used in our explanatory models (resolver, pull request, and project dimensions).}
	\footnotesize
	\begin{tabular}{|>{\centering\arraybackslash}m{0.45in}|m{0.60in}|m{0.4in}|m{2.5in}|}
		\hline
		\textbf{Dimension} & \textbf{Attributes} & \textbf{Type} & \textbf{Definition (d) | Rationale (r)} \\ \hline
		\multirow{12}{*}{Resolver} & \multirow{7}{2cm}{Contributor Experience}
		& \multirow{7}{*}{Numeric} & \textbf{d:} The number of
		previously released PRs that were submitted by the contributor of a
		particular PR. We consider the author of the PR to be its
		contributor. \\ \cline{4-4}
		& &  & \textbf{r:} The greater
		the experience and participation of a user within a specific
		open-source project, the greater his/her chance of having
		his/her PR reviewed and integrated into the codebase of such a project
		by its core integrators \citep{shihab2010predicting}.   \\
		\cline{ 2- 4}
		
		& \multirow{5}{2cm}{Contributor Integration} & \multirow{5}{*}{Numeric} & \textbf{d:} The average in days of the previously released PRs that were submitted by
		a particular contributor.
		\\ \cline{4-4}
		& &  & \textbf{r:} If a particular contributor usually submits PRs that are merged and released quickly, his/her future PR might be merged and released quickly as well \citep{Da_Costa2016-cb}.  \\ \hline
		
		
		\multirow{11}{1.2cm}{Pull Request} & \multirow{6}{2cm}{Stack Trace Attached} & \multirow{6}{*}{Boolean} & \textbf{d:} We verify if the PR report has a stack trace attached in its description. \\ \cline{4-4}
		& &  & \textbf{r:} 
		If the PR provides a bug fix, a stack trace attached may provide useful information regarding the causes of the bug and the importance of the submitted code, which may quicken the merge of the PR and its delivery in a release of the project \citep{schroter2010stack}.\\ \cline{ 2- 4}
		& \multirow{4}{1.5cm}{Description Size} & \multirow{4}{*}{Numeric} & \textbf{d:} The number of characters in the body (description) of a PR.
		\\ \cline{4-4}
		& &  & \textbf{r:} PRs that are well described might be easier to merge and release than PRs that are more difficult to understand  \citep{Da_Costa2016-cb}.
		\\ \hline
		
		
		\multirow{13}{*}{Project} & \multirow{4}{2cm}{Queue Rank} & \multirow{4}{*}{Numeric} & \textbf{d:} The number that represents the moment when a PR is merged compared to other merged PRs in the release cycle. For example, in a queue that contains 100 PRs, the first merged PR has position 1, while the last merged PR has position 100.  \\ \cline{4-4}
		& &  & \textbf{r:} A PR with a high \textit{queue rank} is a recently merged PR. A merged PR might be released faster/slower depending on its queue position \citep{Da_Costa2016-cb}. \\ \cline{ 2- 4}
		
		& \multirow{6}{1.5cm}{Merge Workload} & \multirow{6}{*}{Numeric} & \textbf{d:} The number of PRs that were created and still waiting to be merged by a core integrator at the moment at which a specific PR is submitted. \\ \cline{4-4}
		& &  & \textbf{r:} A PR might be released faster/slower depending of the
		amount of submitted PRs waiting to be merged. The higher the
		amount of created PRs waiting to be analyzed and merged, the
		greater the workload of the contributors to analyze these PRs,
		which may impact their delivery time.
		\\ \hline
	\end{tabular}
	\label{tab_explanatory_variables_1}
\end{table*}
\egroup

\bgroup
\def\arraystretch{1}

\begin{table*}[htbp]
	\caption{Metrics that are used in our explanatory models (process dimension).}
	\footnotesize
	\begin{tabular}{| >{\centering\arraybackslash}m{0.45in}|m{0.60in}|m{0.4in}|m{2.5in}|}			
		\hline
		\textbf{Dimension} & \textbf{Attributes} & \textbf{Type} & \textbf{Definition (d) | Rationale (r)} \\ \hline
		
		\multirow{33}{*}{Process} & \multirow{4}{2cm}{Number of Impacted  Files} & \multirow{4}{*}{Numeric} & \textbf{d:} The number of files linked to a PR submission.  \\ \cline{4-4}
		& &  & \textbf{r:} The delivery time might be related to the high number of files of a PR, because more effort must be spent to integrate it \citep{jiang2013will}.   \\ \cline{ 2- 4}
		
		& \multirow{5}{*}{Churn} & \multirow{5}{*}{Numeric} & \textbf{d:} The number of added lines plus the number of deleted lines to a PR.
		\\ \cline{4-4}
		& &  & \textbf{r:} A higher churn suggests that a great amount of work might be required to verify and integrate the code contribution sent by means of PR \citep{jiang2013will, nagappan2005use}.  \\ \cline{ 2- 4}
		
		& \multirow{3}{*}{Merge Time} & \multirow{3}{*}{Numeric} & \textbf{d:} Number of days between the submission and merge of a PR.
		\\ \cline{4-4}
		& &  & \textbf{r:} If a PR is merged quickly, it is more likely to be released faster.  \\ \cline{ 2- 4}
		
		& \multirow{5}{2cm}{Number of Activities} & \multirow{5}{*}{Numeric} & \textbf{d:} An activity is an entry in the PR's history. 
		\\ \cline{4-4}
		& &  & \textbf{r:} A high number of activities might indicate that much work was required to make the PR acceptable, which may impact the integration of such PR into a release \citep{jiang2013will}.  \\ \cline{ 2- 4}
		
		& \multirow{5}{2cm}{Number of Comments} & \multirow{5}{*}{Numeric} & \textbf{d:} 
		The number of comments of a PR.  \\ \cline{4-4}
		& &  & \textbf{r:} A high number of comments might indicate the
		importance of a PR or the difficulty to understand it
		\citep{giger2010predicting}, which may impact its delivery time \citep{jiang2013will}. \\ \cline{ 2- 4}
		
		& \multirow{5}{2cm}{Interval of Comments} & \multirow{5}{*}{Numeric} & \textbf{d:} 
		The sum of the time intervals (days) between comments divided by the total number of comments of a PR.   \\ \cline{4-4}
		& &  & \textbf{r:} A short \textit{interval of comments} indicates the discussion was held with priority, which suggests that the PR is important, thus, the PR might be delivered faster \citep{Da_Costa2016-cb}. \\ \cline{ 2- 4}
		
		& \multirow{5}{1.5cm}{Commits per PR} & \multirow{5}{*}{Numeric} & \textbf{d:} Number of commits per PR.
		\\ \cline{4-4}
		& &  & \textbf{r:} The higher the number of commits in a
		PR, the greater the amount of contribution to be analyzed by the project integrators, which might impact the delivery time of the PR.  
		\\ \hline
	\end{tabular}
	\label{tab_explanatory_variables_2}
\end{table*}
\egroup

\subsubsection{Research Approach}

\begin{figure}[t]
	\centering
	\includegraphics[width=\columnwidth,keepaspectratio]{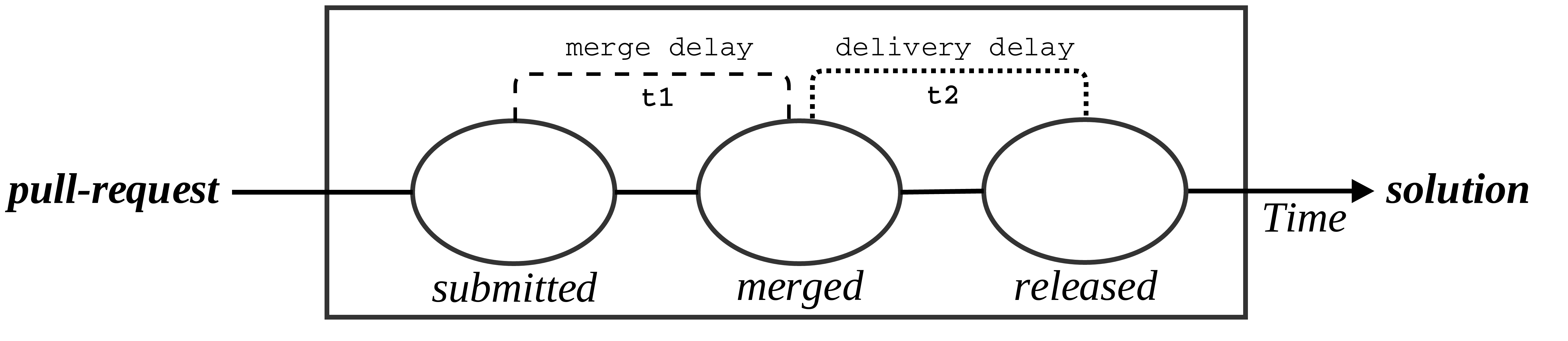}
	\caption{The basic life-cycle of a delivered pull request.}
	\label{fig_released_pull_request_life_cycle}
\end{figure}

Figure \ref{fig_released_pull_request_life_cycle} shows the basic life cycle of a delivered PR, where $t1$ is the merge phase and $t2$ is the delivery phase. We refer to  $t1 + t2$ as the lifetime of a PR. In $RQ1$, we analyze the \textit{merge} and \textit{delivery} phases. The merge phase ($t1$) is the required time for PRs to be merged into the codebase, whereas the \textit{delivery phase} ($t2$) refers to the required time for merged PRs to be released. 

We use Mann-Whitney-Wilcoxon (MWW) tests~\citep{wilks2011statistical} and Cliff's delta effect-size measurements~\citep{cliff1993dominance} to compare different distributions of values. MWW is a non-parametric test whose null hypothesis is that two distributions come from the same population ($\alpha = 0.05$). The Cliff's delta is a non-parametric effect-size metric to verify the magnitude of the difference between the values of two distributions. 
We use the thresholds provided by \citet{Romano:2006} to interpret the Cliff's delta, i.e. $delta<0.147$ (\textit{negligible}), $delta<0.33$ (\textit{small}), $delta<0.474$ (\textit{medium}), and $delta>=0.474$ (\textit{large}). We analyze the entire life-cycle of a PR \textit{before} and \textit{after} the adoption of \textsc{TravisCI}. First, we analyze the \textit{delivery time} ($t2$) and then, we analyze the {\em merge time} ($t1$). Lastly, we analyze the {\em lifetime} of a PR ($t1 + t2$).

Similar to $RQ1$, in $RQ2$ we use MWW tests \citep{wilks2011statistical}
and Cliff's delta measurements \citep{cliff1993dominance} to analyze the data. We use
box plots \citep{Williamson-1989} to visually summarize the distributions and perform comparisons.
In $RQ2$, we investigate whether the increase in the lifetime of PRs
\textit{after} adopting \textsc{TravisCI} is related to a significant increase in PR
submission, merge, and delivery rates (also \textit{after} the adoption of \textsc{TravisCI}). 

We group our dataset into two time periods: \textit{before} and \textit{after} the
adoption of \textsc{TravisCI}. For each time period, we count the number of PRs that are submitted,
merged, and delivered per release.  We perform three comparisons in $RQ2$.
First, we compare whether the PR submission, merge, and delivery rates per
release significantly increase \textit{after} the adoption of \textsc{TravisCI}. Next, we verify
whether there is a statistical increase in the release frequency of the
projects \textit{after} the adoption of \textsc{TravisCI}. 
We use the Pearson correlation \citep{best1975algorithm},
to test whether two variables are significantly correlated. 

In RQ3, we use multiple regression modeling ({\em Ordinary Least Squares}) to describe the
relationship between $X$ (i.e., the set of explanatory variables, e.g.,
\textit{churn}, \textit{description length}), and the response variable $Y$,
i.e., the \textit{delivery time} of merged PRs in terms of days. We control
covariates that might influence the results. For each project, we build two
explanatory models, one using the PR data \textit{before} the adoption of \textsc{TravisCI}, and another using
PR data \textit{after} the adoption of \textsc{TravisCI}. Tables \ref{tab_explanatory_variables_1} and \ref{tab_explanatory_variables_2} show the definition and rationale for each explanatory variable used in our
models. Our response variable $Y$ is the length of time between when a PR was merged and the time at which the same PR was delivered (i.e., delivery time).

We follow the guidelines of \citet{harrell2015regression} for fitting linear models.  We assess how stable our models are by computing the \textit{optimism-reduced} $R^2$. Finally, we use the Wald $X^2$ maximum likelihood test to evaluate the impact of each explanatory variable in our models. The larger the $X^2$ value for a variable, the larger the impact of a variable \citep{Da_Costa2016-cb}.
Next, we analyze the direction of the relationship between the most influential variables of our models and the delivery time. 
The process we use to build our statistical models is explained in more detail in our earlier publication~\citep{bernardo2018studying}.

\subsection{\textbf{Qualitative Study (Study II)}}
\label{sec:qualitative_study}

In this section, we explain the data collection and research approach of our qualitative study (Study II).

\subsubsection{Subject Projects}
\label{sec:subject_projects}

As the main goal of qualitative analysis is to complement Study I, we select the same 87 GitHub projects used in Study I for Study II. 
The goal of Study II is to better understand the influence CI can have on the delivery time of merged PRs.
We also take the opportunity to better understand the perceived influence of CI on the code review and release processes of our studied projects (i.e., according to the perception of our participants).

\subsubsection{Data Collection}
\label{sec:data_collection}

We first identify contributors who have submitted at least one PR that made into an official release of their project. The release date of the PRs must have fallen between the projects' creation date and November 11th, 2016, i.e., the range used in our search on \textsc{GitHub} for Study I. By inspecting the PR meta-data of the studied projects, we find a total of 20,698 contributors that fulfill our criteria. To prioritize frequent contributors, for each studied project, we select 15\% of contributors that have the highest number of delivered PRs, resulting in 3,105 contributors.

To collect our data, we designed a web-based survey and sent it by email to all 3,105 participants (i.e., the contributors of our subject projects). To encourage participation, we randomly provided six \$50 Amazon gift cards to respondents who explicitly stated their willingness to participate in the draw. To be eligible for the gift cards, the participants needed to answer all questions of the survey. In total, we received 450 responses, resulting in a response rate of 14.5\% (\nicefrac{450}{3105}). Our invitation letter is available in Appendix~\ref{sec:appendix_invitation_latter}.

Our survey has three major \textit{parts}. The first part concerns the influence of CI on the delivery time of merged PRs, whereas the second and third parts concern the potential influence of CI on the release and review processes of the studied projects, respectively. A complete example of our survey is available in Appendix \ref{sec:appendix_project_survey_example}, which shows the questionnaire sent to participants of the \textit{haraka/haraka}\footnote{\url{github.com/haraka/haraka}} project. Because our goal was to provide data specific to the projects of our participants, we designed 87 different questionnaires, aiming to obtain richer information about the project and encourage participants to respond more fully to the survey.

Our questionnaire is organized as follows. The first six questions (\#3--\#8) collect demographic information. Questions \#9--\#13 tap into the general experience of our participants, whereas questions \#14--\#26 present data specific to our participants' projects. 
In terms of questions' goals, questions \#9--\#13 and \#26 capture the potential influence of CI on the delivery time of merged PRs. 
\textit{Question \#14} captures the perceived correctness of our approach to define the \textsc{TravisCI} adoption date in our studied projects. 
\textit{Questions \#17--\#21} capture the potential influence of CI on the code review process of the projects, whereas \textit{Questions \#22--\#25} capture the potential influence of CI on the release process of the projects. 

\subsection{Research Approach}
\label{sec:research_approach}

We use an inductive thematic analysis, which is designed for identifying, analyzing, and reporting themes found within qualitative data~\citep{braun2006using}. In this study, we use the guidelines proposed by \cite{nowell2017thematic} to perform our thematic analysis.

The first step in our thematic analysis is the coding of our data. This step consists of attaching codes to any piece of relevant qualitative data collected from our questionnaire. The first author conducts three sessions of open coding of the responses to open-ended questions. The second author independently conducts three sessions of open coding for 10\% of the responses for each of those questions. Afterward, a new set of codes is generated by the merge of the codes created by each author. 
We use Cohen's Kappa test to verify the agreement rate between authors when coding the responses to 13 open-ended questions of our questionnaire.
We calculate the Kappa value separately for each of the 13 questions. We achieved a median Kappa value of 0.84, indicating substantial agreement~\citep{landis1977kappa}. 
The third author reviews the set of codes to add additional entries and resolve disagreements between the codes from the first and second authors. 
Next, the first author organizes the codes into {\em themes} through axial coding. These themes represent higher conceptual constructs (e.g., a theme might group many codes). 
This categorization was double-checked by the second author. 
We report the codes and themes generated by our thematic analysis in the result section. When reporting the results of $RQ4$---$RQ8$, we indicate (in superscript) the number of quotes citing each code and theme. It is important to highlight that the number in superscript does not necessarily indicate the relevance of a code, e.g., a code may be mentioned in more quotes because the code is more easily remembered by our participants. 
Additionally, when reporting our qualitative results, the frequency with which codes occur across responses can be higher than the total of responses. This is because a response from a participant can be associated with several codes. For example, consider the following quote \textit{``Anything that is considered a critical security fix or major bug fix is generally shipped within 1-2 weeks of submission. This happens frequently"} (C020). We derived two codes from this quote, which are \textit{bug fix} and \textit{security fix}.
We use representative quotes from our participants to aid in the understanding of the interpretation of the codes. 
We omit the participants' names by replacing them with an ID, e.g., {\em participant 01} receives the ``name'' C001. In Appendix \ref{sec:appendix_participants_ids_and_their_projects}, we provide the IDs assigned to our participants and their project.

\section{\textbf{Quantitative Study Results}}
\label{sec_quantitative_study_results}

In this section, we present the results of our quantitative study ($RQ1$---$RQ3$).

\subsection*{\textbf{\RQone}}
\label{RQ1_results}

\textit{\textbf{Only 51.3\% of the projects deliver merged PRs more quickly \textit{after} the adoption of TravisCI}}.  Out of 87 projects, we observe that
82.7\% (\nicefrac{72}{87}) obtained significant \textit{p-values} (i.e., $p<0.05$) when comparing the delivery time of merged PRs {\em before} and {\em after} adopting \textsc{TravisCI}. Surprisingly, we observe that only 51.3\% (\nicefrac{37}{72}) of these projects deliver merged PRs more quickly {\em after} adopting \textsc{TravisCI}. 
Our analyses indicate that 82.7\% (\nicefrac{72}{87}) of the projects have a statistical difference on the
delivery time of merged PRs, but a small median Cliff's delta of $0.304$.                                     

\textit{\textbf{In 73\% (\nicefrac{46}{63}) of the projects, PRs are merged faster \textit{before} adopting TravisCI.}} A total of 72.4\% (\nicefrac{63}{87}) of the projects have a statistical difference on the time to merge PRs with a median Cliff's delta of $0.206$ (\textit{small}). With respect to such
projects, we observe that 73\% (\nicefrac{46}{63}) merge PRs more quickly \textit{before} the adoption of \textsc{TravisCI}. 

\textit{\textbf{Surprisingly, in 54\% of the projects, PRs have a longer lifetime after adopting TravisCI.}}
We observe that in 54\% (\nicefrac{47}{87}) of our projects, PRs have a longer lifetime after the adoption of \textsc{TravisCI}. 71.3\% (\nicefrac{62}{87}) of these projects yield a statistically significant difference (\textit{p-value $<$ 0.05}) and a
$non-negligible$ median $delta$ between the distributions of PR lifetime
($delta>=0.147$). 37.1\% (\nicefrac{23}{62}) of such projects yield a large Cliff's delta (median $0.604$), while 22.6\% (\nicefrac{14}{62}) and 40.3\%
(\nicefrac{25}{62}) of the projects obtained medium and small Cliff's deltas, respectively (medians of $0.362$ and $0.223$).
Regarding the projects that yield a \textit{p-value} $< 0.05$, we observe that 51.6\% (\nicefrac{32}{62}) have a shorter PR lifetime \textit{before} the adoption of \textsc{TravisCI}, while 48.4\% (\nicefrac{30}{62}) have a shorter PR lifetime \textit{after} the adoption of \textsc{TravisCI}. 

\begin{center}
	\begin{tabular}{|p{.96\columnwidth}|}
		\hline
		\textbf{Summary:}
		\textit{			
		Surprisingly, only 51.3\% of the projects deliver merged PRs more quickly after the adoption of \textsc{TravisCI}. In 54\% (\nicefrac{47}{87}) of the projects, PRs have a longer lifetime \textit{after} the adoption of \textsc{TravisCI}. Finally, PRs are merged faster before the adoption of \textsc{TravisCI} in 71.3\% (\nicefrac{63}{87}) of the studied projects.} \\
		\textbf{Implications:}
		\textit{If the decision to adopt a CI service is mostly driven by the goal of quickening the delivery time of merged PRs, this decision must be more carefully considered by development teams.}
		 \\
		\hline
	\end{tabular}
\end{center}

\subsection*{\textbf{\RQtwo}}\label{RQ2_results}

\begin{figure}[!t]
	\centering
	\begin{subfigure}{3.0cm}
		\centering
		\includegraphics[width=2.7cm, height=4.0cm]{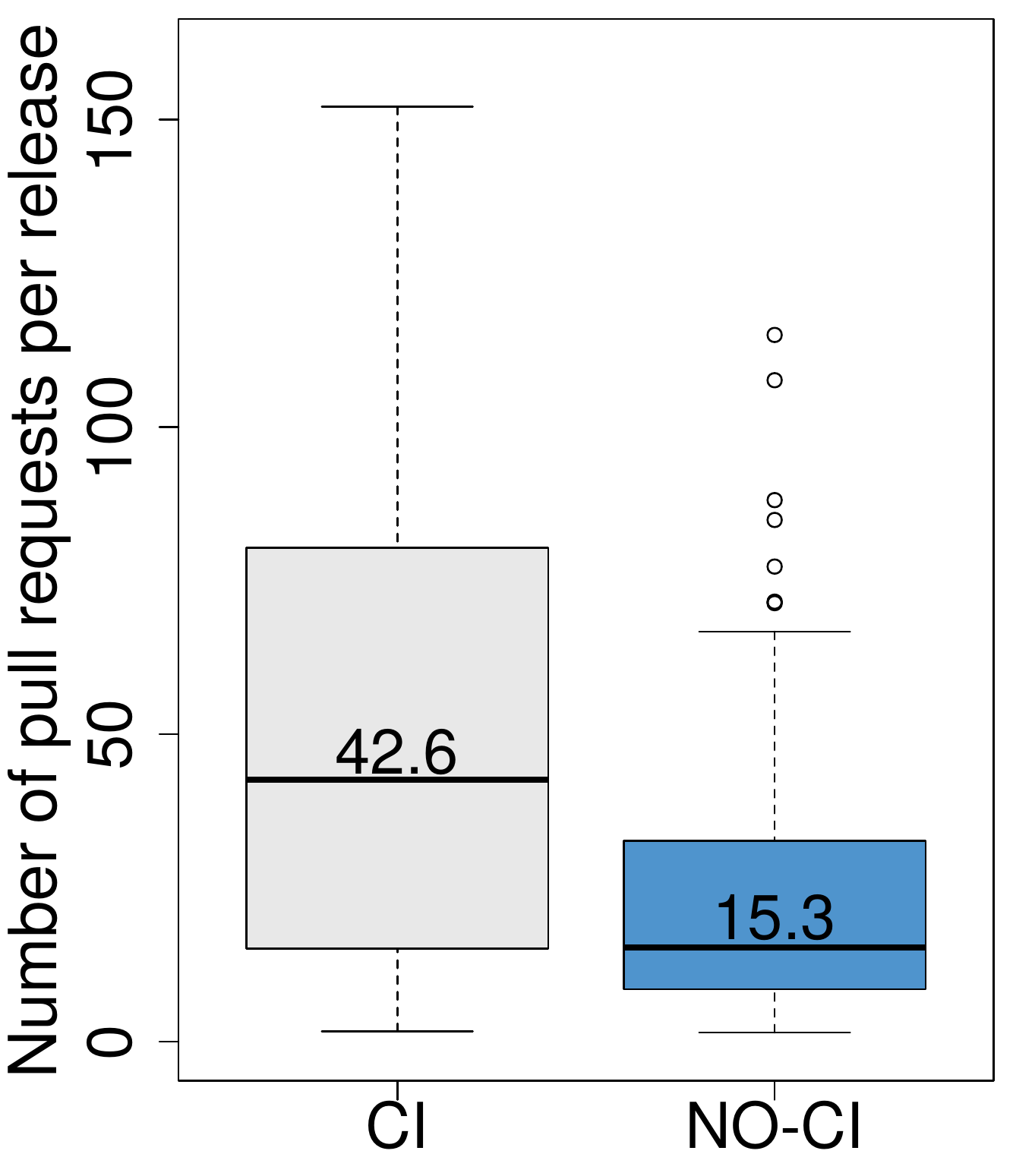}
		\caption{Submitted PRs}
		\label{fig:submitted_prs_per_release}
	\end{subfigure}%
	\begin{subfigure}{2.8cm}
		\centering
		\includegraphics[width=2.5cm, height=4.0cm]{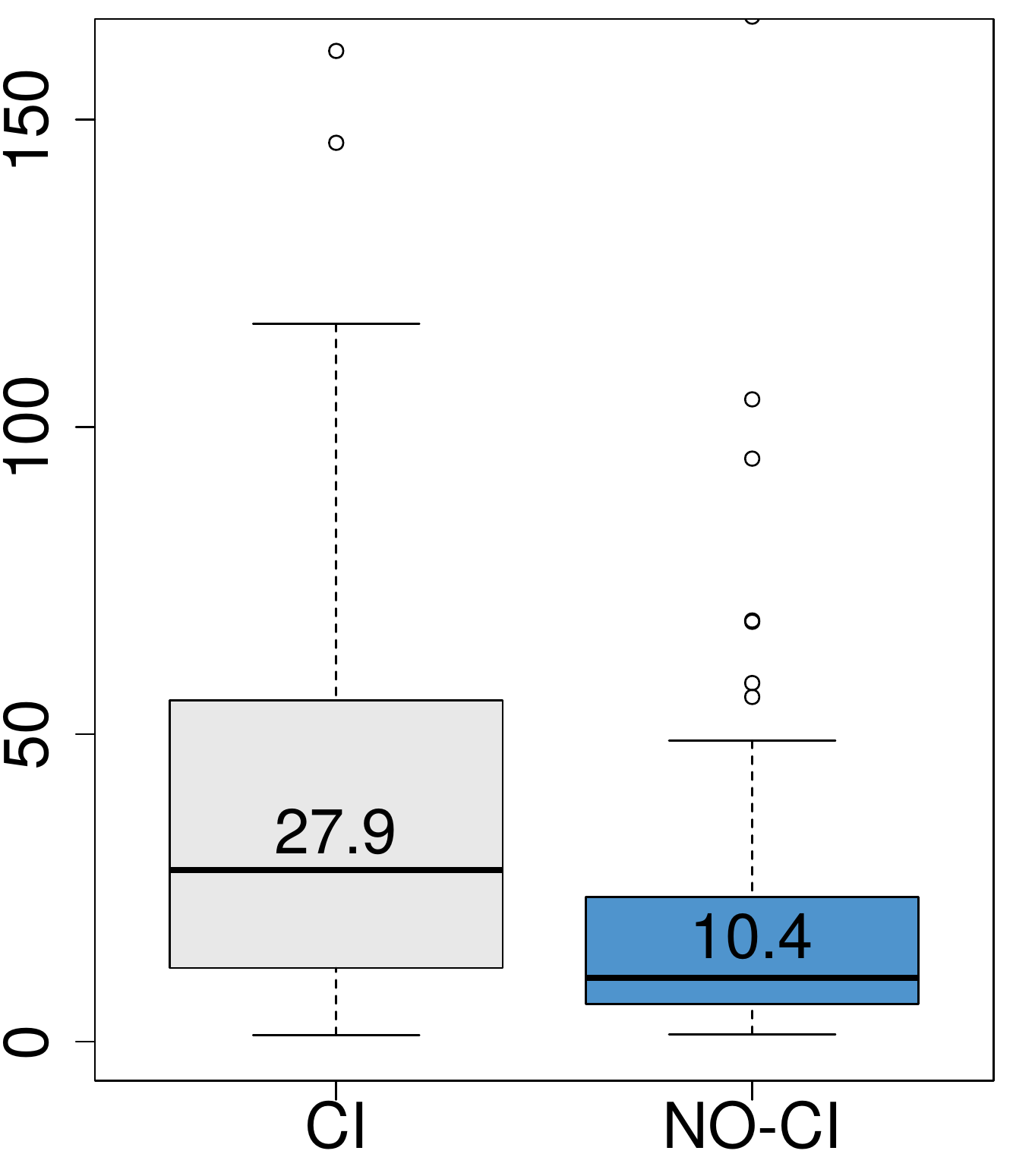}  
		\caption{Merged PRs}
		\label{fig:merged_prs_per_release}
	\end{subfigure}
	\begin{subfigure}{2.8cm}
		\centering
		\includegraphics[width=2.5cm, height=4.0cm]{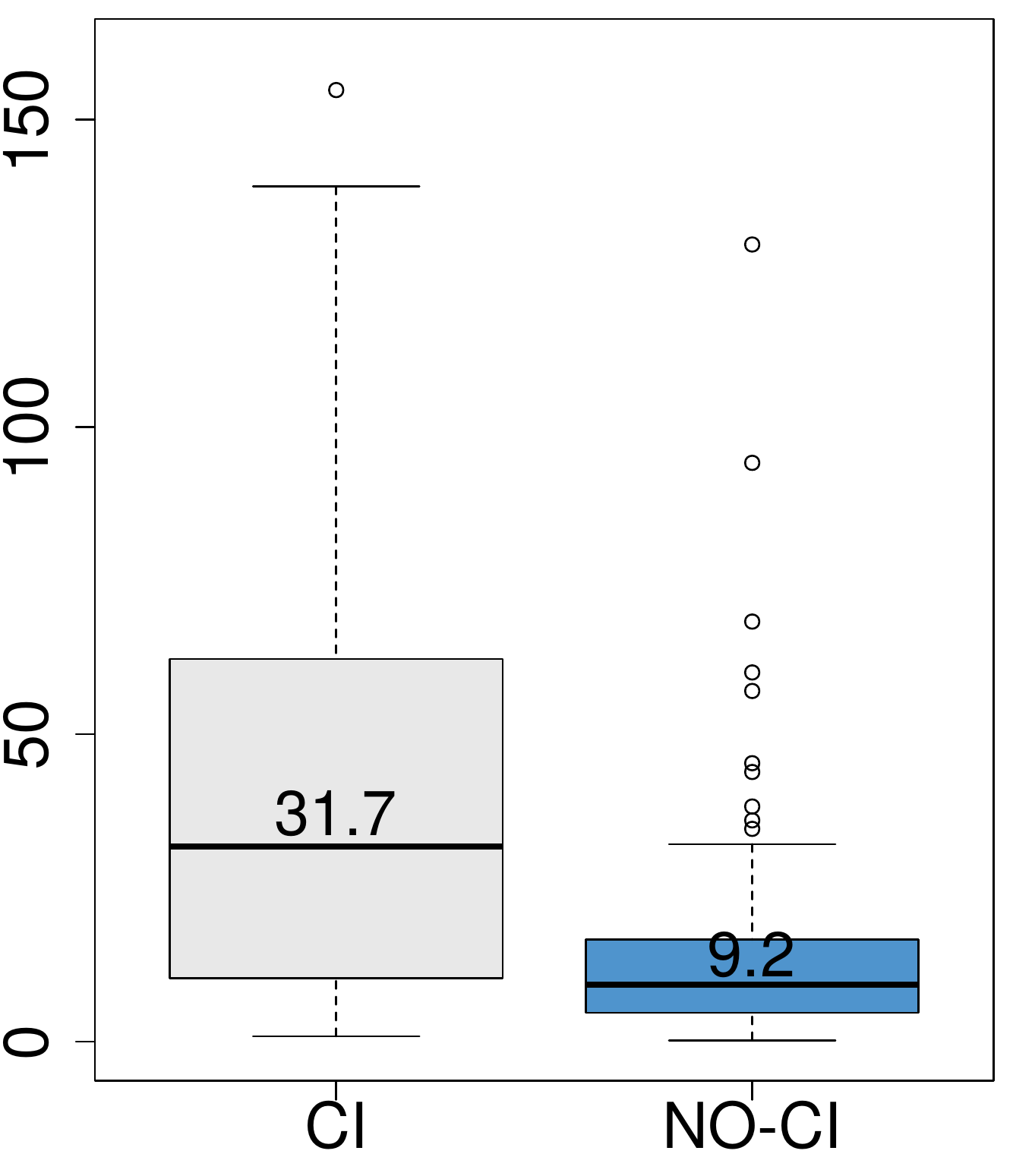}
		\caption{Delivered PRs}
		\label{fig:released_prs_per_release}
	\end{subfigure}
	\caption{PR submission, merge, and delivery rates per release.}
	\label{fig:pr_workflow}
\end{figure}

\textit{\textbf{71.3\% (\nicefrac{62}{87}) of the projects receive more PR
submissions after the adoption of \textsc{TravisCI}.}} Figure \ref{fig:pr_workflow} shows the
distributions of PRs submitted, merged, and delivered per release
for the studied projects. We observe that projects tend to submit a median of
42.6 PRs per release \textit{after} the adoption of \textsc{TravisCI}, while the median number of PRs
submitted per release \textit{before} the adoption of \textsc{TravisCI} is 15.3. A Wilcoxon signed rank test reveals that the increase in the number of PR submissions is statistically significant (\textit{p-value} $= 0.0001547$), with a Cliff's delta of $0.332$ (\textit{medium} effect-size).
We also observe a significant increase in the number of merged PRs per release
\textit{after} the adoption of \textsc{TravisCI} (\textit{p-value} $= 7.897e-05$, with a
\textit{medium} Cliff's delta of $0.347$). The number of merged PRs per release
increases from 10.4 (median) (\textit{before} the adoption of \textsc{TravisCI}) to 27.9 \textit{after} the adoption of \textsc{TravisCI}.
Interestingly, we also observe an increase in PR code churn per
release {\em after} the adoption of \textsc{TravisCI}. We obtain a \textit{p-value} $= 0.002273$ and a Cliff's delta value of $0.27$ (small). This significant increase in PR code churn per release
may help explain the increased lifetime of PRs {\em after} the adoption \textsc{TravisCI}. Given that
more code modifications are performed in PRs \textit{after}
the adoption of \textsc{TravisCI}, they may require more time to be reviewed, merged and delivered.

\textit{\textbf{After the adoption of \textsc{TravisCI}, projects deliver 3.43 times more PRs per
release than before the adoption of \textsc{TravisCI}.}} When we analyze the PR throughput per release, we find
that the number of PRs delivered per release increases significantly
\textit{after} the adoption of \textsc{TravisCI}. The number of PRs delivered increased from 9.2
to 31.7 \textit{after} the adoption of \textsc{TravisCI} (see Figure \ref{fig:released_prs_per_release}).
Furthermore, the increase in the number of PRs delivered per release 
is statistically significant (\textit{p-value} $= 1.366e-05$, with a {\em medium}
Cliff's delta of $0.3819527$).

\begin{figure}[!t]
	\centering
	\includegraphics[width=2.9cm, height=4.0cm]{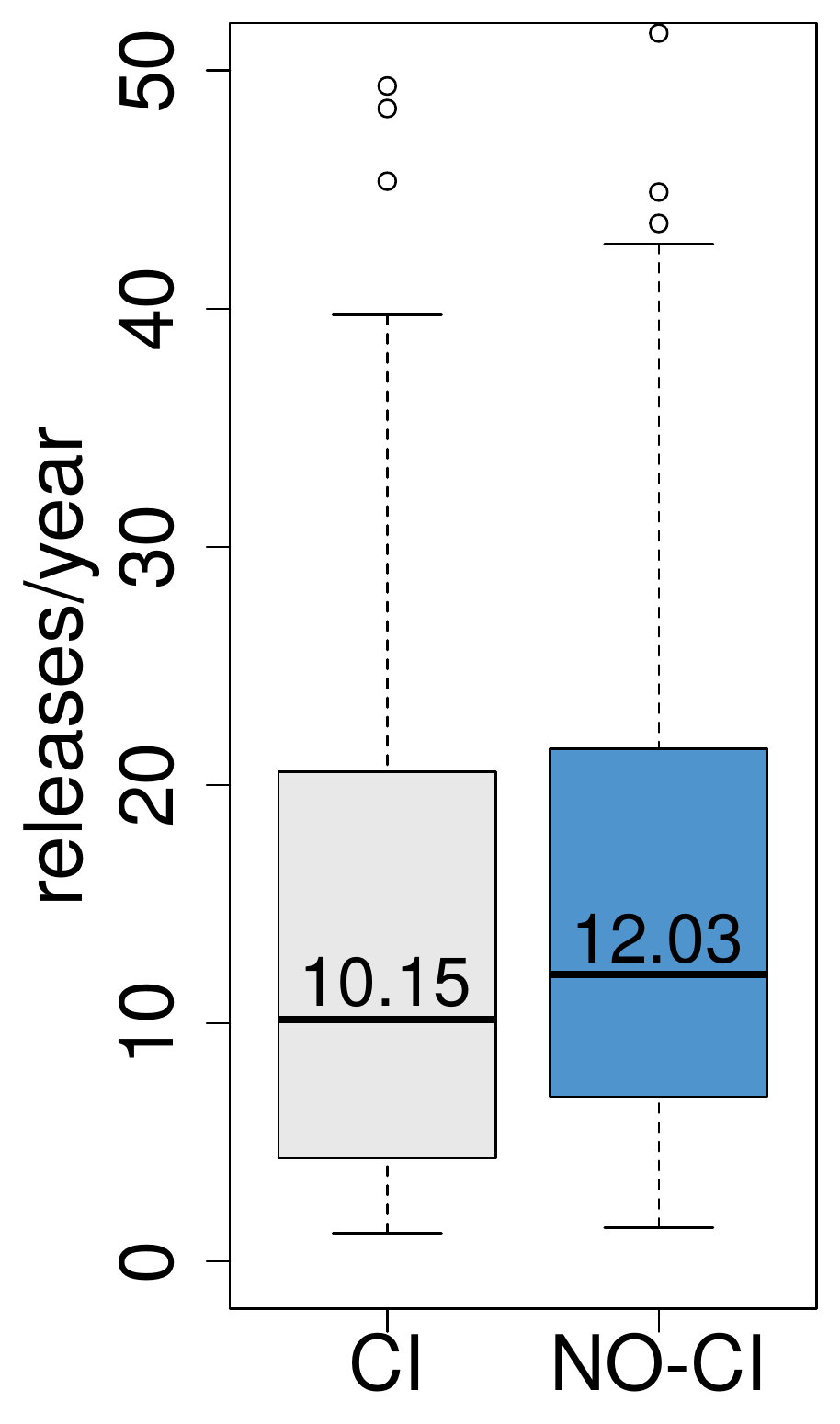}
	\caption{Releases per year \textit{before} and \textit{after} \textsc{TravisCI}.}
	\label{fig:releases_per_year}
\end{figure}

\textit{\textbf{We do not observe a significant difference in release frequency
after the adoption of \textsc{TravisCI}.}} A significant increase in PR submissions may be related to
an increase in release frequency \textit{after} the adoption of \textsc{TravisCI}. Figure
\ref{fig:releases_per_year} shows the distributions of releases
per year \textit{before} and \textit{after} the adoption of \textsc{TravisCI} (for each of the
studied projects). In the median, projects tend to ship 12.03 releases per year
\textit{before} the adoption of \textsc{TravisCI}, whereas the median drops to 10.15 \textit{after} the adoption to \textsc{TravisCI}. However, we obtain a
\textit{p-value} $= 0.146$, indicating that the differences in release frequency 
per year \textit{before} and \textit{after} the adoption of \textsc{TravisCI}
 are statistically insignificant. Our results suggest
that the high increase in the number of PRs delivered per release is unlikely to be linked with an increase in the number of releases. 
We investigate whether the increased number of PRs delivered may be due to an increase in the number of contributors
\textit{after} the adoption of \textsc{TravisCI}. 

\begin{figure}[!t]
	\centering
	\includegraphics[width=2.7cm, height=4.0cm]{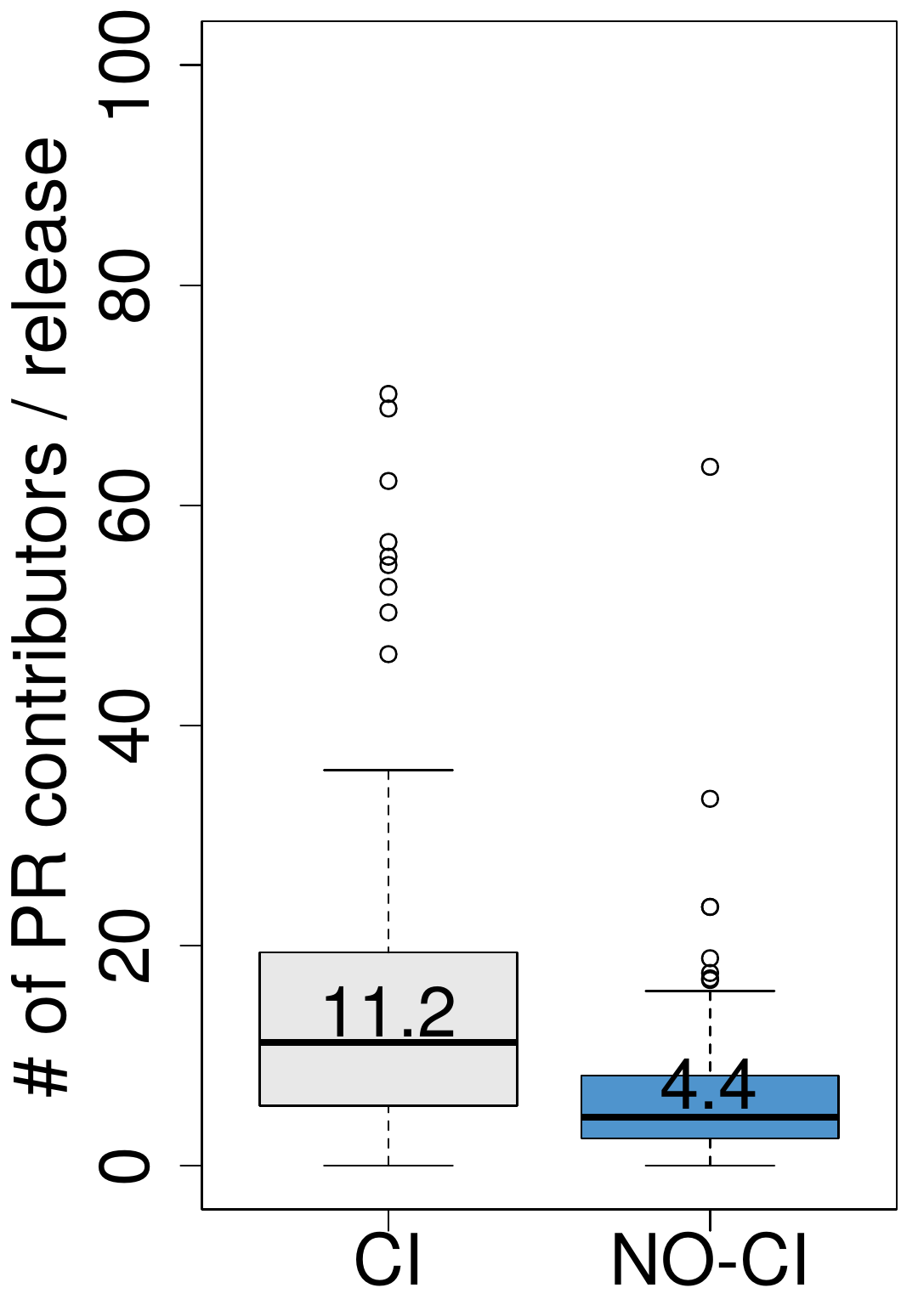}
	\caption{PR contributors per release.}
	\label{fig:contributors_per_release}
\end{figure}

\textbf{\textit{We find that 75.9\% (\nicefrac{66}{87}) of projects
had an increase in the number of contributors per release after the adoption of \textsc{TravisCI}.}}
Figure \ref{fig:contributors_per_release} shows the distributions of contributors per release both \textit{before} and \textit{after} the adoption of \textsc{TravisCI}. The median number of contributors per release increases from $4.4$ to $11.2$ \textit{after} the adoption of \textsc{TravisCI}. We observe that the difference 
is statistically significant (\textit{p-value} $= 2.525e-06$ with a {\em medium} Cliff's delta of $0.413$). 

Despite the increase in contributors and PRs delivered per release \textit{after} the adoption of \textsc{TravisCI},
we did not observe a statistically significant correlation between PRs delivered and number of contributors. Our
results show that the number of PRs delivered per release and the number of contributors in PRs per release have small positive coefficient correlation of $0.1906346$. A Pearson correlation test reveals that this
correlation is not statistically significant (\textit{p-value} $= 0.07695$). 
Our observations suggest that the increase in PRs delivered {\em after}
the adoption of \textsc{TravisCI} is not tightly related to the increase in the number of
contributors per release or release frequency. The increase in PRs delivered per release might be
due to the quicker feedback of automated tests provided by \textsc{TravisCI}. A
qualitative study with developers may shed more light upon this matter. We
further discuss this issue in Section~\ref{sec_threats_to_the_validity}.

\begin{center}
	\begin{tabular}{|p{.96\columnwidth}|}
		\hline
		\textbf{Summary:}
		\textit{After the adoption of \textsc{TravisCI}, projects deliver 3.43 times more PRs
			per release than \textit{before} \textsc{TravisCI}. The increase in
			PRs submitted, merged, and delivered after the adoption of \textsc{TravisCI} is a possible
			reason as to why projects may deliver PRs more quickly
		\textit{before} the adoption of \textsc{TravisCI}.} \\
		\textbf{Implications:}
		\textit{Teams that wish to adopt \textsc{TravisCI} should be aware that their projects will not always deliver merged PRs more quickly or release more often. Instead, a pivotal benefit of a CI service is the ability to process more contributions in a given time frame.}
		\\
		\hline
	\end{tabular}
\end{center}

\subsection*{\textbf{\RQthree}}

\textit{\textbf{Our models achieve a median $R^2$ of 0.64 using pull request
data \textit{before} the adoption of \textsc{TravisCI}, while achieving 0.67 \textit{after} the adoption of \textsc{TravisCI}.}} Moreover, the
\textit{median} bootstrap-calculated optimism is less than $0.069$ for both set
of $R^2$s obtained by our models.\footnote{\url{https://prdeliverydelay.github.io/\#rq3-r-squared-and-optimism}} 
These results suggest that our models are stable enough to perform the
statistical inferences that follow.

\begin{figure}[!t]
	\centering
	\includegraphics[width=.60\columnwidth,keepaspectratio]{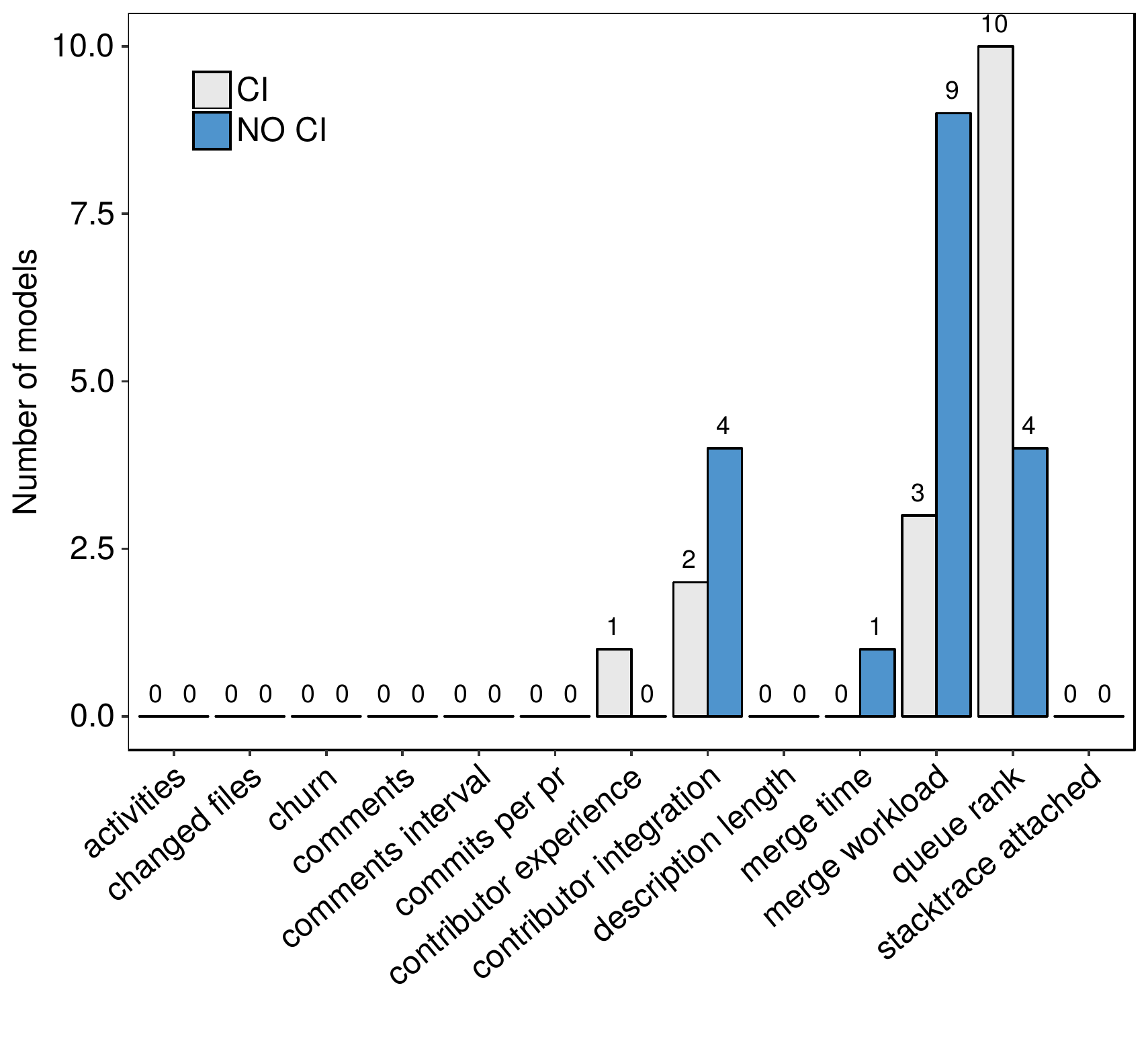}
	\caption{The number of models per most influential variables.}
	\label{fig:number_of_projects_by_influential_variables}
\end{figure}

\begin{figure*}
	\begin{subfigure}{0.4\textwidth}
		\centering
		\includegraphics[width=\linewidth,keepaspectratio]{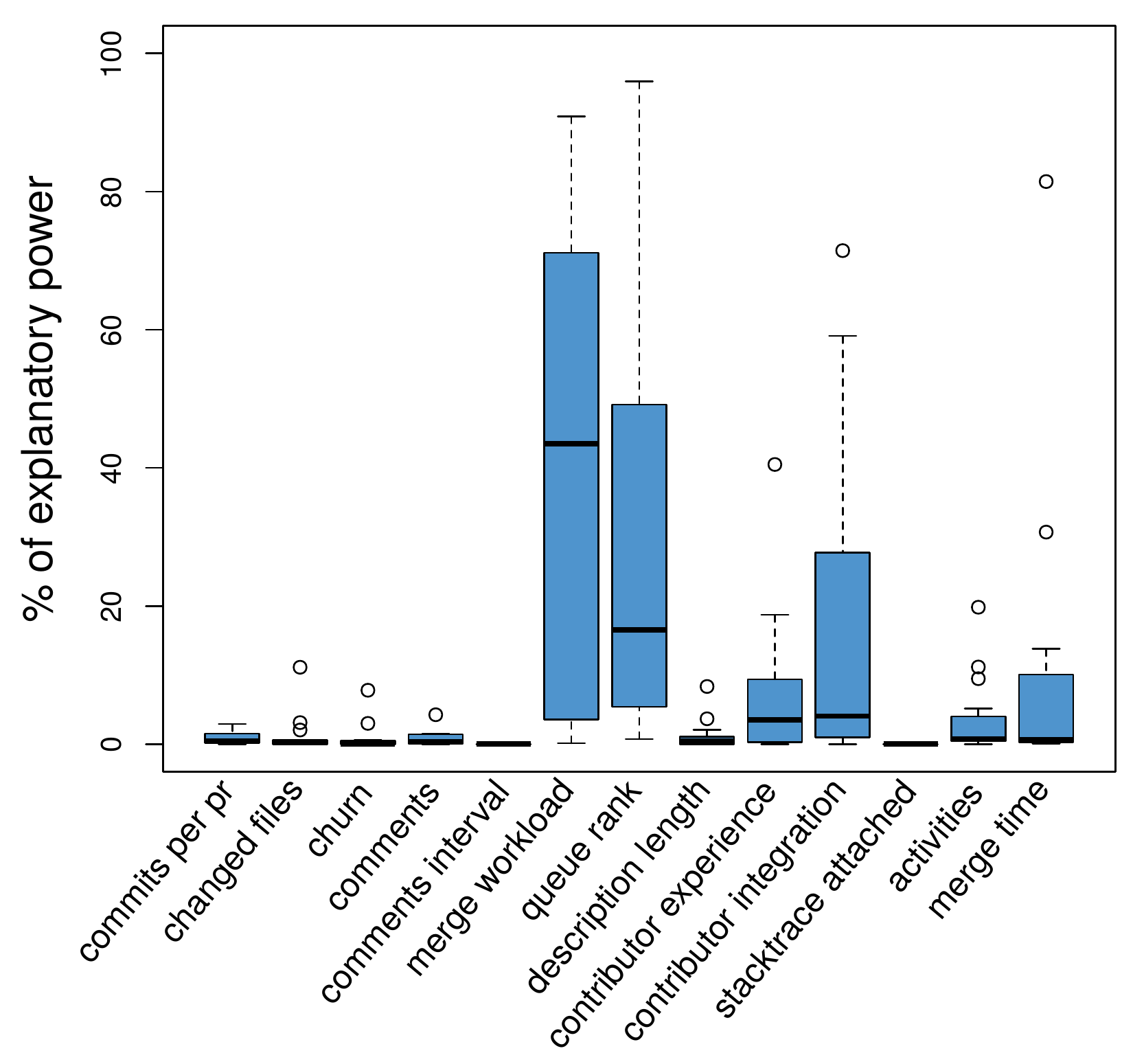}
		\caption{Explanatory power of variables \textit{before} adopting \textsc{TravisCI}.}
		\label{img:variables_importance_before_ci}
	\end{subfigure}\hfill
	\begin{subfigure}{0.4\textwidth}
		\centering
		\includegraphics[width=\linewidth,keepaspectratio]{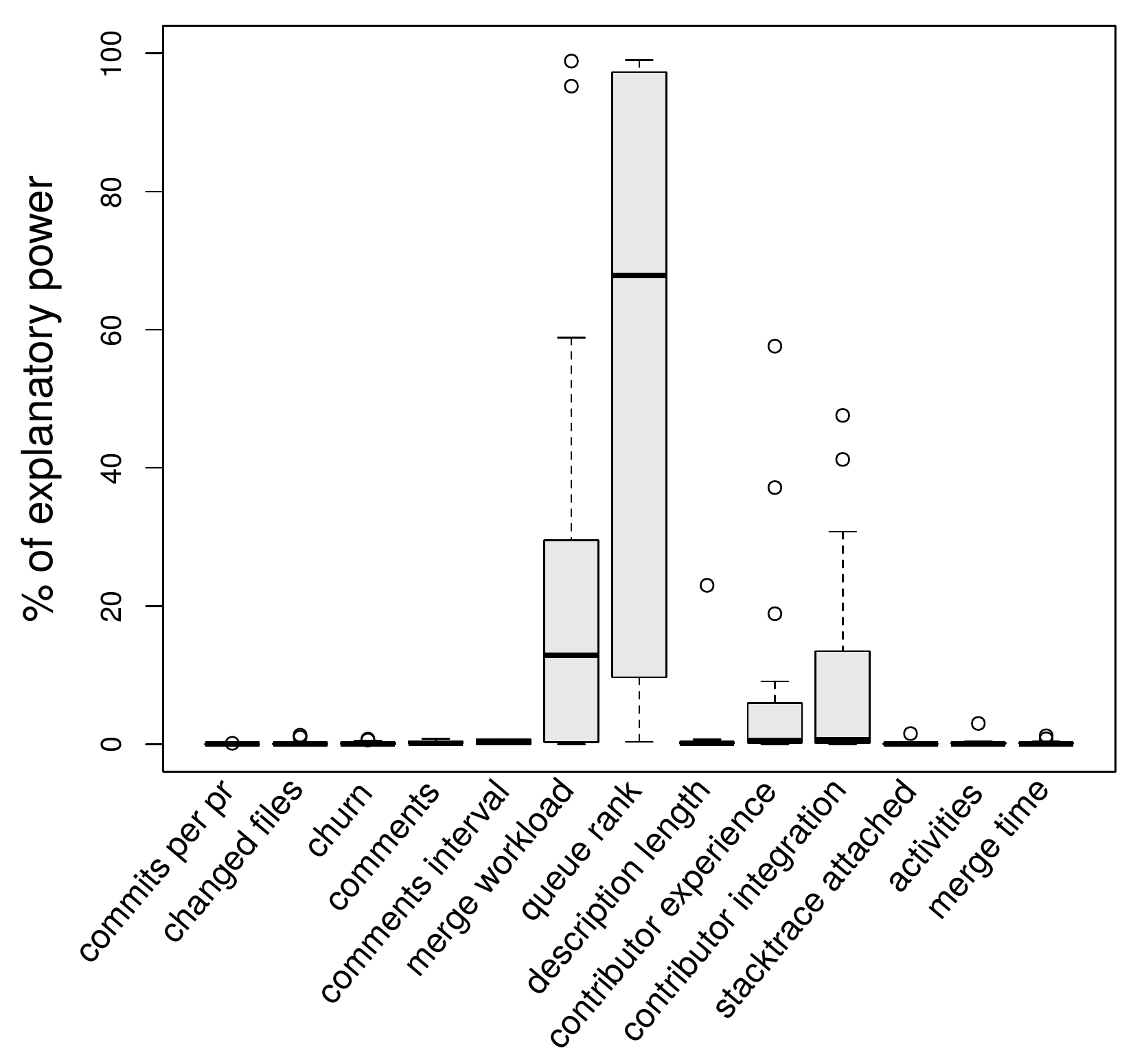}
		\caption{Explanatory power of variables \textit{after} adopting \textsc{TravisCI}.}
		\label{img:variables_importance_using_ci}
	\end{subfigure}\hfill
	\caption{Distributions of the \textit{explanatory power} of each variable of our models.}
	\label{img:variables_importance}
\end{figure*}

\begin{figure}[!t]
	\centering
	\begin{subfigure}{3.4cm}
		\includegraphics[width=3.4cm, height=3.4cm]{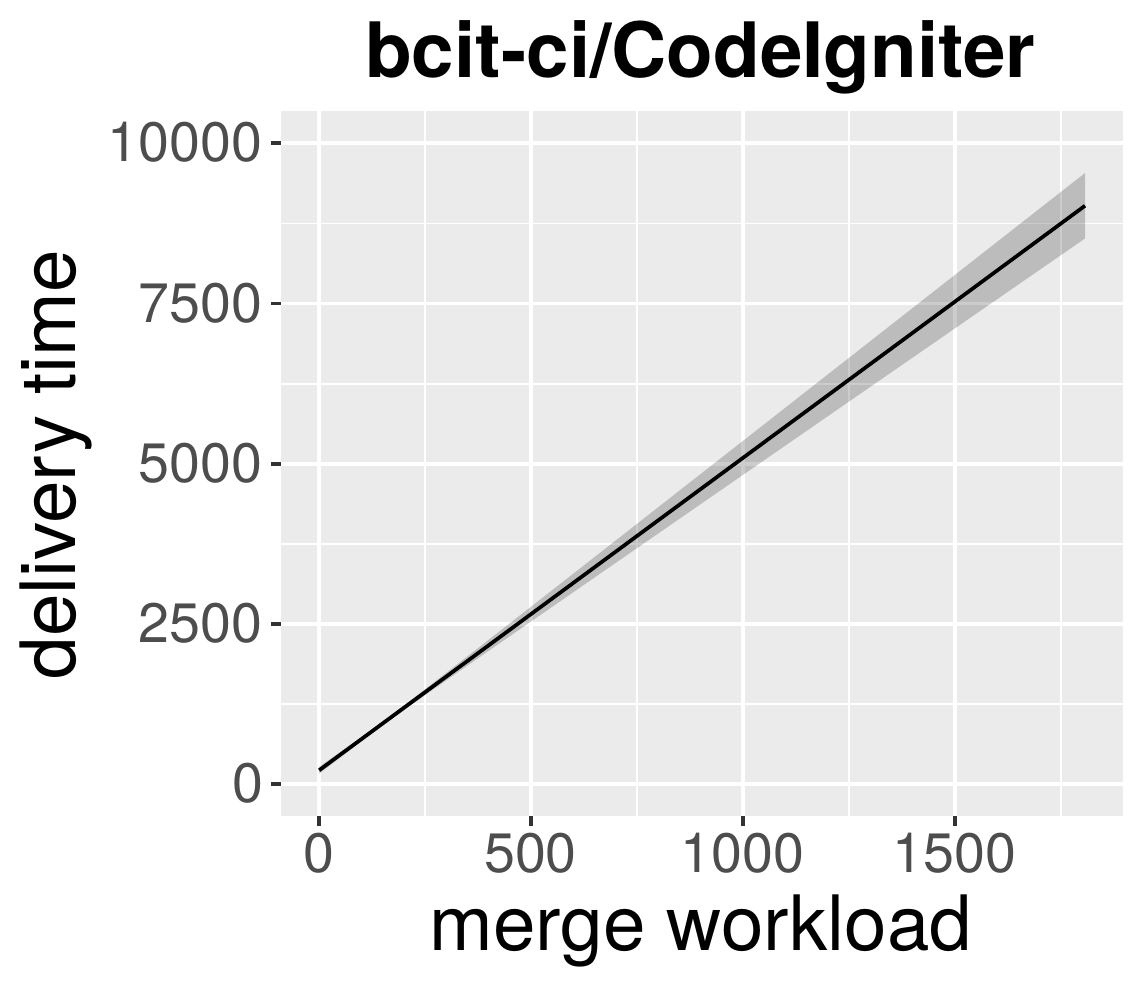}
		\vspace{-1.7em}
		\caption{}
		\vspace{0.4em}
		\label{fig:merge_workload_direction}
	\end{subfigure}%
	\begin{subfigure}{3.4cm}
		\centering
		\includegraphics[width=3.4cm, height=3.4cm]{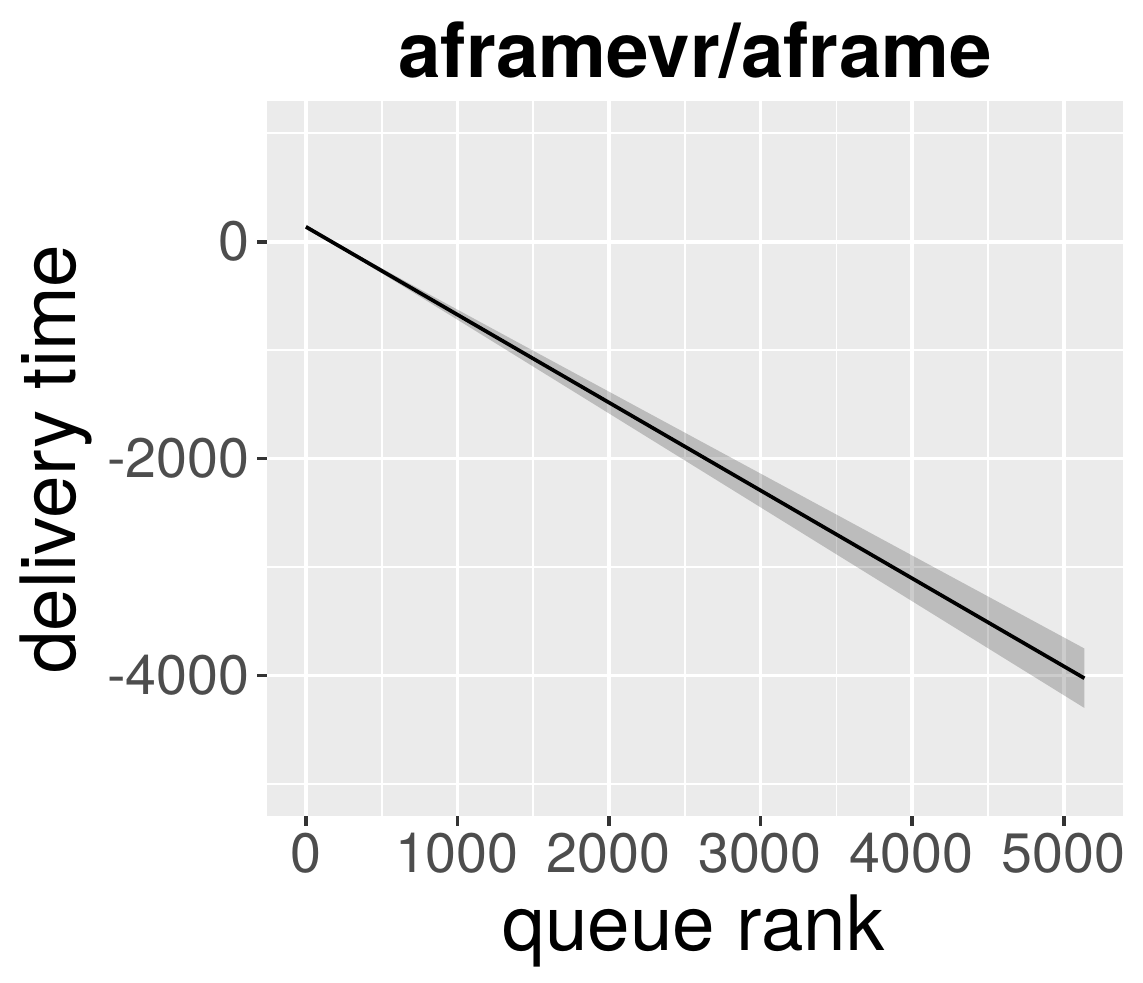}
		\vspace{-1.7em}
		\caption{}
		\vspace{0.4em}
		\label{fig:queue_rank_direction}
	\end{subfigure}
	\begin{subfigure}{3.4cm}
		\includegraphics[width=3.4cm, height=3.4cm]{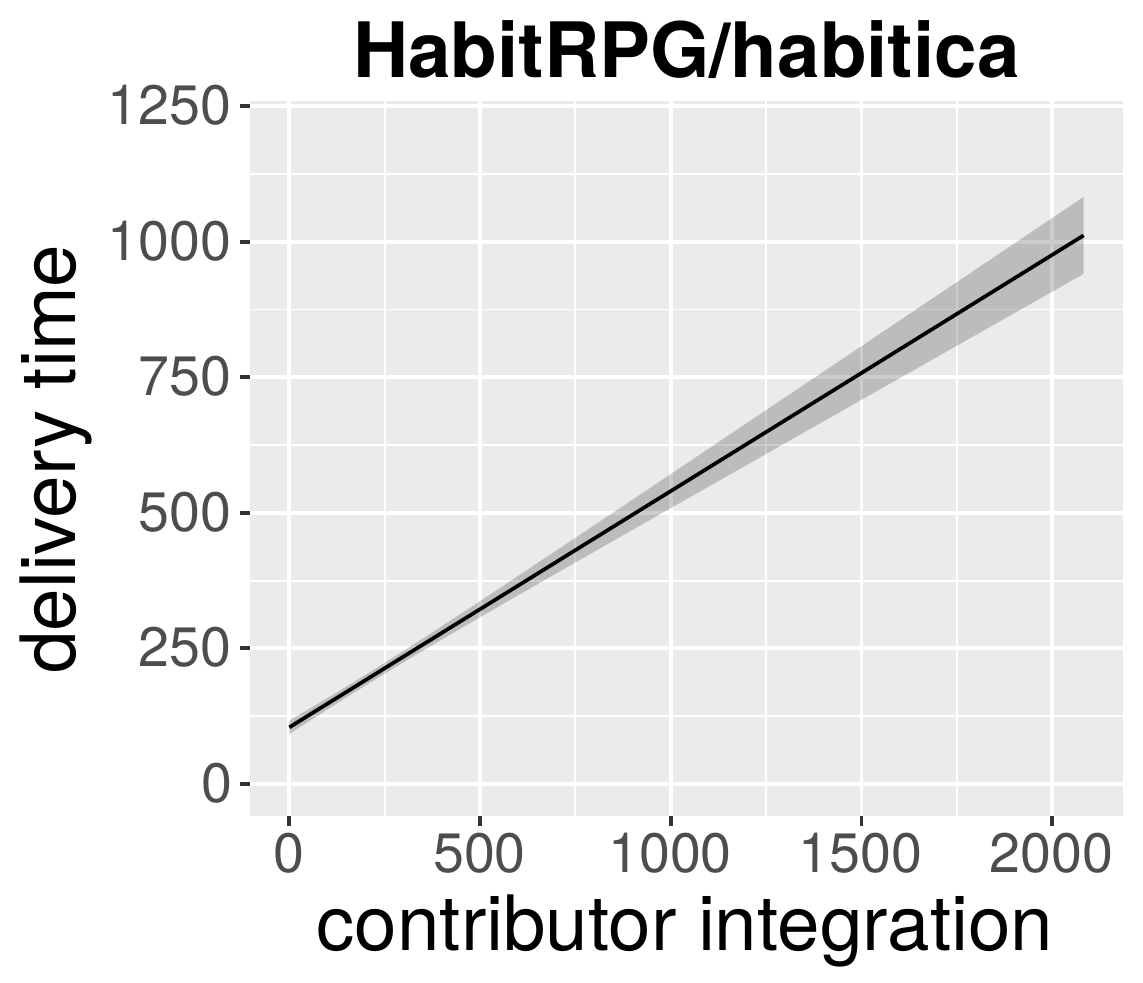}
		\vspace{-1.7em}
		\caption{}
		\label{fig:contributor_integration_direction}
	\end{subfigure}%
	\caption{The relationship between the most influential variables and delivery time.}
	\label{fig:relationship_most_important_variable}
\end{figure}

\textit{\textbf{The ``merge workload'' is the most influential variable in the
	models fit for the time period \textit{before} the adoption of \textsc{TravisCI}.}} \textit{Merge
workload} represents the number of PRs competing to be merged (see Table
\ref{tab_explanatory_variables_2}) at a point in time. Figure \ref{img:variables_importance} shows
the distributions of the explanatory power of each variable of our
models. The higher the median explanatory power for a variable, the
higher the influence of such a variable on the delivery time of PRs. We
observe that \textit{merge workload} has the strongest influence on our models to
explain delivery time {\em before} the adoption of \textsc{TravisCI}. Our models reveal that the
higher the merge workload, the higher the delivery time of 
a PR. Figure~\ref{fig:number_of_projects_by_influential_variables} shows each
explanatory variable and the number of models for which these variables are the
most influential. Indeed, \textit{merge workload} is the most influential
variable in (\nicefrac{9}{18}) of models fit for the time period
\textit{before} the adoption of \textsc{TravisCI}. Figure \ref{fig:relationship_most_important_variable}
shows the relationship between the most influential variables of our models and
delivery time. The relationship between \textit{merge workload} and
delivery time is shown in Figure \ref{fig:merge_workload_direction}. We choose
3 models with the highest $R^2$s out of the 34 models to plot the relationships.
Indeed, the rest of our models reveal a similar
trend.\footnote{\url{https://prdeliverydelay.github.io/\#rq3-variables-explanatory-power}}

\textit{\textbf{The ``queue rank'' variable is the most influential variable in
the models fit for the time period \textit{after} the adoption of \textsc{TravisCI}.}} \textit{Queue rank}
represents the moment at which a PR is merged in relation to other merged PRs within the
release cycle. Figure \ref{fig:queue_rank_direction} shows the relationship
between \textit{queue rank} and delivery time. Our models reveal that
merged PRs have a lower delivery time when they are merged more recently in the
release cycle. In addition, \textit{contributor integration} is the third most
influential variable in our models for both time periods, i.e., \textit{before} and
\textit{after} the adoption of \textsc{TravisCI}. {\em Contributor integration} represents the
average number of days that previously delivered PRs submitted by a
particular contributor took to be merged. Our models also reveal that if a contributor has
their prior submitted PRs delivered quickly, their future PR submissions tend
to be delivered more quickly (Figure
\ref{fig:contributor_integration_direction}).

\begin{center}
	\begin{tabular}{|p{.96\columnwidth}|}
		\hline
		\textbf{Summary:}
		\textit{Our models suggest that ``merge workload'' is the most
			influential variable to model the delivery time of merged PRs
			\textit{before} the adoption of \textsc{TravisCI}. Additionally, our
			models show that after the adoption of \textsc{TravisCI}, merged PRs have a lower delivery time when they are merged more recently in the release cycle.} \\
		\textbf{Implications:}
		\textit{If software development teams plan to deliver their merged PRs more quickly to their end-users, they should consider having shorter release cycles.}
		\\
		\hline
	\end{tabular}
\end{center}

\section{\textbf{Qualitative Study Results}}
\label{sec_qualitative_study_results}

We first discuss the demographics of our participants, focusing on their domain, their main software development activities, and their experience using CI. Afterward, we disclose the findings of each RQ ($RQ4$---$RQ8$).

\begin{figure}[H]
	\centering
	\begin{subfigure}[b]{0.3\textwidth}
		\centering
		\includegraphics[width=2.7cm, height=4.0cm]{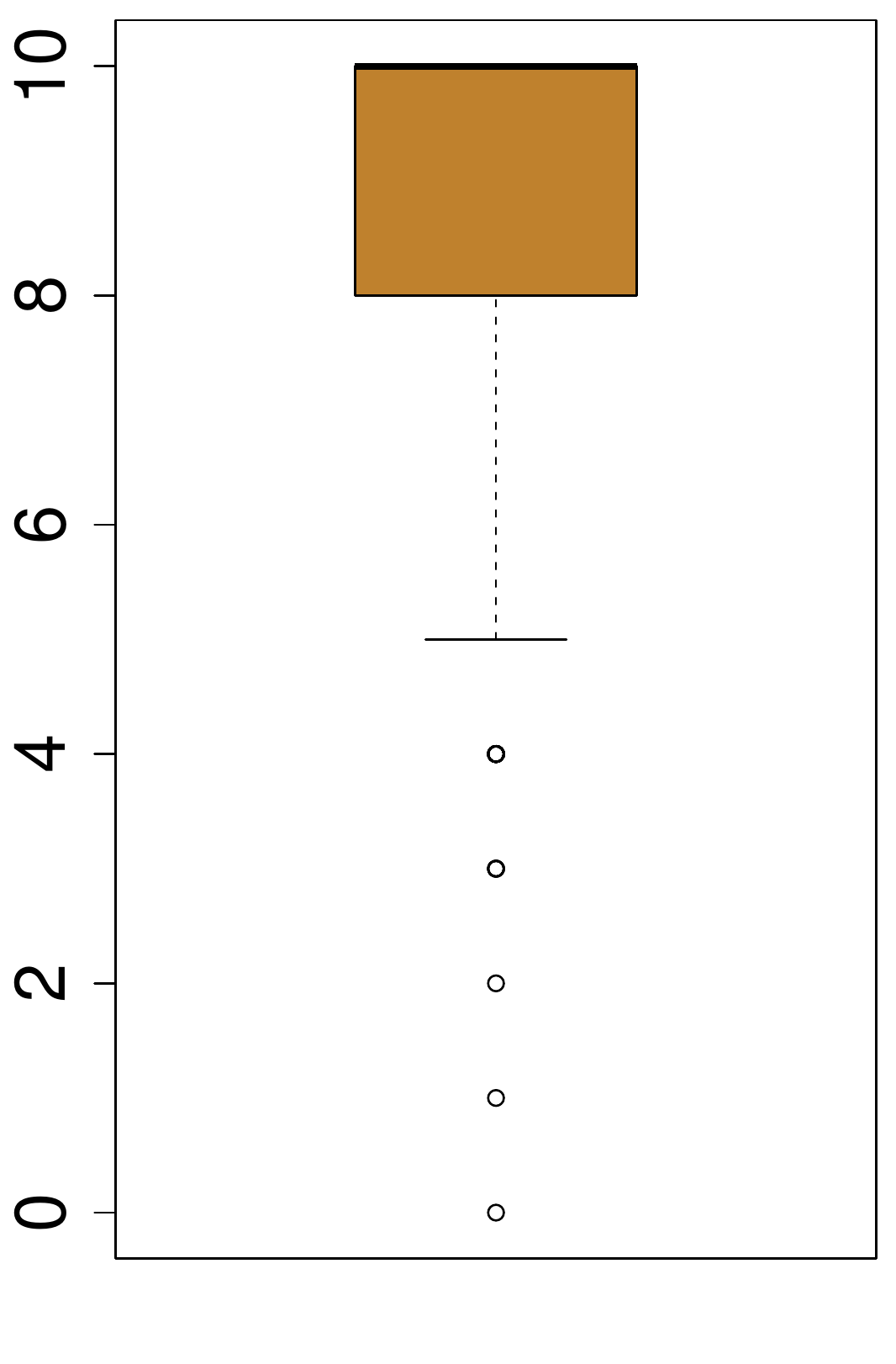}
		\caption{In general.}
		\label{fig:developers_experience_in_general}
	\end{subfigure} %
	\begin{subfigure}[b]{0.3\textwidth}
		\centering
		\includegraphics[width=2.5cm, height=4.0cm]{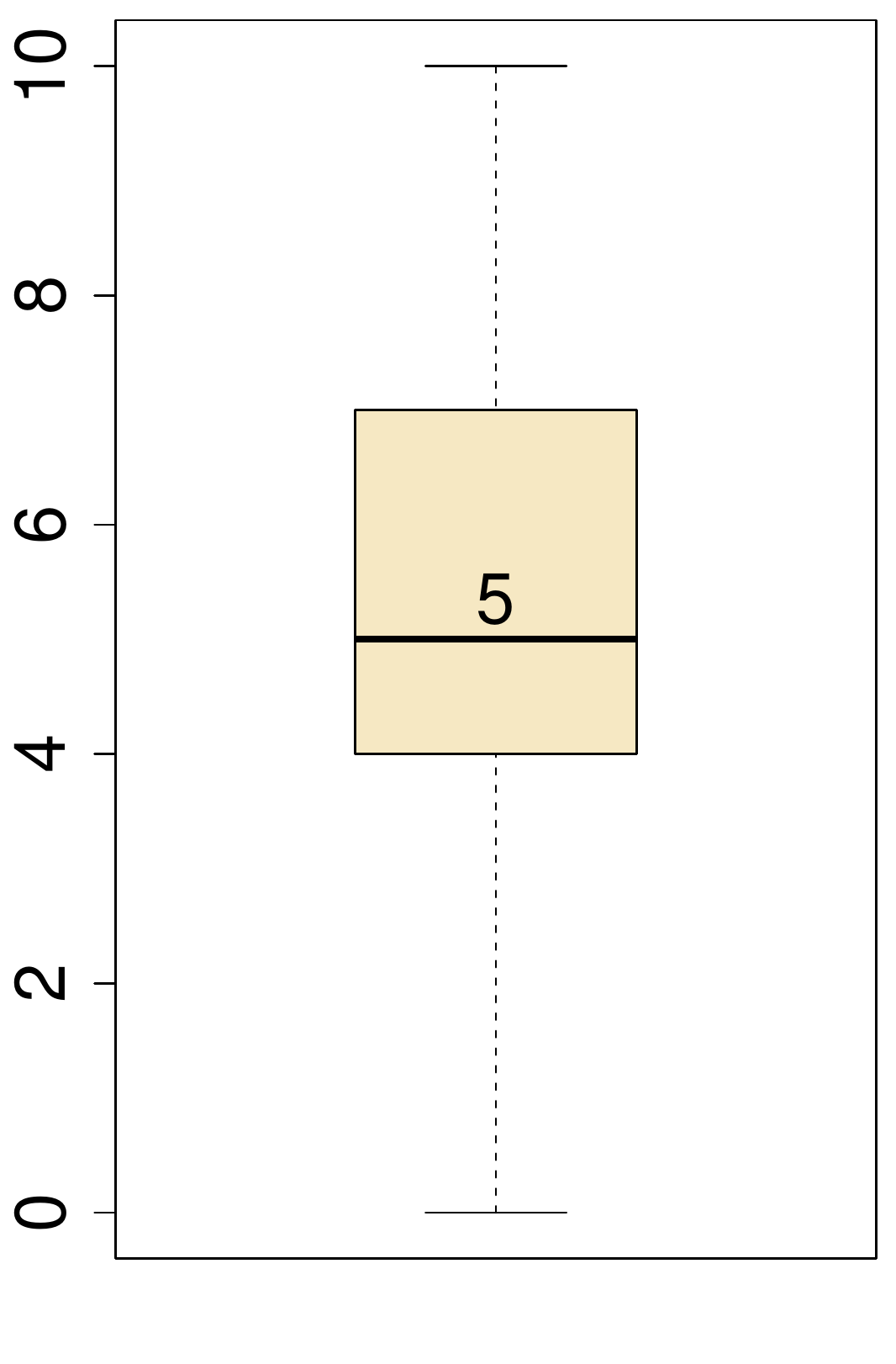}  
		\caption{Using CI.}
		\label{fig:developers_experience_using_ci}
	\end{subfigure}
	\caption{Participants' experience in software development.}
	\label{fig:developers_experience}
\end{figure}

The number of responses to each question may vary as none of the questions are mandatory in our survey (to encourage a higher response rate). Hence, not all participants answered all questions. 
Figure \ref{fig:developers_experience_in_general} shows the participants' experience in software development. 
We collect the data in Figure~\ref{fig:developers_experience_in_general} from \textit{Question \#3} where the options range from ``0 years'' to ``10 or more years''. We observe that 78\% (\nicefrac{348}{444}) of participants have eight or more years of experience in software development. 
Finally, Figure \ref{fig:developers_experience_using_ci} shows the experience of participants with CI. 64.2\% (\nicefrac{282}{439}) of participants have five or more years of experience with CI. Only five (1.3\%) participants report less than one year of experience with CI. 

\begin{figure}[!t]
	\centering
	\includegraphics[ width=9cm]{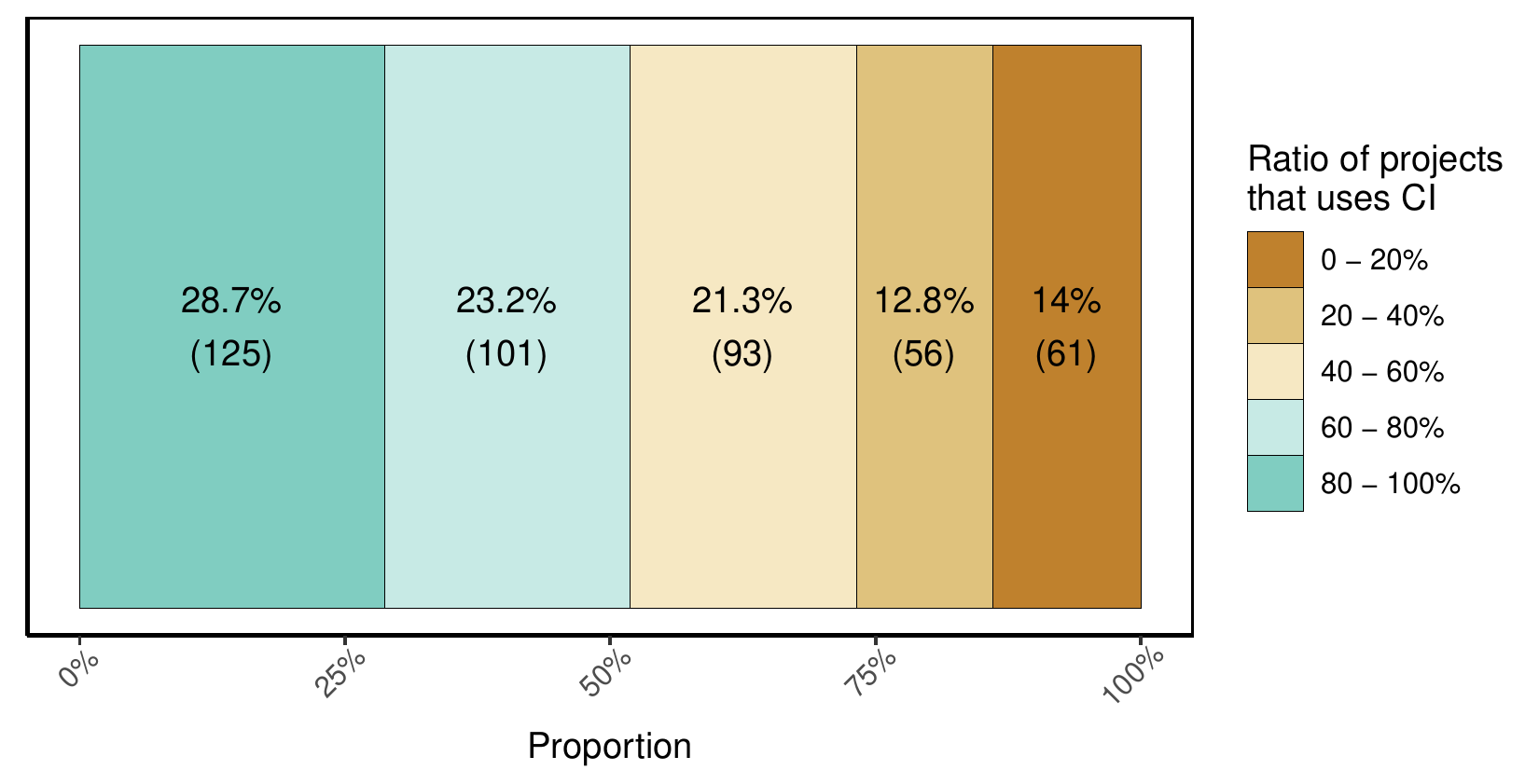}
	\caption{Proportion of participants' projects that use CI (\textit{Question \#5}).}
	\label{fig:ratio_of_projects_using_ci}       
\end{figure}

\begin{figure}[H]
	\centering
	\includegraphics[ width=10cm]{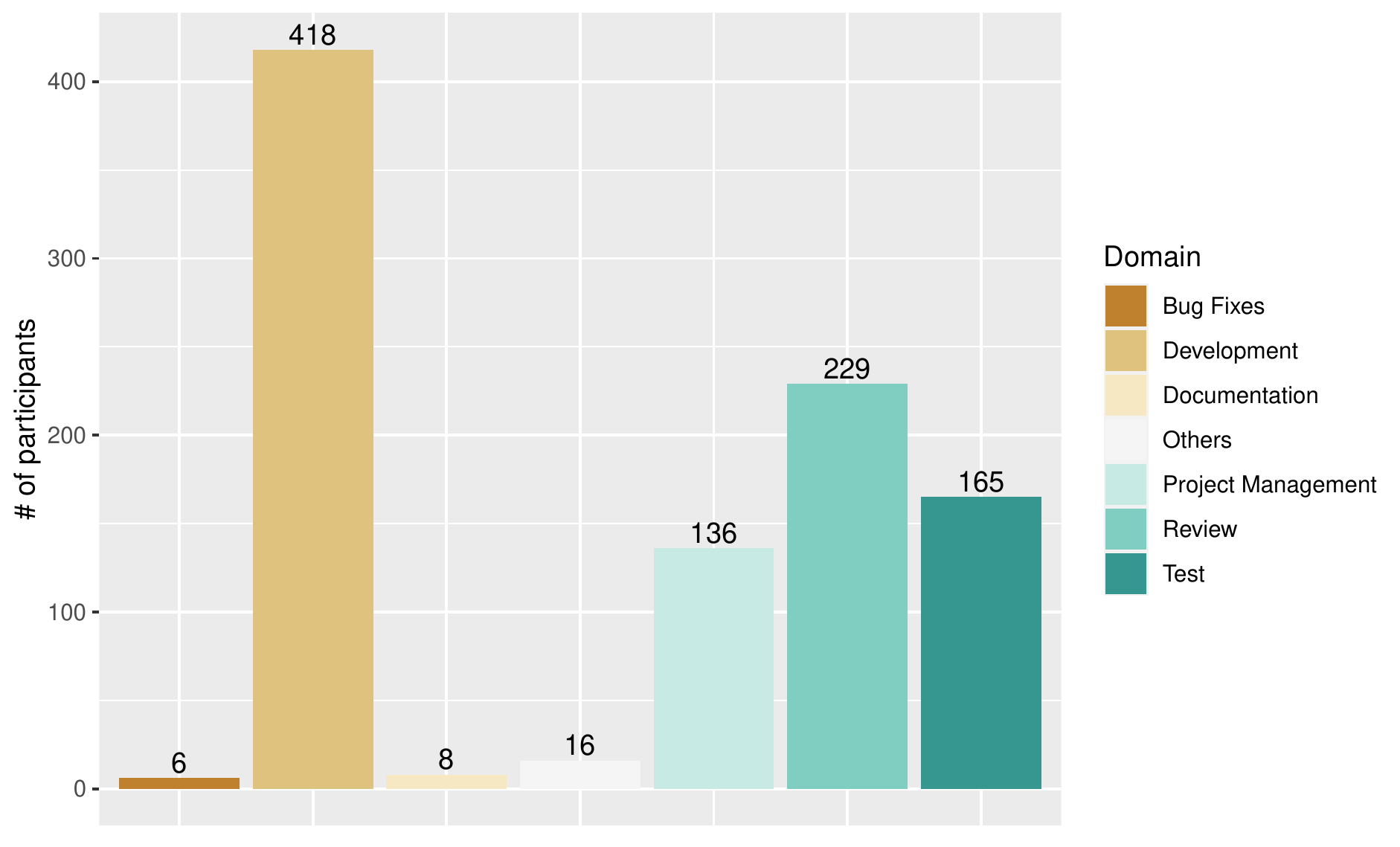}
	\caption{Developers' main activities.}
	\label{fig:main_development_activities}       
\end{figure}

We observe that 51.8\% (\nicefrac{226}{436}) of participants used CI in 60--100\% of their projects (see Figure \ref{fig:ratio_of_projects_using_ci}). 
In terms of participants' main activities, we observe that the development of new features is the most common activity among them, followed by test and review. A total of 92.9\% of participants state that developing new features is one of their main activities in their projects. Another 229 participants (50.9\%) state that code review is another main activity they perform in their projects. Figure \ref{fig:main_development_activities} shows the participants' main activities according to their own classification (\textit{Question \#7}). Given that a participant can take on several roles, the sum of the percentages in Figure \ref{fig:main_development_activities} can be greater than 100\%.
Considering the demographics, we were able to collect a diverse set of participants in terms of experience with CI, domain area, and development activities (e.g., bug fixing, developing new features, or code review). 

With respect to {\em domain expertise}, we observe that {\em web development} is the most common domain of our participants. 77.8\% (\nicefrac{350}{450}) of participants state that web development is one of their main domains (\textit{Question \#8}). The second and third most common domains are business software development (35.3\%, \nicefrac{159}{450}) and mobile applications (32.2\%, \nicefrac{145}{450}). 
Business software is a system used to measure and improve enterprise productivity and to perform other business functions. Document Management Systems, Employee Scheduling Software, and Enterprise Resource Planning (ERP) are examples of business software. Additionally, a significant number of participants are from areas such as scientific development (i.e., those who develop software systems to analyze, visualize, or simulate processes or data) and big data (i.e., those who use scientific software to process and analyze data). This diversity of domains demonstrates that we obtain insights from several roles and development areas.

\begin{figure}[H]
	\centering
	\includegraphics[ width=10cm]{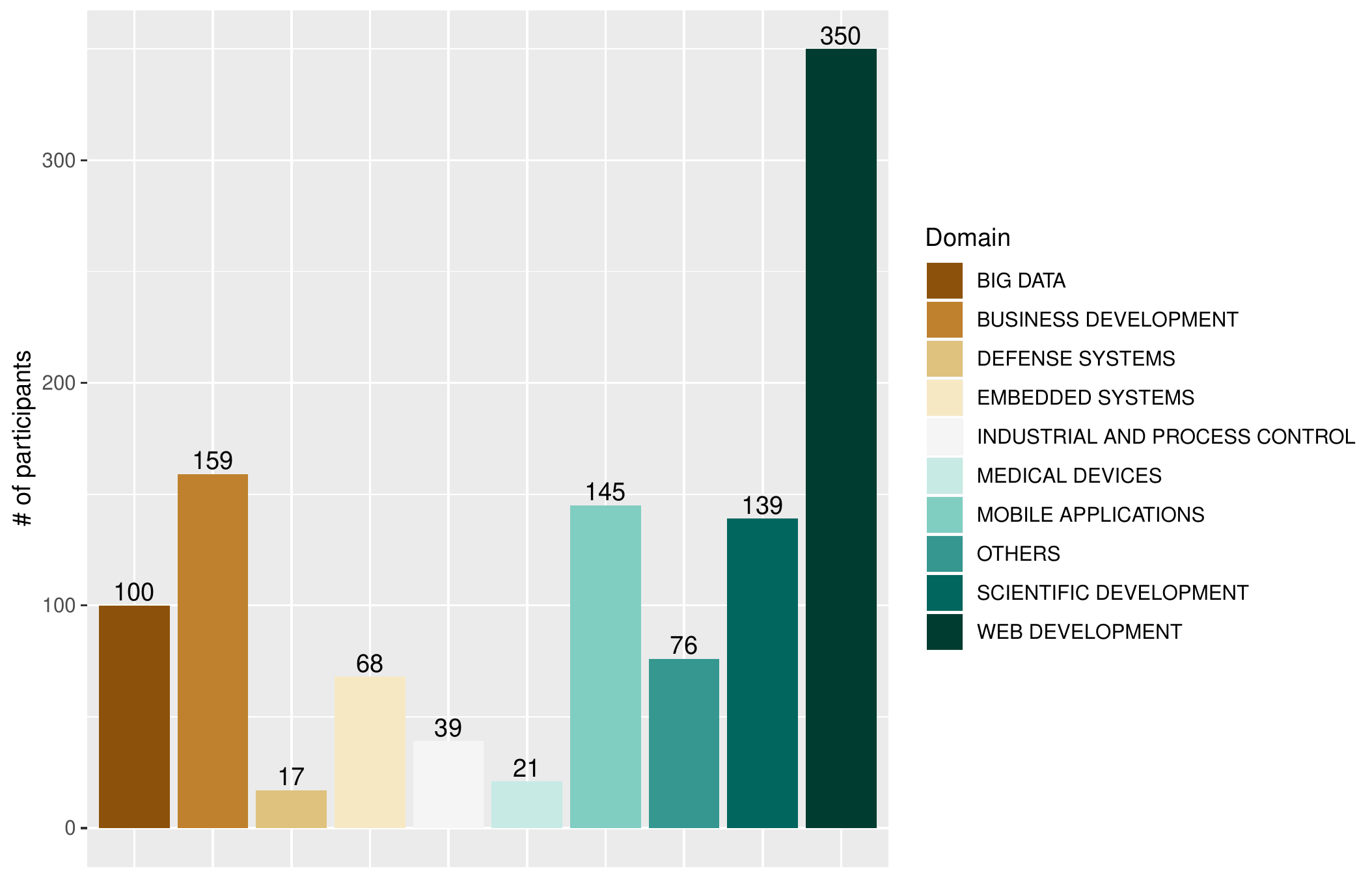}
	\caption{Participants' domain area.}
	\label{fig:developer_domain_area}       
\end{figure}

In Table \ref{tab:themes_per_research_question} we present the summary of Themes that were generated by our thematic analysis for each research question (RQ). 
For instance, in $RQ4$ we investigate the perceived influence of CI on the delivery time of merged PRs, which is associated with the following themes: \textit{automation}, \textit{project quality}, and \textit{release process}. 
Additionally, a theme may emerge in the results of more than one RQ. For instance, the \textit{automation} theme emerges in the results of both $RQ4$ and $RQ6$. This is because participants of our study believe that \textit{automation} impacts the delivery time of merged PRs ($RQ4$) while also impacting the release process of projects ($RQ6$). Table \ref{tab:themes_per_research_question} presents an overview of the findings of the qualitative study. The detailed analysis for each theme is described in the result section of $RQ4$---$RQ8$.

\begin{table}
	\centering
	\caption{High-level Overview of the Themes per RQ of the qualitative study.}
    \begin{tabular}{cp{15em}}
    \hline
    \textbf{Research Question} & \multicolumn{1}{c}{\textbf{Theme}} \bigstrut\\
    \hline
    \multicolumn{1}{l}{\multirow{3}[6]{*}{\textbf{RQ4:} Impact of CI on the PR delivery time}} & \multicolumn{1}{l}{Automation} \bigstrut\\
    \cline{2-2}          & \multicolumn{1}{l}{Project quality} \bigstrut\\
    \cline{2-2}          & Release process \bigstrut\\
    \hline
    \multicolumn{1}{l}{\multirow{6}[13]{*}{\textbf{RQ5:} Themes that impact the PR delivery time in general}} & \multicolumn{1}{l}{PR characteristics} \bigstrut\\
\cline{2-2}          & \multicolumn{1}{l}{Project maintenance} \bigstrut\\
\cline{2-2}          & \multicolumn{1}{l}{Release process} \bigstrut\\
\cline{2-2}          & \multicolumn{1}{l}{Team characteristics} \bigstrut\\
\cline{2-2}          & \multicolumn{1}{l}{Contributors} \bigstrut\\
\cline{2-2}          & \multicolumn{1}{l}{Testing} \bigstrut\\
    \hline
    \multicolumn{1}{l}{\multirow{3}[6]{*}{\textbf{RQ6:} Influence of CI on the project release process}} & Automation \bigstrut\\
\cline{2-2}          & Project Stability \bigstrut\\
\cline{2-2}          & Release characteristics \bigstrut\\
    \hline
    \multicolumn{1}{l}{\multirow{2}[4]{*}{\textbf{RQ7:} Influence of CI on the project review process}} & CI does not impact code review \bigstrut\\
\cline{2-2}          & CI impacts code review \bigstrut\\
    \hline
    \multicolumn{1}{l}{\multirow{2}[4]{*}{\textbf{RQ8:} Impact of CI on attracting more contributors}} & Attractive project characteristics \bigstrut\\
\cline{2-2}          & Lower contribution barrier \bigstrut\\
    \hline
    \end{tabular}%
  \label{tab:themes_per_research_question}%
\end{table}%
\subsection*{\textbf{\RQfive}}

\begin{figure}[h!]
	\includegraphics[height=4cm, width=12cm]{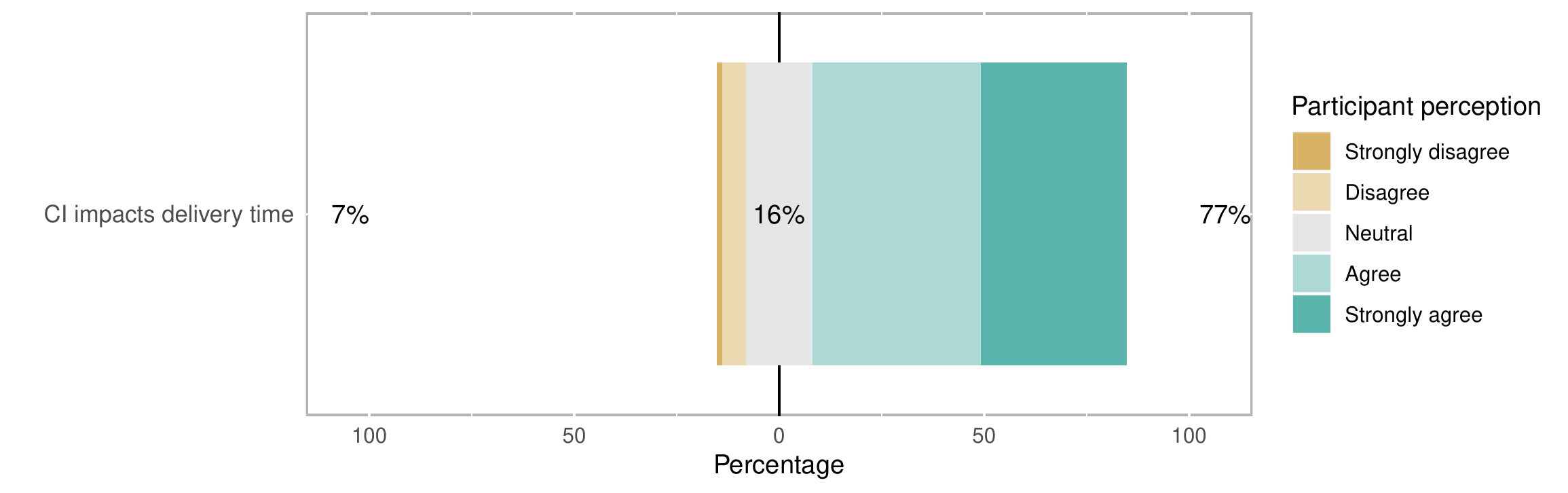}
	\caption{Developer's perception about the influence of continuous integration on the delivery time of merged pull requests (\textit{Question \#12}).}
	\label{fig:developer_perception_about_impact_of_ci_on_delivery_time}       
\end{figure}

\vspace{3mm}	
\noindent\textbf{77\% (\nicefrac{338}{441}) of participants agree with the statement that CI shortens the delivery time of merged PRs.}
In question \#12 of our survey, we ask participants to express the extent to which they agree with the following statement: ``the adoption of CI shortens the time to deliver merged PRs to end users.'' Most of our participants (77\%, \nicefrac{338}{441}) agree with the statement, while 16\% (\nicefrac{72}{441}) are neutral, and 7\% (\nicefrac{31}{441}) disagree or strongly disagree with the statement (see Figure \ref{fig:developer_perception_about_impact_of_ci_on_delivery_time}). However, it is not clear whether these latter participants perceive CI as not having any influence on the delivery time of merged PRs or whether they perceive CI as having a negative influence on the delivery time of merged PRs.

After analyzing our participants' responses, the influence of CI on the delivery time of merged PRs was captured through the following themes: \textit{release process}, \textit{project quality}, and \textit{automation}. The description and examples of mentions for each theme are presented in the following. Additionally, Table \ref{tab:frequency_citation_CI_factors_impact} shows the frequency of mentions for each code and theme related to how CI may influence the delivery time of merged PRs. 

\begin{table}
	\centering
	\caption{Frequency of mentions in participants' responses for each code and theme related to how CI may influence the delivery time of merged PRs.}
	\begin{tabular}{cp{10.75em}cc}
		\hline
		\multirow{2}[4]{*}{\textbf{Theme}} & \multirow{2}[4]{*}{\textbf{Code}} & \multicolumn{2}{c}{\textbf{Frequency}} \bigstrut\\
		\cline{3-4}          & \multicolumn{1}{c}{} & \multicolumn{1}{p{5.415em}}{\textbf{Frequency per code}} & \multicolumn{1}{p{5.585em}}{\textbf{Frequency per theme}} \bigstrut\\
		\hline
		\multirow{7}[14]{*}{\textbf{Automation}} & Automated testing & 39    & \multirow{7}[14]{*}{99} \bigstrut\\
		\cline{2-3}          & Earlier feedback & 24    &  \bigstrut\\
		\cline{2-3}          & Reduced testing time & 10    &  \bigstrut\\
		\cline{2-3}          & Reduced burden on reviewers & 9     &  \bigstrut\\
		\cline{2-3}          & Automated building & 7     &  \bigstrut\\
		\cline{2-3}          & Less manual work & 6     &  \bigstrut\\
		\cline{2-3}          & Improved automation & 4     &  \bigstrut\\
		\hline
		\multirow{5}[10]{*}{\textbf{Project quality}} & Code quality & 14    & \multirow{5}[10]{*}{130} \bigstrut\\
		\cline{2-3}          & Code stability & 23    &  \bigstrut\\
		\cline{2-3}          & Higher test coverage & 6     &  \bigstrut\\
		\cline{2-3}          & Better confidence & 83    &  \bigstrut\\
		\cline{2-3}          & Reduced regression risk & 4     &  \bigstrut\\
		\hline
		\multicolumn{1}{c}{\multirow{3}[6]{*}{\textbf{Release process}}} & Faster release cycle & 29    & \multirow{3}[6]{*}{49} \bigstrut\\
		\cline{2-3}          & Automated deployment & 17    &  \bigstrut\\
		\cline{2-3}          & Smaller release & 3     &  \bigstrut\\
		\hline
	\end{tabular}%
	\label{tab:frequency_citation_CI_factors_impact}%
\end{table}%

\vspace{0.6mm}
\noindent\textbf{Release process.\textsuperscript{(49)}} Several responses to our questionnaire indicate that the adoption of CI influences the delivery time of merged PRs because CI promotes \textit{faster release cycles}.\textsuperscript{(29)} For instance, 
C346 declares that \textit{``Good use of CI could help in faster release cycle, because you can be more confident in shipping something that works.''} Some practitioners of CI \citep{goodman2008s} have claimed that some benefits of using CI are the improved release frequency and predictability. However, our quantitative study ($RQ2$) does not support such a claim. We do not observe a significant difference in release frequency \textit{after} the adoption of a CI service (e.g., \textsc{TravisCI}) for the studied projects. Furthermore, we found that \textit{after} the adoption of \textsc{TravisCI}, projects delivered 3.43 times more PRs per release than \textit{before} the adoption of \textsc{TravisCI}. We observe that although the release frequency was not significantly affected by the adoption of a CI service, projects process substantially more PRs per release than \textit{before} the adoption of \textsc{TravisCI}.
Furthermore, \textit{automated deployment}\textsuperscript{(17)} can be another step in the adoption of CI which helps projects to rapidly deliver software changes to end users \citep{Humble2010-ca}. 
Automated deployment refers to the process of making developers' code available to end users automatically \citep{rahman2015synthesizing}.  C347 explains that \textit{``CI is the only way to automate deployment, thus speeding up customer delivery''}.
 Finally, developers also mentioned \textit{smaller releases}\textsuperscript{(3)} as an influencing factor of CI on the delivery time of merged PRs. For instance, C076 states that it \textit{``makes sense for the releases to be smaller and more frequent as a project reaches a level of stability.''} 

\vspace{0.6mm}
\noindent\textbf{Project Quality.\textsuperscript{(130)}} This is the theme most mentioned by our participants. According to participants, CI influences the delivery time of merged PRs by increasing \textit{code quality},\textsuperscript{(14)} providing \textit{code stability},\textsuperscript{(23)} and reducing \textit{regression risks}.\textsuperscript{(4)} For example, C280 states that CI promotes a \textit{``much better quality of contributions and therefore much shorter release cycles.''} Also, CI influences code stability, as declared by C270: \textit{``it makes the required time to deliver shorter because the maintainer can be relatively sure the change does not break other use cases.''} 
According to \cite{Vasilescu2015-tn}, core developers using CI can discover more bugs than developers in projects not using CI. \textit{Better quality confidence}\textsuperscript{(83)} is the most mentioned code when it comes to the adoption of CI. According to participants, the delivery time of merged PRs is positively influenced by CI because 
\textit{``it's much easier to trust a PR that was built and tested at a CI environment than having to do everything manually on my own machine''} (C361). This trust in CI is also related to a \textit{higher test coverage}.\textsuperscript{(6)} For instance, C151 states that
\textit{``CI requires a good test suite, which gives confidence in the correctness of the project.''} Indeed, poor test coverage may make successful builds misleading~\citep{felidre2019continuous} (i.e., the builds may still contain unidentified bugs). 

\vspace{06.mm}
\noindent\textbf{Automation.\textsuperscript{(99)}} 
According to our participants, CI also influences the delivery time of merged PRs by \textit{improving automation}.\textsuperscript{(4)} Automation is a key aspect of CI. Projects implementing proper CI must at least automate their build and testing processes. Automation leads to \textit{less manual work},\textsuperscript{(6)} as explained by C203 when stating that \textit{``Yes, the amount of manual work involved in a release is much less when you use CI.''}	
The \textit{automated testing}\textsuperscript{(39)} code is mentioned several times as influencing the delivery time of merged PRs, which is in accordance with the study by \cite{rahman2015synthesizing}.
For example, C361 states that \textit{``I worked on the Bokeh project both before and after the full integration of \textsc{TravisCI}. Previously, major releases took months of planning due to a lack of manpower needed for running tests. With the adoption of CI, new versions can be released semi-monthly thanks to CI greatly reducing the number of man-hours needed for testing.''} 

The automated test execution provides \textit{earlier feedback},\textsuperscript{(24)} which is also recurrently mentioned by our participants. For instance, C439 states that \textit{``early errors are identified by CI, so it facilitates the delivery process.''} 
Also according to our participants, CI contributes to a \textit{reduced testing time}.\textsuperscript{(10)} C062 states that \textit{``when you have CI (+ a strong test suite) you can in some cases shorten a lot the manual test of that bug fix, or in some cases even skip it entirely (when the fix is simple enough).''} Finally, it was also mentioned that CI contributes to a \textit{reduced burden on reviewers}.\textsuperscript{(9)} For instance, C026 declares that \textit{``relying on reviewers to build to test will increase time to ship massively and will be a drain on the already scarce resource of reviewers.''} This observation is interesting as it may explain our results in $RQ2$ related to the higher number of PRs delivered per release (see Section~\ref{sec_quantitative_study_results}), i.e., as there is less burden to reviewers, they have more capacity to review PRs that, otherwise, would have waited longer in the delivering queue. Previous studies observed that there is less discussion in PRs \textit{after} the adoption of CI \citep{cassee2020silent}, which could also save reviewers' time. The use of CI automates several tasks in software development (i.e., build and test), thus saving time from maintainers, so they can focus on the content of the proposed software changes and launch software releases. Indeed, several participants stated that \textit{automated building}\textsuperscript{(7)} quickens the delivery time of merged PRs. For instance, C400 declares that \textit{``it [automated building] reduced the required time [to deliver] since CI built code, run tests etc.''} 

\begin{figure}[hbt!]
	\includegraphics[height=6cm, width=12cm]{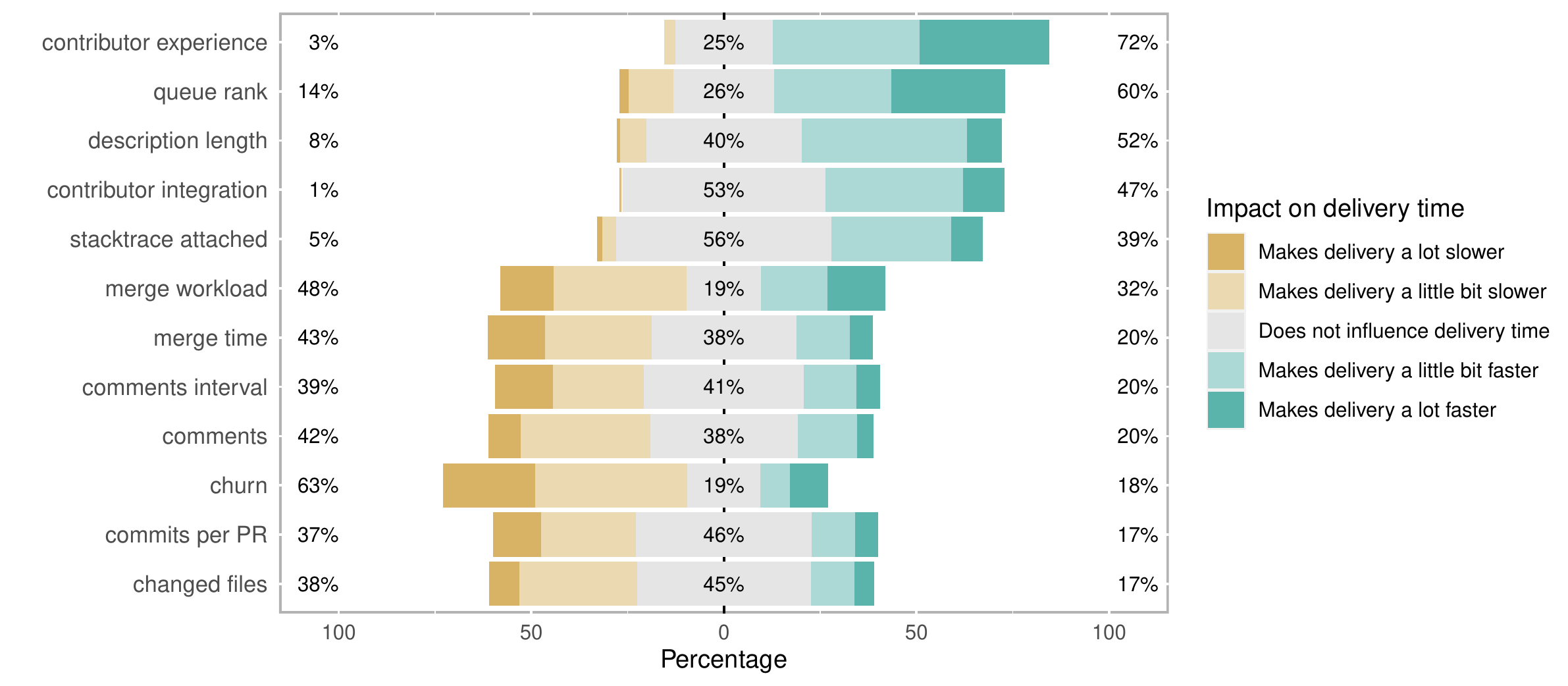}
	\caption{Developers' perception about factors' impact on the delivery time of merged PRs (\textit{Question \#12}).}
	\label{fig:factors_impact_on_delivery_time}       
\end{figure}

\vspace{06.mm}
\noindent\textbf{According to our participants, \textit{PR churn} and the number of PRs waiting to be merged (\textit{merge workload}) are the most important variables of our models in Study I}. 
In question \#12, we request our participants to rate the degree to which 12 variables used in our regression models (see $RQ3$, Tables \ref{tab_explanatory_variables_1} and \ref{tab_explanatory_variables_2}) may influence the delivery time of merged PRs. We present the following variables: (i) a number that represents the moment at which a PR is merged compared to other merged PRs within the release cycle (queue rank); (ii) contributor integration; (iii) stack-trace attached; (iv) description size; (v) contributor experience; (vi) merge workload; (vii) changed files; (viii) churn; (ix) merge time; (x) number of commits per PR; (xi) number of comments; and (xii) interval of comments. 

\begin{figure}[tbh]
	\includegraphics[height=8cm, width=12cm]{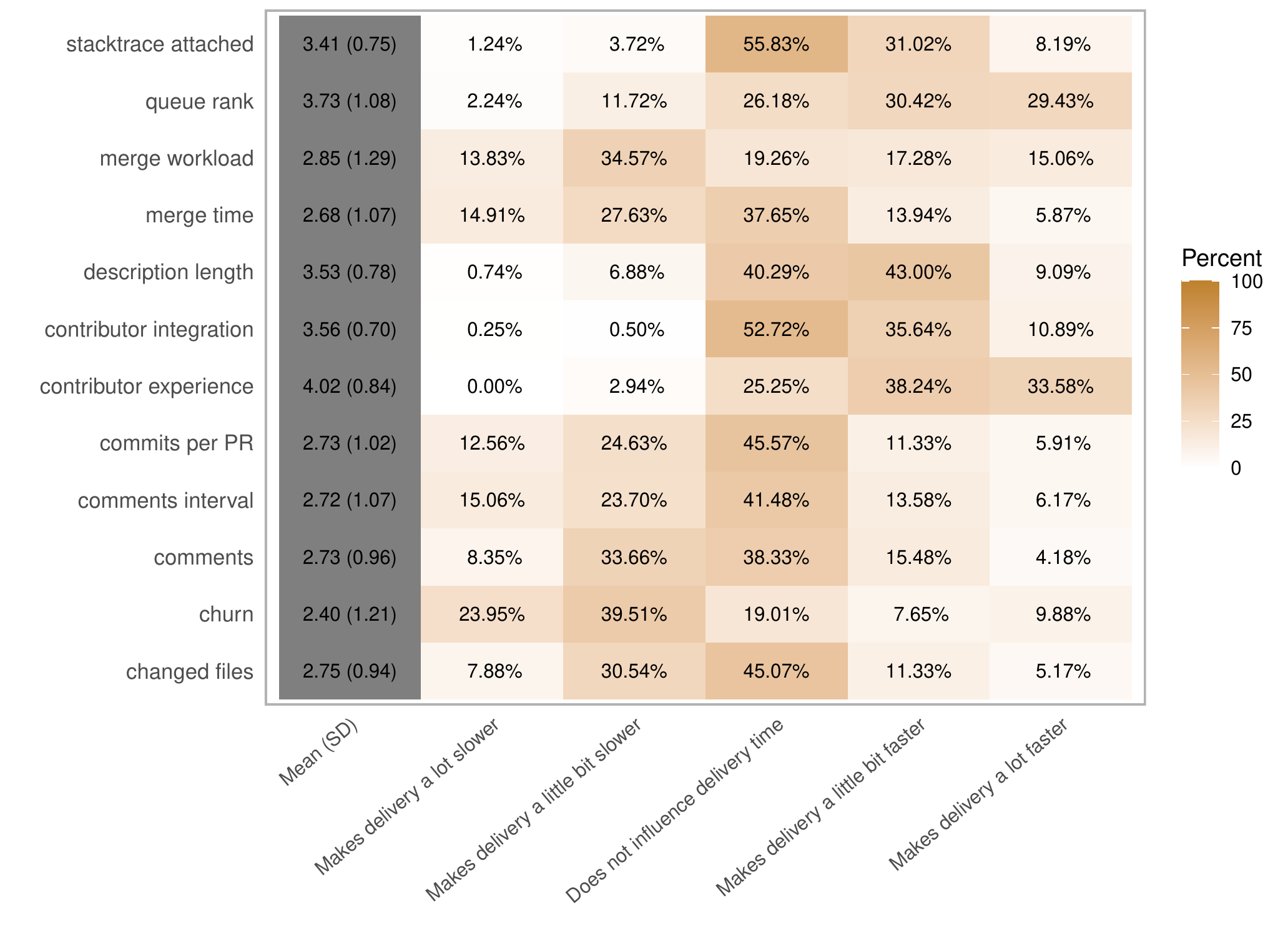}
	\caption{Rating of codes related to delivery time of merged PRs (\textit{Question \#13}).}
	\label{fig:factors_impact_on_delivery_time_heating}       
\end{figure}

Our participants were invited to rate their perception regarding the impact of each above-mentioned variable on a 5-point Likert scale. The options were the following: (i) makes delivery a lot slower; (ii) makes delivery a little bit slower; (iii) does not influence the delivery time; (iv) makes delivery a little bit faster; and (v) makes delivery a lot faster.
Figure \ref{fig:factors_impact_on_delivery_time} shows the perception of participants regarding the influence of each variable on the delivery time of merged PRs. Figure \ref{fig:factors_impact_on_delivery_time_heating} shows the frequency of each rating per variable. The lower the percentage of \textit{``does not influence delivery time,''} the higher the perceived influence of a variable on the delivery time of merged PRs.

The variables that are most rated as having influence on the delivery time are \textit{churn} and \textit{merge workload}. This finding is partially in agreement with our regression models (see $RQ3$). The \textit{merge workload} is also one of the most influential variables in our models. According to our regression models, the higher the merge workload the higher the delivery time of merged PRs. This is in agreement with the perception of 81\% of participants of our survey. 
The influence of \textit{merge workload} is explained by 
C093, when they mention that a longer delivery time can be due to \textit{``not enough developers and too many pull requests \& issues to manage"}. Contrasting our models, participants rated \textit{PR churn} as one of the most influential codes on delivery time. When asked about examples of merged PRs that took long to be delivered, 
C109 mentioned PRs that have a \textit{``big change for a core function''}.
However, our regression models ($RQ3$) do not rate \textit{code churn} as an influential variable to explain delivery time. Another agreement between our models and participants is that 73.8\% of participants rate \textit{queue rank} as influential to explain the delivery time of merged PRs. This corroborates the results from $RQ3$, which reveals that \textit{queue rank} is influential when explaining delivery time. 

Although the responses from participants confirm the influence of certain variables used in our regression models ($RQ3$), in Figure \ref{fig:factors_impact_on_delivery_time_heating}, we observe that the option \textit{``Does not influence delivery time"} is frequently chosen for many other variables. 6 out of 12 variables (\textit{stacktrace attached}, \textit{description length}, \textit{contributor integration}, \textit{commits per PR}, \textit{comments interval} and \textit{changed files}) were ranked above 40\% (2 being over 50\%), which means that a substantial number of participants is skeptical regarding the influence of such variables on delivery time. This finding supports our regression model's results, except for the contributor integration metric. The contributor integration is the third most influential variable in our models for time periods \textit{before} and \textit{after} the adoption of \textsc{TravisCI}. Our models suggest that if a contributor has their prior PR delivered quickly, their future PRs are more likely to be delivered quickly.

\vspace{06.mm}
\noindent\textbf{When presented with the median delivery time of merged PRs \textit{before} and \textit{after} the adoption of \textsc{TravisCI}, 42.9\% (\nicefrac{140}{326}) of participants state that \textsc{TravisCI} has little influence on the delivery time, attributing the change
in delivery time to other unrelated factors.} 
According to responses to Question \#26 of our survey, 
42.9\% of participants attribute the delivery time of PRs to other factors not directly related to the adoption of \textsc{TravisCI} (i.e., project maintenance, release strategy, and PR characteristics).
As an example, we observe that the median delivery time of PRs in the \textit{rails/rails} project increased from 120 to 184 days \textit{after} the adoption of \textsc{TravisCI}. However, according to 13 participants of the \textit{rails/rails} project, \textit{``this is definitely not related to CI. \textit{rails/rails} has its own release schedules''} (C063). Furthermore, when we presented the data for participants of the \textit{ansible/ansible} project, where the median delivery time increased from 36 to 121 days (\textit{after} the adoption of \textsc{TravisCI}), C355 explained that \textit{``all depends on the project owners decision when to release.''} Additionally, in $RQ5$, we further discuss the factors that participants argue to not being directly related to CI, but can influence the delivery time of merged PRs.

\begin{center}
    \begin{tabular}{|p{.96\columnwidth}|}
        \hline
        \textbf{Summary:}
        \textit{The general perception is that CI influences the delivery time by improving \textit{automation}, the \textit{release process}, and the \textit{project quality}. Furthermore, \textit{code churn} and \textit{merge workload} are the variables with the highest perceived influence on the delivery time of merged PRs.  
        However, when showing specific project data to the respective participants, 42.9\% of participants are skeptical regarding the influence of CI on the delivery time of the merged PRs.} \\
        \textbf{Implications:}
        \textit{Our participants consider that the key benefit of CI is to improve the mechanisms by which project contributions are processed (e.g., facilitating decisions related to PR submissions), without compromising the quality or overloading the reviewers and maintainers of the projects.}
        \\
        \hline
    \end{tabular}
\end{center}
\subsection*{\textbf{\RQfour}}

The following themes are generated by our thematic analysis of the potential causes of delivery delay in merged PRs: \textit{\textbf{PR characteristics}}; \textit{\textbf{project maintenance}}; \textit{\textbf{release process}}; \textit{\textbf{team characteristics}}; \textit{\textbf{contributors}}; \textit{\textbf{testing}} and \textit{\textbf{automation}}. Table \ref{tab:general_factors_impact_ci} shows the frequency of mentions in our participants' responses for each of the codes and themes that influence the delivery time of merged PRs. We describe all themes and the most mentioned codes in the following.

\begin{table}
	\centering
	\caption{Frequency of mentions in the developer responses for each code and theme that impact the delivery time of merged PRs.}
	\begin{tabular}{clcc}
		\hline
		\multirow{2}[4]{*}{\textbf{Theme}} & \multicolumn{1}{c}{\multirow{2}[4]{*}{\textbf{Code}}} & \multicolumn{2}{c}{\textbf{Frequency}} \bigstrut\\
		\cline{3-4}          &       & \multicolumn{1}{p{5.335em}}{\textbf{Frequency per code}} & \multicolumn{1}{p{5.665em}}{\textbf{Frequency per theme}} \bigstrut\\
		\hline
		\multirow{14}[28]{*}{\textbf{PR characteristics}} & Bug fix & 91    & \multirow{14}[28]{*}{330} \bigstrut\\
		\cline{2-3}          & \multicolumn{1}{p{10.085em}}{PR position in the release cycle} & 32    &  \bigstrut\\
		\cline{2-3}          & Change complexity & 86    &  \bigstrut\\
		\cline{2-3}          & PR prioritization & 57    &  \bigstrut\\
		\cline{2-3}          & PR size & 13    &  \bigstrut\\
		\cline{2-3}          & Security fix & 12    &  \bigstrut\\
		\cline{2-3}          & Code quality & 11    &  \bigstrut\\
		\cline{2-3}          & Guideline adherence & 7     &  \bigstrut\\
		\cline{2-3}          & Backward compatibility & 6     &  \bigstrut\\
		\cline{2-3}          & Feature dependence & 4     &  \bigstrut\\
		\cline{2-3}          & Good PR description & 4     &  \bigstrut\\
		\cline{2-3}          & Feature improvement & 4     &  \bigstrut\\
		\cline{2-3}          & Breaking change & 3     &  \bigstrut\\
		\hline
		\multicolumn{1}{c}{\multirow{6}[12]{*}{\textbf{Project maintainance}}} & Maintainers availability & 41    & \multirow{6}[12]{*}{112} \bigstrut\\
		\cline{2-3}          & Maintainers activeness & 34    &  \bigstrut\\
		\cline{2-3}          & Volunteer based & 12    &  \bigstrut\\
		\cline{2-3}          & Interest gauging of maintainers & 11    &  \bigstrut\\
		\cline{2-3}          & \multicolumn{1}{p{10.085em}}{Maintainer  responsiveness} & 7     &  \bigstrut\\
		\cline{2-3}          & Maintainers workload & 7     &  \bigstrut\\
		\hline
		\multicolumn{1}{c}{\multirow{7}[12]{*}{\textbf{Release process}}} & Release cycle & 86    & \multirow{7}[12]{*}{137} \bigstrut\\
		\cline{2-3}          & Automated deployment & 16    &  \bigstrut\\
		\cline{2-3}          & Batching & 11    &  \bigstrut\\
		\cline{2-3}          & Business rules & 10    &  \bigstrut\\
		\cline{2-3}          & Manual release process & 5     & \bigstrut\\
		\cline{2-3}          & Misuse of CI & 5     &  \bigstrut\\
		\cline{2-3}          & \multicolumn{1}{p{10.085em}}{Release early, release often culture} & 4     &  \bigstrut\\
		\hline
		\multirow{4}[10]{*}{\textbf{Team characteristics}} & Team size & 21    & \multirow{4}[10]{*}{40} \bigstrut\\
		\cline{2-3}          & Small project & 13    &  \bigstrut\\
		\cline{2-3}          & Open source & 4     &  \bigstrut\\
		\cline{2-3}          & Paid staff & 2     &  \bigstrut\\
		\hline
		\multirow{3}[10]{*}{\textbf{Contributors}} & Contributor trustworthiness & 4     & \multirow{3}[10]{*}{9} \bigstrut\\
		\cline{2-3}          & Contributor experience & 3     &  \bigstrut\\
		\cline{2-3}          & \multicolumn{1}{p{10.085em}}{Contributor and maintainers relationship} & 2     &  \bigstrut\\
		\hline		
		\multirow{6}[11]{*}{\textbf{Testing}} & Test coverage & 12     & \multirow{6}[11]{*}{35} \bigstrut\\
		\cline{2-3}          & Testing time & 9    &  \bigstrut\\
		\cline{2-3}          & Lacking tests & 7     &  \bigstrut\\
		\cline{2-3}          & Broken tests & 3     &  \bigstrut\\
		\cline{2-3}          & Manual testing & 2     &  \bigstrut\\
		\cline{2-3}          & Build duration & 2     &  \bigstrut\\
		\hline
	\end{tabular}%
	\label{tab:general_factors_impact_ci}%
\end{table}%

\vspace{0.6mm}
\noindent\textit{\textbf{PR Characteristics.\textsuperscript{(330)}}} The {\em characteristics of merged PRs} theme is the theme most mentioned by our participants. According to participants' responses, \textit{PR prioritization}\textsuperscript{(57)} is one of the main factors that can shorten or lengthen the time to deliver merged PRs. The priority of a PR is recurrently associated with \textit{bug fixes}\textsuperscript{(91)} and \textit{security fixes}.\textsuperscript{(12)} According to C020, \textit{``Anything that is considered a critical security fix or major bug fix is generally shipped within 1-2 weeks of submission. This happens frequently.''} In contrast, several participants\textsuperscript{(22)} state that PRs with a longer delivery time are frequently associated with non-urgent features. As stated by C111, \textit{``The most frequent reason [for a longer delivery time] is that the PR is not business-critical.''} In a similar vein, the study by \cite{gousios2015work} found that integrators commonly prioritize contributions by examining their criticality or urgency, e.g., bug fixes or new important features are commonly assigned a higher priority. 

Additionally, the \textit{code quality}\textsuperscript{(11)} of the PR, the \textit{change complexity}\textsuperscript{(10)} and the \textit{guideline adherence}\textsuperscript{(7)} are commonly mentioned codes related to the delivery time of merged PRs.
For example, C168 exemplified that \textit{``adherence to pull request guidelines. Small fix. Clearly defined solution''} are factors that quicken the delivery time of merged PRs. Along the same lines, \cite{gousios2015work} argue that contributions conforming to project style and architecture, source code quality, and test coverage are top priorities for integrators. Finally, the \textit{change complexity}\textsuperscript{(86)} is also mentioned to help PRs to be quickly evaluated and delivered. For instance, C431 states that \textit{``the PR I issued to Crafty was integrated very quickly, mainly because it was a trivial, absolutely non-breaking change.''} The study by \cite{Yu2016-cy} also identified that the complexity of PRs is a factor that influence the PR latency (i.e., the time taken for a PR to be merged). \cite{weissgerber2008small} observed that smaller PRs are more likely to be accepted. Indeed, according to our survey participants, the less complex or the more trivial a PR is, the greater the more likely that the PR will be quickly delivered as less effort is needed. 

\vspace{0.6mm}
\noindent\textit{\textbf{Project maintenance.\textsuperscript{(112)}}} Project maintenance is associated with the project maintainers' activities. When considering the influence of the maintainers on the delivery time of merged PRs, most participants mentioned \textit{maintainers' availability}\textsuperscript{(41)}.
The study by \cite{Yu2016-cy} found that integrators' availability has a significant effect on PR latency. Additionally, our study suggests that \textit{maintainers' availability} influences the delivery of PRs. For instance, C115 considers that a longer delivery time is associated with \textit{``long times between releases, mostly due to maintainer availability''}.
Another important and frequently mentioned cause of delay is that open-source projects are \textit{volunteer based},\textsuperscript{(12)} i.e., contributors are often volunteers \citep{alexander2002working}. For instance, C336 stated that \textit{``OSS projects are staffed by volunteers who come and go and then the priorities of the project shift and some feature become less important''}. 
Furthermore, \textit{maintainers' workload},\textsuperscript{(7)} \textit{maintainer activeness}\textsuperscript{(34)} and \textit{maintainers' engagement}\textsuperscript{(11)} are believed to influence the delivery time of merged PRs. For instance, C419 states that \textit{``if the maintainers of the project are interested, it [the PR] will get processed quickly.''} Previous work also observed that \textit{workload} is a factor that plays a key role in the delivery time of addressed issues of three large open-source projects \citep{daCosta2018impact}, which corroborates our results. The \textit{maintainer responsiveness}\textsuperscript{(7)} also influences the delivery time of merged PRs. C090 states that \textit{``it also helps [to deliver PRs more quickly] if maintainers can be easily contacted (IRC/Slack/Twitter)''}. Furthermore, codes related to project maintenance should be carefully considered in project management, since they might influence not only the delivery time of merged PRs, but also project success. The study by \cite{coelho2017modern} elucidates that lack of maintainers' time and interest are factors that might lead open-source projects to fail. 

\vspace{0.6mm}
\noindent\textit{\textbf{Team Characteristics.\textsuperscript{(40)}}} Team characteristics are also believed to influence the delivery time of merged PRs. Several participants explained that a long delivery time may occur due to 
 \textit{team size}.\textsuperscript{(21)} As stated by C393 \textit{``the team doing reviews were (and is still) understaffed.''} 
The study by \cite{Vasilescu2015-tn} also identified that larger teams can process more PRs (i.e. merge or reject PRs).
This is common in \textit{open-source}\textsuperscript{(4)} projects, as explained by C238 when stating that \textit{``Open source projects tend to delay the publication. Private projects suffer from this problem with much less impact. I wonder if it is due to the lack of a dedicated team in the open-source project, or maybe the focus isn't necessary in the part of the software I contributed.''} \textit{Small project}\textsuperscript{(13)} is also believed to quicken the delivery of merged PRs, as explained by C321, \textit{``on small projects some PRs might be released as a hotfix release very quickly. So I think the speed of delivery is usually in direct proportion to the size of the project.''} 

Overall, the codes related to project characteristics should be carefully observed for projects attempting to decrease the time to deliver their merged PRs. Our participants believe that small projects tend to deliver their PRs more quickly, as they can manage the incoming contributions more easily. With project growth (e.g., an increased number of PRs and project complexity), a small core \textit{team size} can become a bottleneck when delivering merged PRs. In open-source projects, the bottleneck may be exaggerated as projects are volunteer-based \citep{alexander2002working} with most developers working in their free time. To overcome these barriers, projects may consider adopting strategies to deal with the large increase in contributions, as well as to deal with maintainers' inattentiveness, like \textit{transfer the project to new maintainers} or \textit{accept new core developers} \citep{coelho2017modern}. Additionally, adding \textit{paid staff}\textsuperscript{(2)} to the project could be an alternative to deal with the project workload and quicken the delivery time of the merged PRs. For instance, C225 explained the importance of paid staff on the software project by stating the following: \textit{``I have submitted PRs to very large open-source projects, like sklearn or AWS CLI. These projects typically get released frequently on established schedules by maintainers who are, in part, employed to release the projects.''}

\vspace{0.6mm}
\noindent\textit{\textbf{Contributors.\textsuperscript{(9)}}} This theme reflects the potential influence that contributors (i.e., those who submit PRs) have on the delivery time of PRs. The contributors' social status is important when it comes to the delivery of their PRs. When analyzing the time length between the submission and merge of a PR, \cite{Yu2016-cy} found that open-source projects prefer to quickly accept PRs originating from trusted contributors. After PRs are merged, the contributors' social status also influences the time to deliver PRs. For example, \textit{contributor trustworthiness}\textsuperscript{(4)} is often evaluated before a PR is delivered. C421 states that \textit{``after you work with people for a while, you recognize and trust those that have proved to be good at what they do.''} According to our participants, the greater the \textit{contributor experience},\textsuperscript{(3)} the more likely the contributor will have their PR delivered more quickly. 
Furthermore, \cite{soares2018factors} observed that the social relationship between contributors and reviewers influences the evaluation of a PR. In fact, we also identify that the \textit{contributor and maintainer relationship}\textsuperscript{(2)} positively influences the delivery time of merged PRs (i.e., contributors that are socially closer to a core team member have a higher chance to have their PRs delivered more quickly). C403, for example, states that \textit{``some [PRs] were shipped quickly because one of maintainers is my friend then he merged immediately.''}

\vspace{0.6mm}
\noindent\textit{\textbf{Testing.\textsuperscript{(35)}}} The delivery time of merged PRs can also be associated with testing. When asked about factors that might cause a PR to be delayed, our participants listed \textit{testing time},\textsuperscript{(9)} \textit{lacking tests},\textsuperscript{(7)} \textit{broken tests},\textsuperscript{(3)} and \textit{manual testing}.\textsuperscript{(2)} For instance, in relation to testing time, C030 states that \textit{``release time can be quite long but not due to a specific PR, but to an overall review and testing process of a whole software release.''} 
In this respect, reducing manual steps is a must for projects wishing to release code more frequently \citep{neely2013continuous}. Test planning is very important in the Quality Assurance (QA) process. For instance, with automated test suites, the QA team no longer needs to manually execute the tests for the majority of the system, which would be more error-prone and slow \citep{neely2013continuous}. However, beyond the automated test execution, projects must also be concerned with \textit{test coverage}\textsuperscript{(12)}. Automated test execution for projects with low test coverage potentially leads to certain bugs not being identified during build time \citep{felidre2019continuous}. Several participants mention \textit{test coverage}\textsuperscript{(12)} as influencing the delivery time of PRs. C347 declares that \textit{``code that had a large test coverage and small PR are generally deployed safely and quickly.''}
Additionally, C287 mentions issues related to building duration when testing their PRs: \textit{``NixOS nixpkgs stable chanel: CI for deep dependencies take a huge amount of computation time.''} Indeed, keeping the build and test process short (ideally by not taking more than 10 minutes) is one of the prerequisites for CI adoption~\citep{Humble2010-ca}. A long build duration may lead to a set of problems, for example, developers may check-in their code less often, as they have to sit around for a long time while waiting for the build (and tests) to run. 

\vspace{0.6mm}
\noindent\textit{\textbf{Release process.\textsuperscript{(132)}}} The delivery time of merged PRs may also be associated with the release process of projects. The \textit{release cycle}\textsuperscript{(86)} code is the most cited code of this theme. For instance, C029 states that \textit{``code was merged quickly, but had to wait for the test, review, and release cycle to complete. So had to wait for months to see the code I needed released publicly.''} 
Indeed, a shorter release cycle has been mentioned to shorten the time-to-market and quicken the users' feedback loop~\citep{da2016agility}. \cite{daCosta2018impact} also found that traditional release cycles (which are longer cycles) could actually deliver new functionalities more quickly by using minor releases. Therefore, it seems that the higher the frequency of {\em user-intended} releases, the quicker the delivery of merged PRs. 

When explaining potential reasons for the quick delivery of PRs, our participants recurrently mentioned \textit{automated deployment}.\textsuperscript{(16)} For instance, C100 states \textit{``we also implemented continuous deployment so that when a change is merged, it is automatically deployed.''} 
Automated deployment is mandatory for the adoption of continuous deployment (CD). The goal of CD is to automatically deploy every change to the production environment \citep{shahin2017continuous}. In the context of the pull-based development model, CD is said to have a substantial influence on the delivery time of the proposed changes, since each merged PR is automatically deployed.	

Additionally, \textit{batching}\textsuperscript{(11)} and \textit{business rules}\textsuperscript{(10)} are also important codes when it comes to the delivery time of PRs. The \textit{batching} code is associated with the process of queuing up PRs to launch bigger releases. For instance, C092 states that a PR that experiences a long delivery time is linked to the fact that \textit{``Usually they [project maintainers] wait for a good amount of fixes or an important fix like the ones involving security [to launch a release].''} 
However, project managers should be careful about the risks of bigger releases. Usually, big releases are a result of a longer release cycle, which may delay the delivery of PRs for a longer time. In contrast, smaller batch sizes would help the production environment to have fewer defects as smaller code changes may lead to faster feedback from the CI system \citep{neely2013continuous}.
Regarding the business impact on delivery time, C353 declared that \textit{``If the change introduces a new feature that has an impact on end users, the marketing and customer success team need to communicate to them in advance and it takes weeks.''} Indeed, the work of \cite{daCosta2018impact} also found that the delivery time may also be associated with the collaboration with other teams. They also mention that the marketing team is recurrently cited when delays to release occur due to other teams' collaboration. For instance, a PR may be delayed due to the need of aligning the software release with external events for marketing reasons. 
Additionally, the \textit{misuse of CI}\textsuperscript{(5)} is mentioned as a factor that negatively impacts the delivery time of PRs, i.e., C118 states the following: \textit{``I've worked in a project that hadn't CI nor automated tests, so each release took at least one week to be deployed. Once, we had some issues with our deploy system that made us delay the deploy for one month.''}
Lastly, the release culture also impacts delivery time. For example, C090 stated that \textit{``projects with \textit{release early, release often}\textsuperscript{(4)} culture are usually the fastest to deliver.''} Overall, our results suggest that projects should reduce manual processes to quicken the release process and increase the release frequency. The adoption of continuous software engineering practices \citep{shahin2017continuous}, i.e., CI, Continuous Delivery (CDE), and Continuous Deployment (CD), should be considered in this regard. 

\begin{center}
	\begin{tabular}{|p{.96\columnwidth}|}
		\hline
		\textbf{Summary:}
		\textit{The delivery time of merged PRs is impacted by several factors. 87.3\% (\nicefrac{579}{663}) of the mentions associate the delivery time of PRs with their characteristics, the project release process, and project maintenance. According to our survey responses, simple PRs and PRs that fix bugs are delivered more quickly. The PR delivery time is also often linked to the availability of maintainers and the size of the release cycle.} \\
		\textbf{Implications:}
		\textit{Teams that wish to deliver their merged PRs more quickly to their users should also be concerned with other aspects beyond CI, such as encouraging their contributors to submit simple PRs and maintaining short release cycles.}
		\\
		\hline
	\end{tabular}
\end{center}
\subsection*{\textbf{\RQsix}}

We report the results of $RQ6$ in two subsections. First, we report on the release processes of our participants' projects in general (i.e., not considering influences of CI yet). This is important to obtain an overview of the variety of release processes in our data and will provide us with more context when interpreting the influence of CI on release processes. 
Afterward, we show the perceived influence of CI on the release processes of our participants' projects.

\vspace{2mm}
\noindent\textbf{Release processes in general.}
\vspace{2mm}

According to our participants, the release process of their project are \textit{goal oriented},\textsuperscript{(97)} \textit{maintainer oriented},\textsuperscript{(18)} follow a specific \textit{release strategy},\textsuperscript{(94)} such as \textit{continuous delivery},\textsuperscript{(28)} and are driven by \textit{business}\textsuperscript{(18)} and \textit{user demand}.\textsuperscript{(7)} In addition, some participants perceive an \textit{ad hoc}\textsuperscript{(7)} approach to the release process of their project. Table~\ref{tab:freq_citations_project_releasing_process} shows the citation frequency of each theme and code related to the release process of projects.

\begin{table}
	\centering
	\caption{Frequency of themes and codes as captured from our participants' responses.}
	\begin{tabular}{p{8.915em}rrc}
		\hline
		\multirow{2}[4]{*}{\textbf{Theme}} & \multicolumn{1}{c}{\multirow{2}[4]{*}{\textbf{Code}}} & \multicolumn{2}{p{9.33em}}{\textbf{Frequency}} \bigstrut\\
		\cline{3-4}    \multicolumn{1}{c}{} &       & \multicolumn{1}{p{4.665em}}{\textbf{Frequency per Code}} & \multicolumn{1}{p{4.665em}}{\textbf{Frequency per Theme}} \bigstrut\\
		\hline
		\multirow{7}[14]{*}{\textbf{Goal oriented}} & \multicolumn{1}{p{11.915em}}{Code stability} & \multicolumn{1}{c}{28} & \multirow{7}[14]{*}{97} \bigstrut\\
		\cline{2-3}    \multicolumn{1}{c}{} & \multicolumn{1}{p{11.915em}}{New feature} & \multicolumn{1}{c}{22} &  \bigstrut\\
		\cline{2-3}    \multicolumn{1}{c}{} & \multicolumn{1}{p{11.915em}}{Tested code} & \multicolumn{1}{c}{13} &  \bigstrut\\
		\cline{2-3}    \multicolumn{1}{c}{} & \multicolumn{1}{p{11.915em}}{Feature completeness} & \multicolumn{1}{c}{10} &  \bigstrut\\
		\cline{2-3}    \multicolumn{1}{c}{} & \multicolumn{1}{p{11.915em}}{Enough content} & \multicolumn{1}{c}{10} &  \bigstrut\\
		\cline{2-3}    \multicolumn{1}{c}{} & \multicolumn{1}{p{11.915em}}{Project roadmap} & \multicolumn{1}{c}{7} &  \bigstrut\\
		\cline{2-3}    \multicolumn{1}{c}{} & \multicolumn{1}{p{11.915em}}{Project milestone} & \multicolumn{1}{c}{7} &  \bigstrut\\
		\hline
		\multirow{5}[10]{*}{\textbf{Release strategy}} & \multicolumn{1}{p{11.915em}}{Fixed periods} & \multicolumn{1}{c}{49} & \multirow{5}[10]{*}{94} \bigstrut\\
		\cline{2-3}    \multicolumn{1}{c}{} & \multicolumn{1}{p{11.915em}}{Continuous delivery} & \multicolumn{1}{c}{28} &  \bigstrut\\
		\cline{2-3}    \multicolumn{1}{c}{} & \multicolumn{1}{p{11.915em}}{Release schedule} & \multicolumn{1}{c}{10} &  \bigstrut\\
		\cline{2-3}    \multicolumn{1}{c}{} & \multicolumn{1}{p{11.915em}}{Release early, release often practice} & \multicolumn{1}{c}{4} &  \bigstrut\\
		\cline{2-3}    \multicolumn{1}{c}{} & \multicolumn{1}{p{11.915em}}{Time-based release} & \multicolumn{1}{c}{3} &  \bigstrut\\
		\hline
		\textbf{Maintainer oriented} &       &       & 18 \bigstrut\\
		\hline
		\textbf{Business-driven} &       &       & 18 \bigstrut\\
		\hline
		\textbf{User demand} &       &       & 7 \bigstrut\\
		\hline
		\textbf{Ad hoc} &       &       & 7 \bigstrut\\
		\hline
	\end{tabular}%
	\label{tab:freq_citations_project_releasing_process}%
\end{table}%

\textit{Goal oriented}\textsuperscript{(97)} is the theme that emerged the most when it comes to the release process of our participants' projects. An example of a goal that projects strive to achieve is \textit{code stability}.\textsuperscript{(28)} As C370 explains, their project creates a release \textit{``when the API is stable and extensively tested.''} Other participants also explain that their projects produce a release when a desired \textit{new feature} has been developed,\textsuperscript{(22)} the \textit{code has been tested},\textsuperscript{(13)} or \textit{enough content}\textsuperscript{(10)} has been developed to launch a new version of the software. For example, C178 explains that \textit{``it [the release] is usually done when enough new features and patches have been made.''} 
Therefore, we observe that some participants perceive that their projects adopt a more traditional release strategy (as opposed to rapid releases~\citep{Da_Costa2016-cb}) to deliver new versions of their project. This strategy may also be called \textit{feature-based releases}, where a project launches a new release when a set of bug fixes and new features are ready~\citep{michlmayr2015and}. However, {\em feature-based releases} may not be ideal in a volunteer-based open-source project. For example, there is a risk that projects have an unpredictable release schedule, where releases may take a long time to be launched or never happen due to certain features never being completed \citep{michlmayr2015and}.

In a different vein, when explaining the release process of certain projects, our participants perceive that they adopt a more modern \textit{release strategy}.\textsuperscript{(94)} For example, in certain projects, there exist \textit{fixed periods}\textsuperscript{(49)} and a predictable \textit{release schedule}\textsuperscript{(10)} for launching new releases. Other projects use \textit{continuous delivery}\textsuperscript{(28)} or follow the \textit{release early, release often}\textsuperscript{(4)} practice. C032 explains that in the \textit{saltstack/salt} project there is a \textit{fixed period}\textsuperscript{(49)} for its releases, \textit{``feature release every 6 months, bug fix releases when necessary"}. 
Practices such as \textit{release early, release often} are well established in open source development, which leads to benefits related to quality and consistency as errors can be detected sooner \citep{fitzgerald2017continuous}.   

Some participants have recurrently mentioned that the release process of their project is \textit{maintainer oriented},\textsuperscript{(18)} \textit{business driven},\textsuperscript{(18)} or dependent on the \textit{user demand}.\textsuperscript{(7)} We also observe seven citations that state that the process is not clearly defined or is \textit{ad hoc}.\textsuperscript{(7)} For instance, C444 explains that the \textit{bokeh/bokeh} project \textit{``is volunteer based, so [the release is launched] when we can and think it's reasonable.''} C117 also states that the \textit{fog/fog} project produces its release \textit{``when the maintainer decided it was time to do so.''} Additionally, C003 states that the act of launching a release in \textit{grails/grails-core} \textit{``is a decision between business and the development team.''} The release process may also depend on the outreach of the release. For instance, C302 explains that an important factor to produce a release in \textit{boto/boto} is \textit{``when [they] need a wider audience.''}

\begin{center}
	\begin{tabular}{|p{.95\columnwidth}|}
		\hline
	\textit{Our participants perceive different release strategies for their project. Some projects follow a feature-based release strategy, whereas others adopt a time-based release approach (e.g., release based on business needs or user demands).} \\
		\hline
	\end{tabular}
\end{center}

\vspace{2mm}
\noindent\textbf{The influence of CI on release processes.}
\vspace{2mm}

\begin{figure}[H]
	\centering
	\includegraphics[ width=12cm]{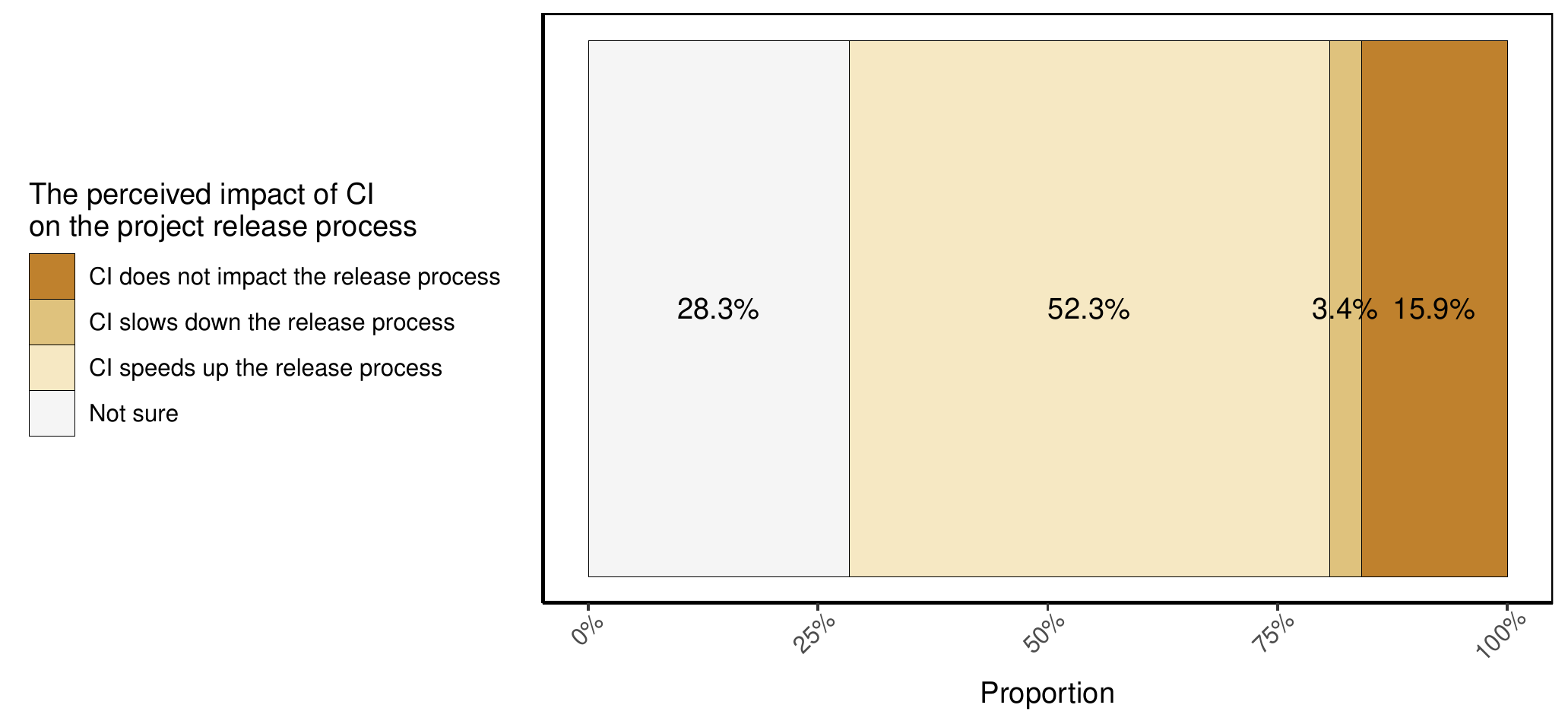}
	\caption{Percentage of the perceived influence of CI on release processes (\textit{Question \#18}).}
	\label{fig:perc_perceived_impact_of_ci_releasing_process}       
\end{figure}

\noindent\textbf{52.3\% (\nicefrac{168}{321})  of the developers perceive an increase in release frequency after the adoption of CI.} 15.9\%  (\nicefrac{51}{321})  of the developers do not perceive any influence from CI on release processes. Also, 3.4\% (\nicefrac{11}{321}) agree that CI leads to a decrease in the number of releases while 28.3\%  (\nicefrac{91}{321}) refrain from stating an opinion (i.e., the participants do not know of or are unsure about an influence of CI on release processes, see Figure \ref{fig:perc_perceived_impact_of_ci_releasing_process}).

\noindent\textbf{CI increases the release frequency by improving \textit{automation},\textsuperscript{(59)} \textit{project stability},\textsuperscript{(47)} and \textit{release characteristics}.\textsuperscript{(13)}} Table \ref{tab:CI_impacts_on_the_releasing_process} shows the frequency of citations identified for each theme related to the use of CI that may impact the project release process. Each theme is described in the following.

\begin{table}
	\centering
	\caption{Frequency of citations for each theme and code related to CI that might impact the project release process.}
	\begin{tabular}{cp{9.25em}cc}
		\hline
		\multicolumn{1}{c}{\multirow{2}[4]{*}{\textbf{Theme}}} & \multirow{2}[4]{*}{\textbf{Code}} & \multicolumn{2}{p{10em}}{\textbf{Frequency}} \bigstrut\\
		\cline{3-4}          & \multicolumn{1}{c}{} & \multicolumn{1}{p{5em}}{\textbf{Frequency per code}} & \multicolumn{1}{p{5em}}{\textbf{Frequency per theme}} \bigstrut\\
		\hline
		\multicolumn{1}{c}{\multirow{6}[12]{*}{\textbf{Automation}}} & Automated testing & 24    & \multirow{6}[12]{*}{59} \bigstrut\\
		\cline{2-3}          & Release automation & 17    &  \bigstrut\\
		\cline{2-3}          & Earlier feedback & 11    &  \bigstrut\\
		\cline{2-3}          & Easier to produce a release & 3     &  \bigstrut\\
		\cline{2-3}          & Automated building & 2     &  \bigstrut\\
		\cline{2-3}          & Earlier integration & 2     &  \bigstrut\\
		\hline
		\multicolumn{1}{c}{\multirow{4}[8]{*}{\textbf{Project Stability}}} & Confidence & 35    & \multirow{4}[8]{*}{47} \bigstrut\\
		\cline{2-3}          & Code stability & 6     &  \bigstrut\\
		\cline{2-3}          & Releasable master & 3     &  \bigstrut\\
		\cline{2-3}          & Less regressions & 3     &  \bigstrut\\
		\hline
		\multicolumn{1}{c}{\multirow{4}[8]{*}{\textbf{Release characteristics}}} & Minor releases & 5     & \multirow{4}[8]{*}{13} \bigstrut\\
		\cline{2-3}          & Smaller releases & 3     &  \bigstrut\\
		\cline{2-3}          & Bug fix releases & 3     &  \bigstrut\\
		\cline{2-3}          & Security releases & 2     &  \bigstrut\\
		\hline
	\end{tabular}%
	\label{tab:CI_impacts_on_the_releasing_process}%
\end{table}%

\vspace{1mm}
\noindent\textbf{Automation.\textsuperscript{(59)}} In the previous subsection, we identified {\em automation} as one of the themes that may quicken the time to deliver merged PRs. In this subsection, we identify that 59 of our participants draw a relationship between {\em automation} improvements brought by CI and the increase in release frequency. For example, \textit{automated tests}\textsuperscript{(24)} and \textit{release automation}\textsuperscript{(17)} are frequently cited when participants explain the increase in release frequency. As explained by C070, \textit{``The testing becomes much simpler and automated, thus it takes less time to validate a release.''} C79 complements the previous answer when they state that \textit{``CI helps with automated tests, so we can merge and release faster.''} Furthermore, when discussing the benefits of CI in relation to release automation, C252 states: \textit{``we can do release often with CI. It's automated.''} 
CI is seen as a continuous deployment enabler, as explained by C327: \textit{``with CI, you can increase the frequency because CI can also deploy automatically.''} 

\vspace{1mm}
\noindent\textbf{Project stability.\textsuperscript{(47)}} The release process of projects is impacted by project stability. The most cited code related to project stability is \textit{confidence}.\textsuperscript{(35)} For instance, C084 states that \textit{``with the confidence gained from CI jobs, maintainers can reduce their time with testing tasks and focus on releases/new features.''}  C056 also states that \textit{``CI allows to keep a releasable master-branch at all times and allows external parties to rely on the quality of master.''} The feedback from our participants reveals that \textit{code stability}\textsuperscript{(35)} allows releases to be prepared more easily. 
C331 expresses that, because CI reduces the number of blocking issues, creating a release becomes easier: \textit{``it [CI] in general contributes to fewer bugs so there are less blocking issues so it is easier to follow a release schedule.''}

\vspace{1mm}
\noindent\textbf{Release characteristics.\textsuperscript{(13)}} The adoption of CI is also mentioned to have changed certain characteristics of releases, which might lead to a higher frequency of releases over time. Several participants perceive that the adoption of CI started a trend of \textit{smaller}\textsuperscript{(3)} and \textit{minor releases}.\textsuperscript{(5)} Participant C203 from \textit{chef/chef} explains the following: \textit{``Yes it [CI] made releases and deployments much more frequent. It removed a lot of manual testing effort and validation. It also changed the culture of the team in such a way that people delivered changes in a smaller and more incremental way.''} Additionally,  participants also perceived an increase in \textit{bugfix}\textsuperscript{(3)} and \textit{security fix}\textsuperscript{(2)} releases. 

15.9\% (\nicefrac{51}{321})  of participants do not perceive any influence from CI on the release process of their projects. According to such participants, the release frequency is \textit{maintainer oriented}\textsuperscript{(2)}, depends strictly on the project \textit{release policy}\textsuperscript{(1)} or depends on the \textit{project maturity}.\textsuperscript{(1)} According to participants, instead of influencing the release frequency, CI influences the \textit{merge time}\textsuperscript{(1)}, \textit{quickens the testing process}\textsuperscript{(1)}, and provides \textit{better quality}\textsuperscript{(5)}. For instance, C317 states that \textit{``CI may help add more features. It should not increase the frequency of releases essentially. It is a matter of policy I think.''} Additionally, C110 expresses: \textit{``It looks like it depends on wish of repository owners and their plan''}. On another note, C169 perceives that the increase in release frequency is a side-effect of the maturation of a project, which can occur due to the adoption of CI: \textit{``Only insofar as adoption of CI indicates the professionalization of a project, which can be correlated with more frequent releases.''} 

Finally, only 3.4\%  (\nicefrac{11}{321}) of participants perceive that CI decreases the release frequency in their projects. According to 11 participants, the decrease in release frequency can be related to the influence of CI on code \textit{stability}\textsuperscript{(2)} and \textit{quality}.\textsuperscript{(2)} For instance, C094 explains that in \textit{ipython/ipython} \textit{``The releases are bit less frequent. The use of CI makes the code more tested and lowers a need for bugfix-minor releases.''} Additionally, C257 declares that \textit{``CI made the number of releases smaller and less frequent, because it helped to catch errors in the code during review. While this made releasing slower, it made the quality of those releases much better.''}  

\begin{center}
\begin{tabular}{|p{.96\columnwidth}|}
    \hline
    \textbf{Summary:}
    \textit{52.3\% (\nicefrac{168}{321})  of participants perceive an increase in the release frequency after the adoption of CI. This increase is related to improvements brought by CI, such as better \textit{automation}, \textit{project stability}, and changes in \textit{release characteristics} (e.g., smaller releases). Furthermore, only 3.4\%  (\nicefrac{11}{321}) of participants agree that CI decreases the release frequency of their projects.} \\
    \textbf{Implications:}
    \textit{Teams planning to improve their release process should consider the adoption of CI, which will not always release more often. However, the automation provided by CI fosters more confidence in releasing the software.}
    \\
    \hline
\end{tabular}
\end{center}
\subsection*{\textbf{\RQseven}}

We first request our participants to detail the review process of their project before we gauge their perceptions about how CI may influence the code review process.
	Our participants highlight that the projects have specific \textit{review strategies}\textsuperscript{(184)} that are followed. Examples of these strategies are \textit{peer review}\textsuperscript{(109)} and \textit{expert review}.\textsuperscript{(11)} Additionally, review processes have to be mindful of \textit{quality assurance metrics}.\textsuperscript{(130)} Table \ref{tab:projects_reviewing_process} shows the frequency of codes generated from the responses of our participants.

	\begin{table}
		\centering
		\caption{Frequency of citations for each theme and code related to review processes.}
		\begin{tabular}{cp{13.0em}cc}
			\hline
			\multicolumn{1}{c}{\multirow{2}[4]{*}{\textbf{Theme}}} & \multirow{2}[4]{*}{\textbf{Code}} & \multicolumn{2}{p{10em}}{\textbf{Frequency}} \bigstrut\\
			\cline{3-4}          & \multicolumn{1}{c}{} & \multicolumn{1}{p{5em}}{\textbf{Frequency per code}} & \multicolumn{1}{p{5em}}{\textbf{Frequency per theme}} \bigstrut\\
			\hline
			\multicolumn{1}{c}{\multirow{9}[18]{*}{\parbox{3cm}{\centering \textbf{Project review strategy}}}} & Peer review & 109   & \multirow{9}[18]{*}{184} \bigstrut\\
			\cline{2-3}          & GitHub standard review & 37    &  \bigstrut\\
			\cline{2-3}          & Expert review & 11    &  \bigstrut\\
			\cline{2-3}          & Ad-hoc code review process & 9     &  \bigstrut\\
			\cline{2-3}          & Review checklist & 5     &  \bigstrut\\
			\cline{2-3}          & Development branch review & 5     &  \bigstrut\\
			\cline{2-3}          & Project goals & 3     &  \bigstrut\\
			\cline{2-3}          & Mailing list discussion & 3     &  \bigstrut\\
			\cline{2-3}          & Pair programming & 2     &  \bigstrut\\

			\hline
			\multicolumn{1}{c}{\multirow{11}[22]{*}{\parbox{3cm}{\centering \textbf{Quality assurance metrics}}}} & CI check & 44    & \multirow{11}[22]{*}{130} \bigstrut\\
			\cline{2-3}          & Tests verification & 38    &  \bigstrut\\
			\cline{2-3}          & Code style & 20    &  \bigstrut\\
			\cline{2-3}          & Proper documentation & 7     &  \bigstrut\\
			\cline{2-3}          & Linter check & 6     &  \bigstrut\\
			\cline{2-3}          & Security & 3     &  \bigstrut\\
			\cline{2-3}          & Data coverage check & 3     &  \bigstrut\\
			\cline{2-3}          & Efficiency check & 3     &  \bigstrut\\
			\cline{2-3}          & Avoiding conflicts & 2     &  \bigstrut\\
			\cline{2-3}          & Code line inspection & 2     &  \bigstrut\\
			\cline{2-3}          & Error detection & 2     &  \bigstrut\\
			\hline
			NA    &     &     & 92 \bigstrut\\
			\hline
		\end{tabular}%
		\label{tab:projects_reviewing_process}%
	\end{table}%

	\textit{Peer review}\textsuperscript{(109)} and \textit{CI check}\textsuperscript{(44)} are the most frequently mentioned codes. 
	\textit{Peer review} is the process of manually checking violations of code standards and logical errors in a patch submitted by developers~\citep{rahman2017impact}. In the existing literature, peer review has been demonstrated to be effective for improving design quality and the overall quality of software projects \citep{rahman2017impact}. 
	In a similar vein, many quotes\textsuperscript{(109)} from our participants highlight the use of peer review in their projects. For example, C420 states that in the \textit{dropwizard} project, \textit{``All changes are reviewed by one or two peers depending on self-assessed complexity of the change''}.
	Other developers state that their projects follow the \textit{GitHub standard review}\textsuperscript{(37)} flow. For instance, C308 states: \textit{``We use GitHub flow, open PR, require review(s), review with suggestions or concerns, discuss and revise if needed and review again, then merge.''} 

	We also identify more specific review strategies in some projects, e.g., \textit{review checklist},\textsuperscript{(5)} \textit{development branch review},\textsuperscript{(5)} \textit{project goals},\textsuperscript{(3)} \textit{mailing list discussion}\textsuperscript{(3)} and \textit{pair programming}.\textsuperscript{(2)} For instance, when talking about the \textit{review checklist}, C203 states that \textit{``Code review occurs when a PR is opened. We have a checklist for both the reviewer and committer to go through for each change. Tests run on each PR and the tests need to pass before a change can be merged.''} Additionally, the submitted PR should be aligned with the project goals, as explained by 
	C129 when declaring that \textit{``The reviewer should read and understand all of the changes, and the changes should be in-line with the project's conventions and goals.''}	We also receive 9 responses in which participants do not identify a specific review process in their projects. These participants state that their projects follow an \textit{ad-hoc code review process}.

	The \textit{quality assurance metrics}\textsuperscript{(130)} theme has also been frequently cited by our participants when explaining the review process of their projects. Indeed, improving the quality of patches to software projects is one of the main motivators of modern code review \citep{bacchelli2013expectations, bavota2015four}. When observing quality assurance metrics, reviewers often use the results of \textit{CI checks}\textsuperscript{(44)} to ensure the quality of submitted PRs. Other verifications are frequently mentioned by participants when checking code quality, e.g., \textit{tests verification},\textsuperscript{(38)} \textit{code style},\textsuperscript{(20)} and \textit{proper documentation}.\textsuperscript{(7)} Project maintainers, in general, rely on automated tools to support the process of code review \citep{Vasilescu2015-tn}. CI is often used in popular open-source projects to check whether the PR breaks the build. Moreover, CI verifies whether the tests pass and automatically checks whether the PR matches the project style guide \citep{cassee2020silent}. According to C160, \textit{``If CI is green and PR looks sane, merge.''} Additionally, C408 explains that, in their project, the review process also has to \textit{``check CI jobs (lint, test, build).''} 

	\begin{center}
		\begin{tabular}{|p{.95\columnwidth}|}
			\hline
			\textit{Our participants perceive that their projects have specific code review strategies (e.g., peer review and expert review). Moreover, the review process of our participants' projects rely on quality assurance verification (e.g., CI check, code style, proper documentation).}\\
			\hline
		\end{tabular}
	\end{center}

\vspace{2mm}
\noindent\textbf{The perceived impact of CI on the project review process}
\vspace{2mm}

	\textit{\textbf{Most of our participants' quotes (58\%, \nicefrac{335}{578}) agree that CI influences the review process of software projects.}} Among the 578 quotes related to the influence of CI on the code review process, we observe that the majority (58\%, \nicefrac{335}{578}) state that CI has some influence on the code review process. 15\% (\nicefrac{87}{578}) of quotes state that CI has no influence on code review, while 27\% (\nicefrac{156}{578}) of quotes did not express a clear position.

	\textit{Automation}\textsuperscript{(139)} is the most cited code when it comes to how CI influences code review. As expressed by C100, \textit{``It made it [the code review process] more efficient because the amount of manual testing that needs to happen reduced a lot. It also democratized the process, so the whole team is able to get started doing reviews.''} 
	Indeed, automated processes (as brought by CI) are often combined with manual code reviews made by the quality assurance team \citep{rahman2017impact}. The infrastructure of CI is frequently used with automated builds and quality checks, involving static analysis tools and automated testing  \citep{zampetti2019study}. In that respect, our participants argue that CI influences code reviews because reviewers enjoy a \textit{better focus}\textsuperscript{(25)} during the review, e.g., a better focus on the code logic, security, and design. C352 elaborates on this focus when stating the following: \textit{``We try to use linters/style checkers to remove the style nit part of the code review process. It means we spent more time thinking about the architecture and logic vs. the formatting''}.

	In addition, reviewers can have a better focus on specific checks because of the \textit{PR filtering}\textsuperscript{(51)} process promoted by the use of CI. The process of PR filtering reduces the reviewers' burden by filtering out PRs that break the build. For instance, 
	C059 highlights that \textit{``we do not even start reviews before CI is green.''} The study by \cite{zampetti2019study} reveals that PRs with green builds have slightly more chances to get merged than broken builds, although other process-related factors have a stronger correlation with the merge process. Table \ref{tab:impact_of_CI_on_reviewing_process} shows the complete list of codes related to how CI influences code review processes.

	\begin{table}
		\centering
		\caption{Frequency of citations for each theme and code related to the impact of CI on review processes.}
		\begin{tabular}{p{11.0em}p{12.5em}cc}
			\hline
			\multirow{2}[4]{*}{\textbf{Theme}} & \multirow{2}[4]{*}{\textbf{Code}} & \multicolumn{2}{p{11.25em}}{\textbf{Frequency}} \bigstrut\\
			\cline{3-4}    \multicolumn{1}{c}{} & \multicolumn{1}{c}{} & \multicolumn{1}{p{5.915em}}{\textbf{Frequency per code}} & \multicolumn{1}{p{5.335em}}{\textbf{Frequency per theme}} \bigstrut\\
			\hline
			\multirow{2}[2]{*}{\parbox{3cm}{\centering \textbf{CI does not influence code review}}} & \multirow{2}[2]{*}{} & \multirow{2}[2]{*}{} & \multirow{2}[2]{*}{87} \bigstrut[t]\\
			\multicolumn{1}{c}{} & \multicolumn{1}{c}{} &       &  \bigstrut[b]\\
			\hline
			\multirow{11}[22]{*}{\parbox{3cm}{\centering \textbf{CI impacts code review}}} & Automation & 139   & \multirow{11}[22]{*}{335} \bigstrut\\
			\cline{2-3}    \multicolumn{1}{c}{} & PR filtering & 51    &  \bigstrut\\
			\cline{2-3}    \multicolumn{1}{c}{} & Higher confidence & 43    &  \bigstrut\\
			\cline{2-3}    \multicolumn{1}{c}{} & Better focus & 25    &  \bigstrut\\
			\cline{2-3}    \multicolumn{1}{c}{} & Faster review & 15    &  \bigstrut\\
			\cline{2-3}    \multicolumn{1}{c}{} & Earlier feedback & 38    &  \bigstrut\\
			\cline{2-3}    \multicolumn{1}{c}{} & Less review workload & 7     &  \bigstrut\\
			\cline{2-3}    \multicolumn{1}{c}{} & Regression identification & 4     &  \bigstrut\\
			\cline{2-3}    \multicolumn{1}{c}{} & Easier to reference failure & 3     &  \bigstrut\\
			\cline{2-3}    \multicolumn{1}{c}{} & Improved testability & 4     &  \bigstrut\\
			\cline{2-3}    \multicolumn{1}{c}{} & Smaller PR granularity & 6     &  \bigstrut\\
			\hline
			{\centering NA}    &     &    & 156 \bigstrut\\
			\hline
		\end{tabular}%
		\label{tab:impact_of_CI_on_reviewing_process}%
	\end{table}%

	\textit{Higher confidence}\textsuperscript{(43)} is recurrently cited when it comes to how CI influences code review.  
	C370 argues that, with CI, there is a \textit{``reduced time of the code review because you are more confident the PR works and focus more on code quality during the review than checking the logic.''} Another important point offered by CI within the reviewer tasks is the \textit{earlier feedback}\textsuperscript{(26)} of the proposed PRs. 
	C316 argues that CI \textit{``increased the speed [of code review], as certain issues are pointed out immediately''}.

	\textit{\textbf{Although most of the quotes related to RQ7 agree that CI influences code review, still, 15\% (\nicefrac{87}{578}) of quotes state that CI has no influence on code review processes.}} 
	For instance, C215 perceives that CI influences the release process of software projects, but does not influence code review, \textit{``I think CI does not have much to do with code review speed. Reviewers only receive one bit of information from CI, and they are expected to look at the code carefully, not the CI results. But the project manager tends to be more confident in releasing new versions with CI enabled''}.

	\begin{center}
		\begin{tabular}{|p{.96\columnwidth}|}
			\hline
			\textbf{Summary:}
			\textit{Most participants agree that CI influences code review processes. 58\% (\nicefrac{335}{578}) of the quotes related to CI and code review agree that CI has an impact on code review. According to participants, automation is a key aspect of CI that speeds up code review.}\\
			\textbf{Implications:}
			\textit{CI may quicken the process of sorting which PRs are worth reviewing, e.g., a PR with green build status, which may improve the decision-making process of software projects.}
			\\
			\hline
		\end{tabular}
	\end{center}
\subsection*{\textbf{\RQeight}}
	
	\noindent\textbf{59\% (\nicefrac{227}{383}) of quotes in RQ8 argue that projects using CI are more attractive to receive external contributors.} The most recurrent themes in RQ8 reveal that projects using CI have more  \textit{attractive characteristics}\textsuperscript{(73)} (e.g., easier PR acceptance) and a \textit{lower contribution barrier}.\textsuperscript{(154)}
	Conversely, 26\% (\nicefrac{98}{383}) of the quotes argue that CI does not influence the number of project contributors, attributing the increase of contributors to other factors, such as \textit{project growth},\textsuperscript{(35)} \textit{maturity}\textsuperscript{(14)} and \textit{popularity}.\textsuperscript{(22)} The remaining 15\% (\nicefrac{58}{383}) of quotes refer to answers, such as ``No Answer (NA)''. Table \ref{tab:ci_impact_on_attracting_contributors} shows the complete list of themes and codes related to the influence of CI on attracting contributors to open-source projects.
	
	\begin{table}
		\centering
		\caption{Frequency of citations per theme and code related to the influence of CI on attracting contributors to open-source projects.}
    \begin{tabular}{cccc}
    \hline
    \multicolumn{1}{c}{\multirow{2}[4]{*}{\textbf{Theme}}} & \multicolumn{1}{c}{\multirow{2}[4]{*}{\textbf{Code}}} & \multicolumn{2}{p{10em}}{\textbf{Frequency}} \bigstrut\\
\cline{3-4}          &       & \multicolumn{1}{p{5em}}{\textbf{Frequency per code}} & \multicolumn{1}{p{5em}}{\textbf{Frequency per theme}} \bigstrut\\
    \hline
    \multicolumn{1}{c}{\multirow{11}[22]{*}{\parbox{3.2cm}{\centering \textbf{Attractive project characteristics with CI}}}} & \multicolumn{1}{p{14em}}{Reduced review effort/time} & 22    & \multirow{11}[22]{*}{73} \bigstrut\\
\cline{2-3}          & \multicolumn{1}{p{14em}}{Clear development process} & 12    &  \bigstrut\\
\cline{2-3}          & \multicolumn{1}{p{14em}}{Project quality} & 7     &  \bigstrut\\
\cline{2-3}          & \multicolumn{1}{p{14em}}{Project stability} & 7     &  \bigstrut\\
\cline{2-3}          & \multicolumn{1}{p{14em}}{Easier PR acceptance} & 7     &  \bigstrut\\
\cline{2-3}          & \multicolumn{1}{p{14em}}{Actively maintained project} & 5     &  \bigstrut\\
\cline{2-3}          & \multicolumn{1}{p{14em}}{Faster delivery} & 4     &  \bigstrut\\
\cline{2-3}          & \multicolumn{1}{p{14em}}{Welcoming for contributions} & 3     &  \bigstrut\\
\cline{2-3}          & \multicolumn{1}{p{14em}}{Regular releases} & 2     &  \bigstrut\\
\cline{2-3}          & \multicolumn{1}{p{14em}}{Projects with CI seems more mature} & 2     &  \bigstrut\\
\cline{2-3}          & \multicolumn{1}{p{14em}}{Best industrial practices followed} & 2     &  \bigstrut\\
    \hline
    \multicolumn{1}{c}{\multirow{5}[10]{*}{\parbox{3.2cm}{\centering \textbf{Lower contribution barrier with CI}}}} & \multicolumn{1}{p{14em}}{CI confidence} & 60    & \multirow{5}[10]{*}{154} \bigstrut\\
\cline{2-3}          & \multicolumn{1}{p{14em}}{Build status awareness} & 57    &  \bigstrut\\
\cline{3-3}          & \multicolumn{1}{p{14em}}{Build and test automation} & 28    &  \bigstrut\\
\cline{2-3}          & \multicolumn{1}{p{14em}}{Engagement to contribute} & 6     &  \bigstrut\\
\cline{2-3}          & \multicolumn{1}{p{14em}}{Lowered entry barrier} & 3     &  \bigstrut\\
    \hline
    \multicolumn{1}{c}{\multirow{5}[10]{*}{\textbf{Not related to CI}}} & \multicolumn{1}{p{14em}}{Project growth} & 35    & \multirow{5}[10]{*}{98} \bigstrut\\
\cline{2-3}          & \multicolumn{1}{p{14em}}{Project popularity} & 22    &  \bigstrut\\
\cline{2-3}          & \multicolumn{1}{p{14em}}{Project maturity} & 14    &  \bigstrut\\
\cline{2-3}          & \multicolumn{1}{p{14em}}{Project activeness} & 3     &  \bigstrut\\
\cline{2-3}          & \multicolumn{1}{p{14em}}{Non-causal correlation} & 24    &  \bigstrut\\
    \hline
    NA    &       &       & 58 \bigstrut\\
    \hline
    \end{tabular}%
		\label{tab:ci_impact_on_attracting_contributors}%
	\end{table}%

\vspace{1mm}
\noindent\textbf{Attractive project characteristics.\textsuperscript{(73)}}
Several quotes state that developers feel more attracted to the characteristics of projects that use CI. They argue that, when using CI, projects tend to have a \textit{reduced review effort/time},\textsuperscript{(22)} which may lead to \textit{PRs being accepted} more easily.\textsuperscript{(7)} As explained by C033, \textit{``It's easier to handle incoming changes from people, so it's possible to have more of them.''} 
Additionally, the \textit{project quality}\textsuperscript{(7)} and \textit{stability}\textsuperscript{(7)} of the projects are frequently mentioned by our participants as a consequence of using CI, which may motivate developers to contribute more. In this regard, 
C178 declares that \textit{``People want to work in high-quality projects. CI is a mark of quality.''} According to our participants, potential contributors also look for stability in projects and CI provides this sense of stability. As explained by C051,
\textit{``Projects that use CI tend to look more stable and serious. It might be a reason for some contributors to be attracted to more serious projects.''}
 Finally, potential contributors prefer projects that follow industry best practices. In this regard, C355 states: \textit{``Maybe because the project looks more professional, following \textit{best industrial practices}.\textsuperscript{(2)} I would not contribute to a project without CI, or I would setup CI first''}.

\vspace{1mm}
\noindent\textbf{Lower contribution barrier.\textsuperscript{(154)}}
The majority of quotes stating that CI attracts more contributors explain that this is due to CI projects having a lower contribution barrier. This lower barrier promotes an \textit{increased confidence}\textsuperscript{(60)} in contributors when a project is using CI. The quotes also argue that CI promotes a \textit{better build status awareness}\textsuperscript{(57)} for contributors. Regarding confidence, 
C146 declares that \textit{``Developers like CI, it adds confidence to your work and it is pleasurable to work in this highly structured and coordinated way.''} These observations corroborate the study by \cite{coelho2017modern}, which also found CI as one of the most important maintenance practices in top open-source projects.	
Regarding CI promoting a better build status awareness, C084 states: \textit{``In my case, I like to have my changeset reviewed and tested soon as possible, and CI jobs are really fast for that.''} 
Moreover, several participants argue that it is \textit{easier to contribute} when a project uses CI. In this matter, 
C100 declares that \textit{``As a contributor not part of the core team, CI makes it easier to understand the code, because you can look at the tests to understand the design. It makes it easier to contribute without a lot of prior knowledge of the project''}.

\vspace{1mm}
\noindent\textbf{Although most quotes agree that CI attracts more contributors, 26\% (\nicefrac{98}{383}) of quotes state that there is no causal relationship between CI and the increase in contributors.} Many participants argue that there is a \textit{non-causal relationship}\textsuperscript{(24)} between the increase in contributors and the adoption of CI. For instance, C352 declares that \textit{``I think that people adopt CI because contributions become difficult to manage due to increasing quantity (probably driven by popularity). I would imagine these two variables are not causally related but simply correlated."}.
Indeed, many of our participants attribute the increase in contributors to \textit{project growth}\textsuperscript{(35)}, \textit{maturity},\textsuperscript{(14)} and \textit{popularity}.\textsuperscript{(22)} 
For instance, C423 states that the increase in contributors is \textit{``probably not related [to CI], but accidentally relates to a hype curve and or maturity level of the project.''} Indeed, the study by \cite{Hilton2016-xy} shows that popular projects are more likely to use CI. \cite{Hilton2016-xy} also found that the first CI build in their investigated projects occurred around 1 year (median) after the project creation. They argue that this is the case because the adoption of CI may not always provide a large amount of value during the very initial phases of the development of a project.

\begin{center}
	\begin{tabular}{|p{.95\columnwidth}|}
		\hline
		\textbf{Summary:}
		\textit{
		Although most quotes from participants (59\%, \nicefrac{227}{383}) argue that projects using CI are more attractive to potential contributors, 26\% (\nicefrac{98}{383}) of quotes argue that there is no causal relationship between CI and the increase in contributors, i.e., other factors play a role instead, such as project growth and increase in project popularity over time.
		}\\
		\textbf{Implications:}
		\textit{Open-source projects intending to attract and retain external contributors should consider the use of CI in their pipeline since CI is perceived to lower the contribution barrier while making contributors feel more confident and engaged in the project.}
		\\
		\hline
	\end{tabular}
\end{center}
\section{Discussion}\label{sec_discussion}

In this section, we outline the implications of our results
to both research and practice in software engineering. 

{\em\bfseries Using a Continuous integration service is not a silver bullet.} Through our
quantitative analyses, we observe that a CI service does not always
reduce the time for delivering merged PRs 
to end users. In fact, analyzing 87 projects, we observe that
only 51\% of the projects deliver merged PRs more quickly \textit{after} the adoption of \textsc{TravisCI} (Section \ref{sec_quantitative_study_results} - $RQ1$). 
Additionally, our qualitative study reveals that there is no consensus regarding the impact of CI on the delivery time of merged PRs.
42.9\% of our participants declared that CI does not have impact on the delivery time of merged PRs (Section \ref{sec_quantitative_study_results} - $RQ4$), instead, factors such as \textit{project release process}, \textit{project maintenance} and \textit{PR characteristics} (i.e., \textit{trivial PRs}) are believed to influence the delivery time of merged PRs (Section \ref{sec_quantitative_study_results} - $RQ5$). 
If the decision to adopt CI is mostly driven by the goal of quickening the
delivery of merged PRs~\citep{Laukkanen2015-ab}, such a decision must be more
carefully considered by development teams. Finally, previous research suggests
that the adoption of CI increases the release frequency of a software
project~\cite{Hilton2016-xy}. However, we did not observe such an increase in
our quantitative analyses (Section \ref{sec_quantitative_study_results} - $RQ2$). Our study only considers user-intended releases, so we do not consider pre, beta, alpha, and rc (release candidate) releases in our analyses. It might be the case that when considering only established releases, the release frequency does not statically increases \textit{after} the adoption of \textsc{TravisCI}.

{\em\bfseries If CI is a CD enabler, why is CD seemingly rare?} CI, Continuous Delivery (CDE), and Continuous Deployment (CD) are complementary practices that can be used in the agile releasing engineering environment \citep{Karvonen201787}. CDE is a practice that automates the software delivery process, and it is often considered to extend CI. Therefore, a project that uses CDE, the delivery can occur at any time, with little manual effort required. Furthermore, CD is a step further from the adoption of CDE, which is a practice where projects release each successful build to end users automatically. 
In this context, several participants of our study consider that 
\textit{automated deployment} is a subsequent step in CI adoption that can help projects rapidly deliver software changes to end users (e.g., CD can deliver merged PRs automatically). However, through the analysis of 9,312 open-source projects that use \textsc{TravisCI}, the study by \cite{gallaba2018use} found that explicit deployment code is rare (2\%), which suggests that developers rarely use \textsc{TravisCI} to implement Continuous Deployment. An interesting future study is to better understand the gap between CI and CD as well as how to bridge this gap.
Furthermore, we observe that \textit{before} the adoption of \textsc{TravisCI}, the merge workload is the most influential variable to model the delivery time of PRs, while \textit{after} the adoption of \textsc{TravisCI}, the most influential variable is the moment at which a PR is merged in the release cycle (i.e., queue rank metric). One possible reason for the change in most influential variables in the time periods \textit{before} and \textit{after} \textsc{TravisCI}, is that \textit{after} the adoption of \textsc{TravisCI}, the merge workload could have been better managed, leading the queue rank to be more influential on the delivery time of merged PRs. This indicates that the delivery time of merged PRs is more dependent on when the PR was merged in the release cycle than whether the project adopts a CI service. Therefore, projects that wish to quicken the delivery of merged PRs need to foster the culture of frequent release instead of solely relying on the adoption of a CI service (i.e., Travis CI) in their pipeline.

{\em\bfseries Automation and confidence are key aspects for the throughput generated by CI.} 
We observe that the adoption of a CI service is associated with many benefits, such as a higher number of contributors, PR submissions, and a higher PR churn per release (Section \ref{sec_quantitative_study_results} - $RQ2$). However, the release frequency is roughly the same as \textit{before} using \textsc{TravisCI}. Therefore, teams that wish to adopt a CI service should be aware that their projects will not always deliver merged PRs more quickly or release them more often, but that the pivotal benefit of a CI service is the ability to process substantially more contributions in a given time frame, which is closely tied to the automation and confidence that release managers (Section \ref{sec_qualitative_study_results} - $RQ6$), reviewers (Section \ref{sec_qualitative_study_results} - $RQ7$), and external contributors (Section \ref{sec_qualitative_study_results} - $RQ8$) feel towards their codebase and project environment. 

{\em\bfseries CI may improve the decision-making process of software projects.} Our results (Section \ref{sec_qualitative_study_results} - $RQ7$) reveal that most contributors' quotes (58\%, \nicefrac{87}{578}) agree that CI impacts the time required to review PRs. According to our participants, CI may quicken the process of sorting which PRs are worth reviewing, e.g., a PRs with green builds. However, project maintainers should be concerned with the test coverage in their project, as it is essential for reviewers to be more confidence in the CI feedback to submitted PRs. We observe that contributors whose previously submitted PRs were merged and delivered quickly, are also likely to have their future PRs delivered quickly (\ref{sec_quantitative_study_results} - $RQ3$).  Hence, we recommend that the first PR submissions of a new contributor should be carefully crafted to maintain a successful track record in their projects (so they can build trust, causing their future PRs to be delivered more quickly).
An interesting future work would be to investigate how CI can influence the decision-making process involved in different development tasks, i.e., from requirements engineering to project delivery~\citep{Sharma2021influence}. 

\section{Threats to the validity and Limitations}\label{sec_threats_to_the_validity} 

In this section, we discuss the threats to the validity and limitations of our studies.

\subsection{\textbf{Threats to Validity -- quantitative study}}

\textit{\textbf{Construct Validity.}} The threats to construct validity are concerned with the extent to which the
operational measures in the study really represent what researchers intended to measure. We define the delivery time of a PR as the time between when a PR is merged and the moment at which a PR is delivered through a release. However, the way we link PRs to releases may not match the actual number of delivered PRs per release. For instance, if a version control system of a project has the following release tags \textit{v1.0, v2.0, no-ver, and v3.0}, we
remove the \textit{no-ver} tag. If there are PRs associated with the
\textit{no-ver release}, such PRs will be associated with release
\textit{v3.0}. However, only 5.36\% (\nicefrac{403}{7519}) of our studied
releases are affected by this bias.

\noindent\textit{\textbf{Internal Validity.}} Internal threats are concerned with the
ability to draw conclusions from the relationship between the dependent
variable (delivery time of merged PRs) and independent variables (e.g.,
release commits and queue rank). 
Regarding our models, we acknowledge that our independent variables
are not exhaustive. Although our models achieve sound $R^2$ values, other variables may
be used to improve performance (e.g., a \textit{boolean} indicating whether a PR
is associated with an issue report and another \textit{boolean} that verifies whether a PR was submitted by a core developer or an external contributor). Nevertheless, our set of independent variables
should be approached as a starting set that can be easily computed rather than a final solution.

At the time the quantitative data was extracted from GitHub, our projects had a median of 5.1 years of life (2 without using \textsc{TravisCI} and 3.1 using \textsc{TravisCI}). As we extracted data from the entire lifetime of the projects, we collected more PRs in the \textit{after-\textsc{TravisCI}} time period. However, the Mann-Whitney-Wilcoxon (MWW) test and Cliff's delta effect size, which were the statistical methods used to perform the comparisons, are suitable for groups of different sizes \citep{mann_whitney_1947}. 
Furthermore, we are aware that factors not related to the adoption of \textsc{TravisCI} may have impacted the delivery time of merged PRs as well (i.e., project maturity and team size). We argue that this concern was alleviated by the conceptual replication of our study conducted by \cite{guo2019studying}. First, they replicated our study using the same subject projects and methodology. Then, they addressed the same research question of our study, by using the Regression Discontinuity Design (RDD), which allows verifying whether there is a trend of PR delivery times over time and whether this trend changes significantly when \textsc{TravisCI} is introduced. Finally, they introduced a control group of comparable projects that never used \textsc{TravisCI}. They found that the results of our quantitative study largely hold in their replication study.

\noindent\textit{\textbf{External Validity.}} External threats are concerned with the extent to which we can generalize our results \citep{Perry:2000:ESS:336512.336586}. 
In this work, we analyzed 162,653 PRs of 87 popular open-source projects from GitHub. All projects adopt \textsc{TravisCI} as their CI service. 
We acknowledge that we cannot generalize our results to any other projects with similar or different settings (e.g., private software projects). Nevertheless, in order to achieve more generalizable results, replications of our study in different settings are required. 
For replication purposes, we publicize our datasets and results to the interested researcher.\footnote{\url{https://prdeliverydelay.github.io/\#datasets}}  

\subsection{\textbf{Limitation of our qualitative study}}

The main limitation of our qualitative study is the thematic analysis of the survey responses. The first author coded all responses to the 13 open-ended questions of the survey, and the second author coded a sub-set of 10\% of the responses to each question, which was selected randomly. Although the systematic analytical process was used to analyze the data, the credibility of the findings is enhanced if more than one researcher completely analyzes the data. Nevertheless, we used the Cohen's Kappa test to verify the agreement level of the two authors when coding the answers. Given that we achieved an almost perfect level of agreement between them (median Kappa's value of $0.84$), we believe that the coding process has been consistent. 

While we achieved theoretical saturation concerning the responses to the research questions of our qualitative study, we are aware that the self-selection of our participants may have biased our results. From the subset of 3,105 individuals, we received 450 responses to the survey (14.5\% response rate). Therefore, the contributors that did not respond to our survey invitation may have different views on the questions, which could lead us to different results.
Additionally, the participants that responded to our survey may have been affected by social desirability bias, which is the tendency to answer questions in a way that will be seen as advantageous by others (i.e., participants responding according to what they think the ``correct'' answer should be, making themselves and their software development look better than it actually is).

In our qualitative study, we investigated the perception of CI practitioners regarding the influence of CI on the delivery time of PRs, and its potential influence on the review and release processes of software projects. The participants of our qualitative study are contributors from 73 out of the 87 GitHub projects studied in our quantitative analysis. 
However, we cannot measure the degree to which the studied projects use CI. Hence, selecting participants of projects that use a CI service, but not necessarily CI as a whole practice, may bias our analysis. This is because we may have received responses from participants that did not fully witness the benefits of CI when it is fully implemented.  
Nevertheless, 73.2\% of the participants reported that they used CI in 60--100\% of their projects, which suggests that most of our participants have a varied experience with CI. Another issue is that we did not distinguish our participants by core and external contributors. 
Had we considered the responses from different types of contributors separately, different conclusions could have been drawn. On the other hand, we were able to collect a diverse set of participants, i.e., while 77.8\% of the participants have web development as one of their main activities, 52\% of participants consider code review as one of their main activities, and 30.9\% are involved in project management (see 
Figure \ref{fig:main_development_activities}), which provides us with insightful feedback.

\section{Conclusion}
\label{sec_conclusions}

Our work consists of two studies that quantitatively and qualitatively investigate the influence of a CI service (e.g., \textsc{TravisCI}), and CI as a practice, on the time-to-delivery of merged PRs, respectively. In our quantitative study, we analyze 162,653 PRs of 87 GitHub projects to understand the factors that influence (and improve) the delivery time of merged PRs. In our qualitative study, we analyze 450 survey responses from participants of 73 projects (out of the initial 87 projects). We investigate the perceived influence of CI on the delivery time of merged PRs. We also study the perceived influence of CI on the code review and release processes.

As a key takeaway, our studies demonstrate that the adoption of \textsc{TravisCI}
will not necessarily deliver or merge PRs more quickly. Instead, the pivotal benefit of a CI service is to improve the mechanisms by which contributions to projects are processed (e.g., facilitating decisions on PR submissions), without compromising the quality of the project or overloading developers. The automation provided by CI and the boost in developers' confidence are key aspects of using CI. For instance, CI may help the process of sorting which PRs are worth reviewing (e.g., PRs with green builds).
Furthermore, open-source projects wishing to attract and retain external contributors should consider the use of CI in their pipeline, since CI is perceived to lower the contribution barrier while making contributors feel more confident and engaged in the project.

\section*{Acknowledgments}
\label{sec_Acknowledgments}

This work is partially supported by INES (\url{www.ines.org.br}), CNPq grants 465614/2014-0 and 425211/2018-5, CAPES grant 88887.136410/2017-00, FACEPE grants APQ-0399-1.03/17, and PRONEX APQ/0388-1.03/14.

\section*{Data Availability}
For replication purposes, we publicize our datasets and results to the interested researcher: \url{https://prdeliverydelay.github.io/#datasets}

\section*{Declarations}

\textbf{Conflict of interests}. The authors declare that they have no conflict of interest.

\appendix
	
	\appendixtitleon
	\appendixtitletocon
	\begin{appendices}

	

## Appendix A  Project Survey Example

## Understanding Delivery Time of Pull-Requests

Once a pull request is merged (i.e., reviewed, tested and ready to be delivered), this pull request may still wait a long time to be delivered to end users of a software project. We call this time delivery time (see Figure 1). In this study, we intend to understand the factors that may accelerate or delay the delivery of merged pull requests. Our research team is from Brazil, New Zealand, and Australia.

This questionnaire is part of broad research effort that studied 87 large open-source projects to investigate the factors that most impact the time that is necessary to deliver merged pull requests to end users of a software project through an official release.

The survey will ask you a few general and specific questions (i.e., we will show you data that is derived from haraka/Haraka project).

Figure 1 - The basic life-cycle of a pull request.

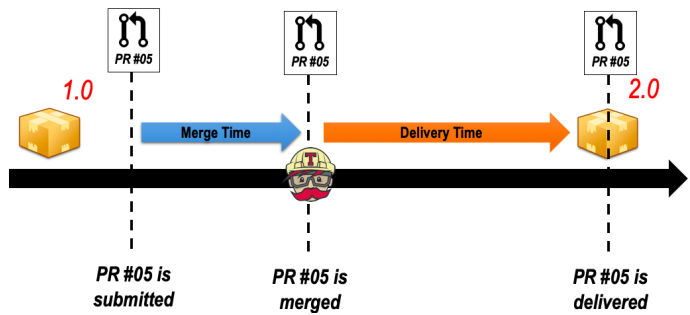

1. By answering this survey, you will be entered into a draw of six $50 Amazon gift cards. If you'd like to participate in the draw, please leave your email address below.

_____

2. Have we correctly identified you as someone who contributes or contributed to a project that uses Continuous Integration? *

*Marcar apenas uma oval.*

○ Yes

○ No    *Pular para a pergunta 31*

## Demographic Questions

3. For how long have you been developing software?

*Marcar apenas uma oval.*

|       | 0 | 1 | 2 | 3 | 4 | 5 | 6 | 7 | 8 | 9 | 10 |         |
|-------|---|---|---|---|---|---|---|---|---|---|----|---------|
| Years | ○ | ○ | ○ | ○ | ○ | ○ | ○ | ○ | ○ | ○ | ○  | or more |

4. Out of your projects which use Continuous Integration (CI), which project have you most contributed over the last years?

_____

5. What is the ratio of projects that you have worked on that used CI?

*Marcar apenas uma oval.*

○ 0 - 20%

○ 20 - 40%

○ 40 - 60%

○ 60 - 80%

○ 80 - 100%

6. For how long have you been developing projects that use CI?

*Marcar apenas uma oval.*

|       | 0 | 1 | 2 | 3 | 4 | 5 | 6 | 7 | 8 | 9 | 10 |         |
|-------|---|---|---|---|---|---|---|---|---|---|----|---------|
| Years | ○ | ○ | ○ | ○ | ○ | ○ | ○ | ○ | ○ | ○ | ○  | or more |

7. What are your main activities on the projects you contribute to?

*Marque todas que se aplicam.*

- [ ] Test
- [ ] Review
- [ ] Development
- [ ] Management

Outro: [ ] _____

8. In which domains of software development have you worked?

*Marque todas que se aplicam.*

- [ ] Mobile applications
- [ ] Web development
- [ ] Scientific development
- [ ] Business development
- [ ] Medical devices
- [ ] Industrial and process control
- [ ] Embedded systems
- [ ] Big Data
- [ ] Defense systems

Outro: [ ] _____

**Perceptions about delivery time of pull-requests**

9. Can you tell us about a pull request that you submitted or reviewed and that eventually got merged, but took a long time to be effectively shipped in an official release? What were the reasons for this delay?

_____

10. Also, can you tell us about a pull request that you submitted or reviewed (and that eventually got merged), which was quickly shipped in an official release? What were the reasons for such a quick shipment?

_____

11. According to your experience, do you think that the adoption of Continuous Integration (CI) may have some influence on the required time to deliver merged pull requests to end users? If so, how does CI influence this required time?

_____

12. To what extent do you agree with the following statement: The adoption of Continuous Integration shortens the time to deliver merged pull requests to end users?

*Marcar apenas uma oval.*

◯ Strongly agree

◯ Agree

◯ Neutral

◯ Disagree

◯ Strongly disagree

13. To what extent do you agree that the factors listed in the table below influence the time necessary to release a merged pull request that is waiting to be delivered?

*Marcar apenas uma oval por linha.*

| | Makes delivery a lot faster | Makes delivery a little bit faster | Does not influence delivery time | Makes delivery a little bit slower | Makes delivery a lot slower |
|---|---|---|---|---|---|
| The contributor experience | ◯ | ◯ | ◯ | ◯ | ◯ |
| The time that previously submitted PRs of a contributor were delivered | ◯ | ◯ | ◯ | ◯ | ◯ |
| A PR has a stacktrace attached | ◯ | ◯ | ◯ | ◯ | ◯ |
| The description size of a PR | ◯ | ◯ | ◯ | ◯ | ◯ |
| The time at which a pull request is merged during a release cycle | ◯ | ◯ | ◯ | ◯ | ◯ |
| The number of PRs waiting to be delivered at the moment that a new PR was merged | ◯ | ◯ | ◯ | ◯ | ◯ |
| The number of comments recorded on a PR | ◯ | ◯ | ◯ | ◯ | ◯ |
| The time interval between the PR comments | ◯ | ◯ | ◯ | ◯ | ◯ |
| The number of files attached to a PR | ◯ | ◯ | ◯ | ◯ | ◯ |
| Number of lines | ◯ | ◯ | ◯ | ◯ | ◯ |

| | | | | | |
|---|---|---|---|---|---|
| of code in a PR | | | | | |
| Number of commits in a PR | ○ | ○ | ○ | ○ | ○ |
| The time that a pull request waited to be merged | ○ | ○ | ○ | ○ | ○ |

| | |
|---|---|
| Specific questions for the project haraka/Haraka | In the next questions of this survey we will show specific data of the project haraka/Haraka that you contribute to. We group such data into two buckets: before and after the adoption of CI. We considered that the project started using CI on the date of its first build on Travis-CI (2014-03-25). Also, we used the GitHub API to collect data of the haraka/Haraka project from its creation on GitHub on March, 10th, 2011 to November 11th, 2016. |

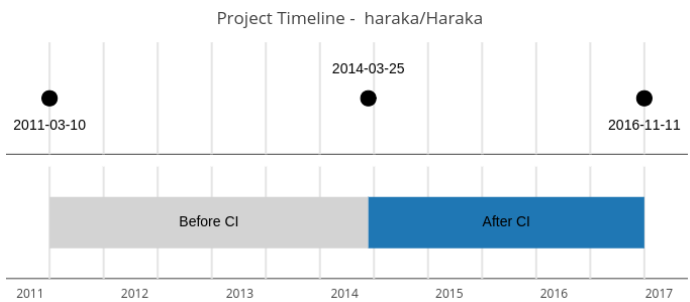

14. Does the way we divided the project timeline into "Before CI" and "After CI" make sense to you?

_____

15. For how long have you worked in the haraka/Haraka project?

*Marcar apenas uma oval.*

|       | 0 | 1 | 2 | 3 | 4 | 5 | 6 | 7 | 8 | 9 | 10 |         |
|-------|---|---|---|---|---|---|---|---|---|---|----|---------|
| Years | ○ | ○ | ○ | ○ | ○ | ○ | ○ | ○ | ○ | ○ | ○  | or more |

16. Did you participate in the adoption of CI in haraka/Haraka?

*Marcar apenas uma oval.*

○ Yes

○ No

17. What is the releasing process/policy of your team? For example, how do you decide when to ship a release of your project?

_____

18. In your opinion, did the adoption of continuous integration have any influence on the frequency of releases? Could you describe this influence to us?

_____

19. When analyzing the data of the project haraka/Haraka, we observed that this project tended to ship 18 releases per year before CI, while it decreased to 6 per year after CI. How would you explain this change?

_____

20. We quantitatively studied 87 popular projects on GitHub and we observed that, after adopting continuous integration, these projects delivered 3 times as many pull requests compared to before. How would you explain this change?

_____

21. In our study, we also observed that, after adopting CI, the projects tended to receive contributions from more developers than before adopting CI. How would you explain this difference?

_____

22. Could you briefly describe to us the code review process adopted by your project?

_____

23. Did the adoption of Continuous Integration influence the code review process of your project? If so, how?

_____

24. In your opinion, did the adoption of Continuous Integration increase or decrease the speed of code reviews? Why?

_____

25. When analyzing the data of the project you contributed to (haraka/Haraka), we observed that before the adoption of travis-CI the submitted pull requests take a median of 3 hours to be merged while this time increased to 8 hours after CI. Do you think the adoption of this CI service had any influence on the time to review and merge a pull request? If so, how?

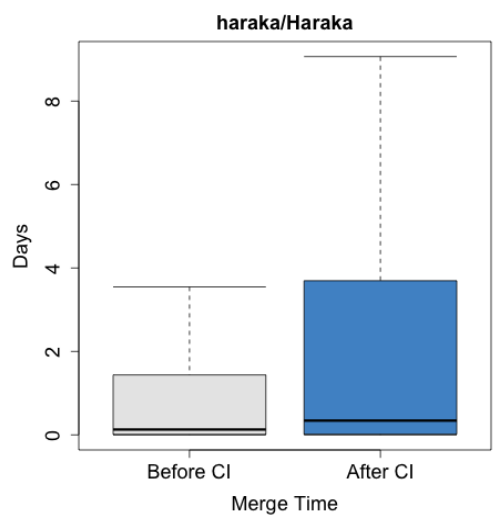

26. According to the data we collected for the project haraka/Haraka, once a pull request is reviewed and merged, this pull request still waited a median of 19 days to be shipped to end users before CI, while after CI this number increases to 68 days. Do these numbers make sense to you? Why?

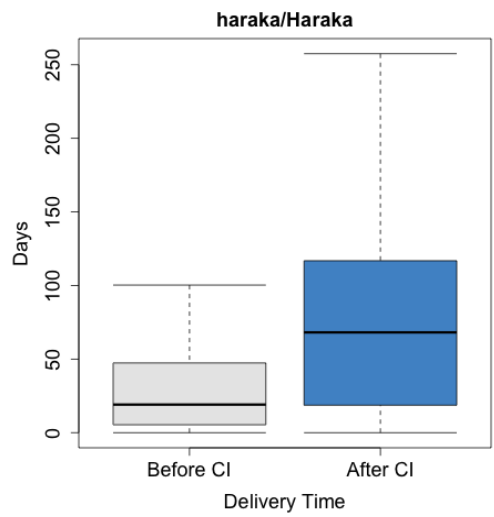

_____

**Ending our questionnaire**

27. Would you like to be informed about our findings?

*Marcar apenas uma oval.*

◯ Yes
◯ No

28. Would you be willing to be contacted for a quick online follow-up interview (at a time convenient for you)?

*Marcar apenas uma oval.*

◯ Yes

◯ No

29. If you checked yes on one of the two questions above and have not yet left your e-mail, please leave it below.

_____

30. Do you have further comments for us?

_____

| Sorry, we are working on our questionnaire respondents selection algorithim | As we are seeking to understand the impact of adopting continuous integration on the delivery time of pull-requests, we are looking for respondents that have used CI in some part of their career. We are trying to improve our questionnaire respondents selection algorithm. You do not have to answer the rest of the questions. Thank you. |
|---|---|

3. If you'd like to be informed about our findings and have not yet left your e-mail, please leave it below.

_____

	
	
	\section{Invitation Letter}
	\label{sec:appendix_invitation_latter}
	
	{\fontfamily{lmss}\selectfont

	\noindent
	MAIL SUBJECT: \textbf{Why do PRs take so long to be delivered? Research survey}
	\vspace{0.3cm} 
	
	\noindent
	Dear \textbf{\${contributor.name}},	
	\vspace{0.3cm} 

	\noindent
	We are a group of researchers from universities based in Brazil, Australia, and New Zealand. We are studying the impact of Continuous Integration on the time to release merged pull requests to end users of open source projects.
	\vspace{0.3cm} 
	
	\noindent
	We have collected public data from the project \textbf{\${project.fullName}} in the period from 
	
	\noindent
	\textbf{\${project.creationDate}} to 2016-11-11. According to our data, you have contributed \textbf{\${contributor.deliveredPRsCount}} pull-requests to \textbf{\${project.fullName}} which were effectively merged and delivered to end users. 
	\vspace{0.3cm} 
	
	\noindent
	As you were a contributor of the project \textbf{\${project.fullName}}, we would appreciate if you shared your experience with us by answering a few questions in the following survey: 
	\vspace{0.3cm} 
	
	\noindent
	Google Form: \textcolor{blue}{Understanding Delivery Time of Pull Requests}
	\vspace{0.3cm} 
	
	\noindent
	The survey has 24 questions (all of them are optional) and will take less than 15 minutes to complete. To compensate you for your time, all participants that answer all questions will be entered into a draw of six \$50 Amazon gift cards.
	\vspace{0.3cm} 
	
	\noindent
	Best Regards,
	\vspace{0.3cm} 
	
	\noindent
	Jo\~{a}o Helis Bernardo.
	
	\noindent
	PhD student at the Federal University of Rio Grande do Norte, Brazil.		
	}
	\newpage
	
	\section{Number of participants per project}
	\label{sec:appendix_participants_ids_and_their_projects}
	
	The number of participants per project are distributed in Tables \ref{tab:participants_per_project_part_i} and \ref{tab:participants_per_project_part_ii}.

	\begin{table*}[htb]
	\centering
	\caption{Number of participants per project and their IDs (PART I)}
	\begin{tabular}{|c|c|c|c|}
		\hline
		\textbf{\#} & \multicolumn{1}{c|}{\textbf{project}} & \textbf{participants IDs } & \textbf{total of participants} \bigstrut\\
		\hline
		1     & grails/grails-core & C001 -- C005 & 5 \bigstrut\\
		\hline
		2     & saltstack/salt & C006 -- C035 & 30 \bigstrut\\
		\hline
		3     & mozilla-b2g/gaia & C036 -- C042 & 7 \bigstrut\\
		\hline
		4     & rails/rails & C043 -- C066 & 24 \bigstrut\\
		\hline
		5     & owncloud/core & C067 -- C079 & 13 \bigstrut\\
		\hline
		6     & cakephp/cakephp & C080 -- C092 & 13 \bigstrut\\
		\hline
		7     & ipython/ipython & C093 -- C097 & 5 \bigstrut\\
		\hline
		8     & ansible/ansible & C098 -- C113 & 16 \bigstrut\\
		\hline
		9     & fog/fog & C114 -- C125 & 12 \bigstrut\\
		\hline
		10    & appcelerator/titanium\_mobile & C126  & 1 \bigstrut\\
		\hline
		11    & TryGhost/Ghost & C127 -- C133 & 7 \bigstrut\\
		\hline
		12    & mozilla/pdf.js & C134 -- C139 & 6 \bigstrut\\
		\hline
		13    & elastic/kibana & C140 -- C143 & 4 \bigstrut\\
		\hline
		14    & AnalyticalGraphicsInc/cesium & C144 -- C147 & 4 \bigstrut\\
		\hline
		15    & twbs/bootstrap & C148 -- C150 & 3 \bigstrut\\
		\hline
		16    & sympy/sympy & C151 -- C157 & 7 \bigstrut\\
		\hline
		17    & matplotlib/matplotlib & C158 -- C169 & 12 \bigstrut\\
		\hline
		18    & scipy/scipy & C170 -- C185 & 16 \bigstrut\\
		\hline
		19    & divio/django-cms & C186 -- C191 & 6 \bigstrut\\
		\hline
		20    & woocommerce/woocommerce & C192 -- C201 & 10 \bigstrut\\
		\hline
		21    & chef/chef & C202 -- C206 & 5 \bigstrut\\
		\hline
		22    & puppetlabs/puppet & C207 -- C211 & 5 \bigstrut\\
		\hline
		23    & Theano/Theano & C212 -- C217 & 6 \bigstrut\\
		\hline
		24    & frappe/erpnext & C218 -- C221 & 4 \bigstrut\\
		\hline
		25    & scikit-learn/scikit-learn & C222 -- C228 & 7 \bigstrut\\
		\hline
		26    & callemall/material-ui & C229 -- C231 & 3 \bigstrut\\
		\hline
		27    & zurb/foundation-sites & C232 -- C240 & 9 \bigstrut\\
		\hline
		28    & laravel/laravel & C241 -- C243 & 3 \bigstrut\\
		\hline
		29    & Leaflet/Leaflet & C244 -- C251 & 8 \bigstrut\\
		\hline
		30    & BabylonJS/Babylon.js & C252 -- C254 & 3 \bigstrut\\
		\hline
	\end{tabular}%
	\label{tab:participants_per_project_part_i}%
\end{table*}%

	\begin{table}[H]
	\centering
	\caption{Number of participants per project and their IDs (PART II)}
	\begin{tabular}{|c|c|c|c|}
		\hline
		\textbf{\#} & \multicolumn{1}{c|}{\textbf{project}} & \textbf{participants IDs } & \textbf{total of participants} \bigstrut\\				
		31    & HabitRPG/habitica & C255 -- C258 & 4 \bigstrut\\
		\hline
		32    & hapijs/hapi & C259 -- C263 & 5 \bigstrut\\
		\hline
		33    & getsentry/sentry & C264 -- C266 & 3 \bigstrut\\
		\hline
		34    & elastic/logstash & C267 -- C268 & 2 \bigstrut\\
		\hline
		35    & kivy/kivy & C269 -- C278 & 10 \bigstrut\\
		\hline
		36    & apereo/cas & C279 -- C283 & 5 \bigstrut\\
		\hline
		37    & jashkenas/underscore & C284  & 1 \bigstrut\\
		\hline
		38    & ether/etherpad-lite & C285 -- C289 & 5 \bigstrut\\
		\hline
		39    & mantl/mantl & C290  & 1 \bigstrut\\
		\hline
		40    & Pylons/pyramid & C291 -- C298 & 8 \bigstrut\\
		\hline			
		41    & boto/boto & C299 -- C309 & 11 \bigstrut\\
		\hline
		42    & request/request & C310  & 1 \bigstrut\\
		\hline
		43    & jhipster/generator-jhipster & C311 -- C318 & 8 \bigstrut\\
		\hline
		44    & refinery/refinerycms & C319 -- C321 & 3 \bigstrut\\
		\hline
		45    & Netflix/Hystrix & C322 -- C323 & 2 \bigstrut\\
		\hline
		46    & square/picasso & C324  & 1 \bigstrut\\
		\hline
		47    & humhub/humhub & C325 -- C326 & 2 \bigstrut\\
		\hline
		48    & bundler/bundler & C327 -- C329 & 3 \bigstrut\\
		\hline
		49    & isagalaev/highlight.js & C330 -- C338 & 9 \bigstrut\\
		\hline
		50    & haraka/Haraka & C339 -- C342 & 4 \bigstrut\\
		\hline
		51    & ReactiveX/RxJava & C343 -- C345 & 3 \bigstrut\\
		\hline
		52    & andypetrella/spark-notebook & C346 -- C348 & 3 \bigstrut\\
		\hline
		53    & TelescopeJS/Telescope & C349 -- C351 & 3 \bigstrut\\
		\hline
		54    & robolectric/robolectric & C352 -- C357 & 6 \bigstrut\\
		\hline
		55    & fchollet/keras & C358 -- C362 & 5 \bigstrut\\
		\hline
		56    & photonstorm/phaser & C363 -- C370 & 8 \bigstrut\\
		\hline
		57    & siacs/Conversations & C371 -- C373 & 3 \bigstrut\\
		\hline
		58    & jsbin/jsbin & C374 -- C381 & 8 \bigstrut\\
		\hline
		59    & buildbot/buildbot & C382 -- C385 & 4 \bigstrut\\
		\hline
		60    & cython/cython & C386 -- C390 & 5 \bigstrut\\
		\hline
		61    & spinnaker/spinnaker & C391  & 1 \bigstrut\\
		\hline
		62    & openhab/openhab & C392 -- C399 & 8 \bigstrut\\
		\hline
		63    & jashkenas/backbone & C400 -- C408 & 9 \bigstrut\\
		\hline
		64    & aframevr/aframe & C409 -- C413 & 5 \bigstrut\\
		\hline
		65    & androidannotations/androidannotations & C414 -- C415 & 2 \bigstrut\\
		\hline
		66    & dropwizard/dropwizard & C416 -- C423 & 8 \bigstrut\\
		\hline
		67    & scikit-image/scikit-image & C424 -- C425 & 2 \bigstrut\\
		\hline
		68    & invoiceninja/invoiceninja & C426 -- C430 & 5 \bigstrut\\
		\hline
		69    & craftyjs/Crafty & C431 -- C433 & 3 \bigstrut\\
		\hline
		70    & serverless/serverless & C434  & 1 \bigstrut\\
		\hline
		71    & bokeh/bokeh & C435 -- C444 & 10 \bigstrut\\
		\hline
		72    & vanilla/vanilla & C445 -- C448 & 4 \bigstrut\\
		\hline
		73    & Yelp/mrjob & C449 -- C450 & 2 \bigstrut\\
		\hline
	\end{tabular}%
	\label{tab:participants_per_project_part_ii}%
\end{table}%

\end{appendices}

\bibliographystyle{spbasic}      
	
\bibliography{references}
	
	
\end{document}